\documentclass[aps,floatfix,twocolumn]{revtex4-2}
\usepackage{bm}
\usepackage{graphicx}
\usepackage{siunitx}
\usepackage[final]{microtype}
\usepackage{amsmath, amsfonts, amssymb}
\usepackage{xcolor}
\usepackage{braket}
\usepackage[normalem]{ulem}
\usepackage{soul}
\usepackage{afterpage}

\DeclareSIUnit\wn{\raiseto{-1}\cm}

\begin{document}

\title{A model for nuclear spin product-state distributions \\ 
of ultracold chemical reactions in magnetic fields}

\author{Goulven Qu{\'e}m{\'e}ner}
\affiliation{Universit\'{e} Paris-Saclay, CNRS, Laboratoire Aim\'{e} Cotton, 91405, Orsay, France}
\author{Ming-Guang Hu, Yu Liu, Matthew A. Nichols, Lingbang Zhu, Kang-Kuen Ni}
\affiliation{Department of Chemistry and Chemical Biology, Harvard University, Cambridge, Massachusetts, 02138,  USA}
\affiliation{Department of Physics, Harvard University, Cambridge, Massachusetts, 02138, USA}
\affiliation{Harvard-MIT Center for Ultracold Atoms, Cambridge, Massachusetts, 02138, USA}

\begin{abstract}
Based on a theoretical model where the nuclear spins remain unchanged during 
a collision, we provide an analytical and general 
expression for the nuclear spin state-to-state distribution 
of an ultracold chemical reaction in a magnetic field, 
for given rotational transitions of the molecules.
It simply requires knowledge of the field-dependent eigenfunctions 
of the molecular reactants and products of the chemical reaction.
The final state-to-state distribution drastically changes with the magnetic field. 
When the distribution is summed over all the final products,
a simplified expression is found 
where only the knowledge of the eigenfunctions of the molecular reactants is required.
The present theoretical formalism has been successfully used to explain the magnetic field behavior
of the product-state distribution in chemical reactions of ultracold KRb molecules
[Hu et al., Nat. Chem. 13, 435 (2021)].
\end{abstract}

\maketitle

\section{Introduction}

The fate of a chemical reaction is a fascinating subject \cite{Herschbach_FDCS_55_233_1973,Herschbach_ACIE_26_1221_1987,Troe_JCSFT_90_2303_1994,Light_DFS_44_14_1967,Zare_S_279_1875_1998}.
It is governed by a potential energy surface and it 
tells about the influence of the numerous 
and complicated, many-body electronic interactions 
when the atoms of the molecules are close to each other, in the region where chemistry prevails.
Ultracold molecules can be used to probe such 
chemical reactions with an unprecedented control at the quantum level
\cite{Krems_PCCP_10_4079_2008,Carr_NJP_11_055049_2009,Quemener_CR_112_4949_2012,Balakrishnan_JCP_145_150901_2016,
Bohn_S_357_1002_2017}.
All the fragments of an ultracold chemical reaction 
can now be observed by ionization spectroscopy and
velocity-map imaging \cite{Hu_S_366_1111_2019,Wolf_S_358_921_2017,Liu_PCCP_22_4861_2020,Liu_NP_16_1132_2020},
from reactants to products, including intermediate complexes.
Molecules possess electronic, vibrational, rotational and spin degrees of freedom
and the way they end up in a chemical reaction via the re-arrangement of the atoms
is complicated.
Full-dimensional potential energy surfaces of heavy tetra-atomic systems 
have recently started to become available \cite{Christianen_JCP_150_064106_2019,Yang_JPCL_11_2605_2020,Klos_SR_11_10598_2021}.
While the dynamics of tri-atomic systems such as ultracold atom-diatom reactions
is numerically tractable using time-independent formalisms, either for heavy systems  \cite{Soldan_PRL_89_153201_2002,Cvitas_PRL_94_033201_2005,Quemener_PRA_71_032722_2005,Quemener_PRA_75_050701_2007,Croft_NC_8_15897_2017,Croft_PRA_96_062707_2017}
and for light systems \cite{Balakrishnan_CPL_341_652_2001},
or using time-dependent formalisms for light systems \cite{Huang_PRL_120_143401_2018},
tetra-atomic systems are more chalenging.
Time-dependent collisional codes based on Jacobi coordinates
are now performant enough to reach the ultracold collision energy regime for light diatom-diatom systems
\cite{Huang_JPCL_11_8560_2020} but are not yet capable of describing heavy systems.
Similarly, time-independent collisional codes for the dynamics of heavy alkali 
diatom-diatom reactions based on hyperspherical coordinates are still lacking. 
Therefore, a full quantum treatment of all degrees of freedom for these heavy alkali systems
is for the moment impractical.
However, a much simpler statistical model \cite{Gonzalez-Martinez_PRA_90_052716_2014,Bonnet_JCP_152_084117_2020}
can shed light on the rotational state-to-state distribution of the products. 
This has been observed in a recent experiment \cite{Liu_N_593_379_2021} where the overall 
agreement indicates the global 
statistical nature of the chemical reaction when the released final kinetic energies of the products
are larger than the ultra-low, initial kinetic energy.  \\

Besides the rotational structure, 
another important point to investigate is the role of the hyperfine structure 
in ultracold collisions \cite{Simoni_LP_16_707_2006,Tscherbul_PRA_75_033416_2007,Gonzalez-Martinez_PRA_84_052706_2011,Tscherbul_PRR_2_013117_2020}
and
up to which point the nuclear spin degrees of freedom are linked to the remaining ones in a chemical reaction
\cite{Quack_MP_34_477_1977}. 
This is the scope of the present paper.
Despite being generally weak compared to the other type of interactions that occur in a chemical reaction, interactions involving the nuclear spins are important in ortho/para conversion of molecules 
\cite{Quack_MP_34_477_1977,Cordonnier_JCP_113_3181_2000,Ren_JCP_125_151102_2006,Lique_IRPC_33_125_2014}.
A recent experimental study showed that they mainly act as spectators
in chemical reactions of ultracold bi-alkali molecules in magnetic fields
\cite{Hu_NC_13_435_2021}, leading to selected values of the rotation parities
of the molecular products.
In that study, the theoretical model 
used to explain the experimental data focused on the specific type of molecule employed
in the experiment and on the specific initial quantum state that was prepared.
Here we provide a complete analytic generalization of the theoretical model. 
The model is mainly based on the knowledge of the eigenfunctions of the reactants and the products
dressed by the magnetic field. \\

This paper is organized as follows. In Sec.~\ref{Theory}, we present the full theory of our model where we define the bare and dressed states of both reactants and products, as well as the unsymmetrized and symmetrized states, 
and provide an expression to compute the nuclear spin state-to-state probability distribution in a magnetic field.
In Sec.~\ref{Application}, we apply our formalism to 
$^6$Li$^{40}$K + $^6$Li$^{40}$K and $^{41}$K$^{87}$Rb + $^{41}$K$^{87}$Rb ultracold reactions. 
We present the nuclear spin distribution from a given initial quantum state of the reactants to the possible final nuclear spin states of the products as a function of an applied magnetic field, for rotational transitions where all molecules are in their ground rotational state.
We conclude in Sec.~\ref{Conclu}.

\section{Theory}
\label{Theory}

We are interested in the general chemical reaction
\begin{eqnarray}
\text{AB$_1$} + \text{AB$_2$} \to \text{AA}+ \text{BB}
\label{cr}
\end{eqnarray}
in a magnetic field.
AB$_1$ and AB$_2$ are the two reactants and AA, BB are the products.
In the following,
the subscripts $1,2$ denote the first and second
entity, whether it is an atom or a molecule:
A$_1$ and A$_2$ will denote the first and second A atoms, 
B$_1$ and B$_2$ the first and second B atoms, 
AB$_1$ and AB$_2$ the first and second reactant molecules.
For sake of simplicity, we will consider diatomic molecules with no 
electronic orbital and spin momentum involved (namely $^1\Sigma$ molecules),
so that $\Lambda_1=\Lambda_2=0$ and $S_1=S_2=0$. 
Similarly, we do not include the vibrational quantum numbers $v_1, v_2$
in the formalism.
The rotational quantum numbers of reactant AB$_1$ (AB$_2$) 
are denoted $n_1, m_{n_1}$ ($n_2, m_{n_2}$)
while for AA (BB) they are denoted
$n_\text{A}, m_{n_\text{A}}$ ($n_\text{B}, m_{n_\text{B}}$).
Finally, the nuclear spins of the atoms 
$\text{A}_{1}, \text{B}_{1}, \text{A}_{2}, \text{B}_{2}$
are denoted respectively
$i_{\text{A}_1}$,  $i_{\text{B}_1}$, $i_{\text{A}_2}$, $i_{\text{B}_2}$.
The numbers
$m_{\text{A}_1}$, $m_{\text{B}_1}$, $m_{\text{A}_2}$, $m_{\text{B}_2}$ 
are short-hand notations for the nuclear spin projection quantum numbers 
$m_i$ of each individual nuclear spins 
onto the magnetic field axis, taken as the quantization axis.
As we consider here identical reactants, we have 
$i_{\text{A}_1} = i_{\text{A}_2} = i_{\text{A}}$
and $i_{\text{B}_1}=i_{\text{B}_2}=i_{\text{B}}$.
The nuclear spin projection quantum numbers of 
molecules AB$_1$, AB$_2$
are respectively denoted
$m_1 = m_{\text{A}_1} + m_{\text{B}_1}$,
$m_2 = m_{\text{A}_2} + m_{\text{B}_2}$,
$m_\text{A} = m_{\text{A}_1} + m_{\text{A}_2}$,
$m_\text{B} = m_{\text{B}_1} + m_{\text{B}_2}$.
The total projection quantum number of 
the first (second) reactant molecule AB$_1$ (AB$_2$) is denoted $M_1$ ($M_2$)
with $M_1 = m_{n_1} + m_{\text{A}_1} + m_{\text{B}_1} $ 
($M_2 = m_{n_2} + m_{\text{A}_2} + m_{\text{B}_2}$).
Similarly, the total projection quantum number of 
the products AA (BB) is denoted $M_\text{A}$ ($M_\text{B}$)
with $M_\text{A} = m_{n_\text{A}} + m_{\text{A}_1} + m_{\text{A}_2}$ 
($M_\text{B} = m_{n_\text{B}} + m_{\text{B}_1} + m_{\text{B}_2}$).
$M_1$, $M_2$, $M_\text{A}$, $M_\text{B}$ are good quantum numbers in a magnetic field.
Depending on the state preparation of the reactants, 
the molecules can have the same values $M_1 = M_2$
or different ones $M_1 \ne M_2$. 
If they are prepared in the same internal state,
they are often called indistinguishable (and necessarily $M_1 = M_2$).
If not, they are prepared in different internal states, they are
called distinguishable ($M_1 = M_2$ or $M_1 \ne M_2$ are both possible). 
In this study, we are interested in finding the state-to-state probabilities
of the products of the chemical reaction Eq.~\eqref{cr}.
For that, we propose a model based on three assumptions.
\\

\paragraph*{First assumption}

Before the collision takes place, the two reactant molecules are quite far apart. 
The magnetic field is then strong enough to polarize them as the molecules feel the field 
via the Zeeman interaction.
As they approach each other and start to collide, 
the molecules will feel the magnetic field less and less 
while they will feel the presence of the other molecule more and more. 
In the short-range region of the tetra-atomic complex where the four atoms are close to each other, 
the nuclear spins are prone to other interactions with the other 
(nuclear and electronic) spins or with the overall rotation.
These interactions can compete with the Zeeman interaction. 
Therefore, the nuclear spins do not remain necessarily polarized throughout the entire reaction and could spin-flip in the short-range region. 
However, including all those spin interactions in the short-range region is difficult
and complicated due to the few-body tetra-atomic aspect of the process.
To circumvent that, we assume that the nuclear spins remain spectators and unchanged during the 
time they spend in the tetramer complex \cite{Quack_MP_34_477_1977}. Then, they do not participate
to the dynamics at short-range, as confirmed by a recent experiment
\cite{Hu_NC_13_435_2021} and as can also be seen in studies of 
ultracold atom-diatom collisions \cite{Simoni_LP_16_707_2006,Tscherbul_PRR_2_013117_2020}.
The atoms of the molecular products then 
simply inherit the nuclear spin projection quantum numbers of the atoms
that are included in the linear combination of the wavefunction of the reactants
in a magnetic field.
This is as if the nuclear spins were spectators in the short-range region 
and then, the processus is only driven by the physics at long range, typically
the interaction of the reactants and products with the magnetic field.
This is what we adopt in the present theoretical model.
The final state-to-state distribution is then mainly governed 
by permutation symmetry considerations for the two identical reactants, 
permutation symmetry considerations for the two identical atoms of the products,
as well as the interaction of the molecules with the magnetic field at long range.
If we consider now two molecules AB$_1$ and AB$_2$
with an orbital angular momentum between 
the reactants denoted by $l_r, m_{l_r}$,
the total projection quantum number of the colliding system is $M = M_1 + M_2 + m_{l_r}$
and is a conserved quantity.
Then for the products, we have $M_\text{A} + M_\text{B} + m_{l_p} = M$,
where the orbital angular momentum of the products is denoted by $l_p, m_{l_p}$.
From the above assumption of the model that the nuclear spins do not change, this implies
$m_{n_1} + m_{n_2} + m_{l_r} = m_{n_\text{A}} + m_{n_\text{B}} + m_{l_p}$.
As it will be done in Sec.~\ref{Application},
simplifications can arise if we consider reactants in the ground rotational state 
$n_1 = n_2 = 0$ so that $m_{n_1} = m_{n_2} = 0$.
\\

\paragraph*{Second assumption} 

We will consider molecular systems in which the couplings between
different rotational quantum numbers does not significantly affect
the nuclear spin structure.
For example, this is the case for bi-alkali molecules where the hyperfine couplings involving rotation, namely the rotation - nuclear spin interaction, the rotation - magnetic field interaction, and the nuclei electric quadrupole moment - gradient of field interaction are weak compared to the rotational constant $B_{\text{rot}}$ of the corresponding molecule \cite{Aldegunde_PRA_78_033434_2008, Aldegunde_PRA_79_013401_2009,Aldegunde_PRA_96_042506_2017}.
Due to these weak couplings, we do not consider couplings between different
rotational quantum numbers of the molecules, 
that is between different values of $n_1$, 
of $n_2$, of ${n_\text{A}}$, or of ${n_\text{B}}$.
However, within a given rotational manifold $n_1$, $n_2$, ${n_\text{A}}$, or ${n_\text{B}}$,
these hyperfine terms give rise to couplings
between different values of $m_{n_1}$, $m_{n_2}$, $m_{n_\text{A}}$, or $m_{n_\text{B}}$.
In the following, it will be implicit that
we focus on a given state-to-state transition of Eq.~\eqref{cr},
from initial quantum numbers
$n_1, m_{n_1}, n_2, m_{n_2}, l_r, m_{l_r}$ 
to final ones
$n_{\text {A}}, m_{n_{\text {A}}}, n_{\text {B}}, m_{n_{\text {B}}}, l_p, m_{l_p}$.
\\

\paragraph*{Third assumption}

We finally consider that we can separate the final state-to-state probabilities
into two independent parts, a rotational one and a nuclear spin one,
and that the rotational part depends on
$n_1$, $n_2$, ${n_\text{A}}$, ${n_\text{B}}$ 
but not on 
$m_{n_1}$, $m_{n_2}$, $m_{n_\text{A}}$, $m_{n_\text{B}}$.
This assumption is somewhat confirmed by
a recent experiment \cite{Liu_N_593_379_2021} which
observed that the rotational state-to-state distribution of the reaction
is globally statistical 
in nature \cite{Gonzalez-Martinez_PRA_90_052716_2014,Bonnet_JCP_152_084117_2020}, 
when the products have a release of the final kinetic energy
larger than the initial one,
and that the probabilities are the same
for any $m_{n_1}$, $m_{n_2}$, $m_{n_\text{A}}$, $m_{n_\text{B}}$
quantum numbers.

\subsection{Bare and dressed states of the reactants and the products. Unsymmetrized states}

We note the unsymmetrized bare state ($b$ for bare) of the reactants AB$_1$ and AB$_2$
\begin{eqnarray}
\big| \, b_{\text {AB}_1} \big\rangle &=&  \big| \, n_{1}  \, m_{n_1}  \, m_{\text{A}_1} \, m_{\text{B}_1} \big\rangle \nonumber \\ 
\big| \, b_{\text {AB}_2} \big\rangle &=&  \big| \, n_{2}  \, m_{n_2} \, m_{\text{A}_2} \, m_{\text{B}_2} \big\rangle 
\label{barestatereac}
\end{eqnarray}
and the unsymmetrized bare state of the products AA and BB
\begin{eqnarray}
\big| \, b_{\text {AA}} \big\rangle &=&  \big| \, n_{\text{A}} \, m_{n_\text{A}} \, m_{\text{A}_1} \, m_{\text{A}_2} \big\rangle \nonumber \\
\big| \, b_{\text {BB}} \big\rangle &=&  \big| \, n_{\text{B}} \, m_{n_\text{B}} \, m_{\text{B}_1} \, m_{\text{B}_2} \big\rangle .
\label{barestateprod}
\end{eqnarray}
In a magnetic field, the unsymmetrized bare states become unsymmetrized dressed states ($d$ for dressed), written for reactant AB$_1$ as
\begin{eqnarray}
\big| \, d_{\text {AB}_1} \big\rangle 
&=& \sum_{ b_{\text {AB}_1}  } \,  \big| \, b_{\text {AB}_1} \big\rangle \, \big\langle b_{\text{AB}_1} \, \big| \, d_{\text{AB}_1} \big\rangle \nonumber \\
&=& \sum_{ m_{n_1}} \, \sum_{m_{\text{A}_1}, m_{\text{B}_1}} \, \big| \, n_{1}  \, m_{n_1}  \, m_{\text{A}_1} \, m_{\text{B}_1} \big\rangle \nonumber \\
& & \big\langle n_{1}  \, m_{n_1}  \, m_{\text{A}_1} \, m_{\text{B}_1}  \, \big| \, d_{\text {AB}_1} \big\rangle 
\label{dresstatereac1}
\end{eqnarray}
and for AB$_2$ as
\begin{eqnarray}
\big| \, d_{\text {AB}_2} \big\rangle 
&=& \sum_{ b_{\text {AB}_2}  } \,  \big| \, b_{\text {AB}_2} \big\rangle \, \big\langle b_{\text{AB}_2} \, \big| \, d_{\text{AB}_2} \big\rangle \nonumber \\
&=& \sum_{ m_{n_2}} \, \sum_{m_{\text{A}_2}, m_{\text{B}_2}} \, \big| \, n_{2}  \, m_{n_2}  \, m_{\text{A}_2} \, m_{\text{B}_2} \big\rangle \nonumber \\
& & \big\langle n_{2}  \, m_{n_2}  \, m_{\text{A}_2} \, m_{\text{B}_2}  \, \big| \, d_{\text {AB}_2} \big\rangle   .
\label{dresstatereac2}
\end{eqnarray}
Similarly for the products, the unsymmetrized dressed states for AA are given by
\begin{eqnarray}
\big| \, d_{\text {AA}} \big\rangle 
&=& \sum_{b_{\text {AA}} } \, \big| \, b_{\text {AA}} \big\rangle  \, \big\langle b_{\text {AA}} \, \big| \, d_{\text {AA}} \big\rangle  \nonumber \\
&=& \sum_{ m_{n_\text{A}}} \, \sum_{m_{\text{A}_1}, m_{\text{A}_2}} \, \big|  \, n_{\text{A}} \, m_{n_\text{A}} \, m_{\text{A}_1} \, m_{\text{A}_2} \big\rangle  \nonumber \\
& & \big\langle  n_{\text{A}} \, m_{n_\text{A}} \, m_{\text{A}_1} \, m_{\text{A}_2} \, \big| \, d_{\text {AA}} \big\rangle 
\label{dresstateprod1}
\end{eqnarray}
and for BB by
\begin{eqnarray}
\big| \, d_{\text {BB}} \big\rangle 
&=& \sum_{b_{\text {BB}} } \, \big| \, b_{\text {BB}} \big\rangle  \, \big\langle b_{\text {BB}} \, \big| \, d_{\text {BB}} \big\rangle  \nonumber \\
&=& \sum_{ m_{n_\text{B}}} \, \sum_{m_{\text{B}_1}, m_{\text{B}_2}} \, \big|  \, n_{\text{B}} \, m_{n_\text{B}} \, m_{\text{B}_1} \, m_{\text{B}_2} \big\rangle  \nonumber \\
& & \big\langle  n_{\text{B}} \, m_{n_\text{B}} \, m_{\text{B}_1} \, m_{\text{B}_2} \, \big| \, d_{\text {BB}} \big\rangle  .
\label{dresstateprod2}
\end{eqnarray}
In all cases, we keep explicitely the rotational quantum numbers in the notations of the kets, which are fixed
once for all for a given rotational state-to-state transition.

\subsection{Symmetrized states}

\subsubsection{For the reactants}

As the two AB molecules are identical, symmetrized states of the combined
reactants AB$_1$ + AB$_2$ have to be built. The bare symmetrized states are
\begin{multline}
\big| \, b_{\text {AB}_1}, b_{\text {AB}_2} ; \eta \big\rangle 
= \frac{1}{\sqrt{\Delta_b}} \, \bigg\{ \big| \, b_{\text{AB}_1} \big\rangle \otimes \big| \, b_{\text{AB}_2} \big\rangle \\
\shoveright{ + \eta \, \big| \, b_{\text{AB}_2} \big\rangle \otimes \big| \, b_{\text{AB}_1} \big\rangle \bigg\} } \\
= \frac{1}{\sqrt{\Delta_b}} \, \bigg\{ \big| \, n_{1}  \, m_{n_1}  \, m_{\text{A}_1} \, m_{\text{B}_1} \big\rangle \otimes \big| \, n_{2}  \, m_{n_2}  \, m_{\text{A}_2} \, m_{\text{B}_2}   \big\rangle \\
+ \eta \, \big| \, n_{2}  \, m_{n_2}  \, m_{\text{A}_2} \, m_{\text{B}_2}  \big\rangle 
\otimes \big| \, n_{1}  \, m_{n_1}  \, m_{\text{A}_1} \, m_{\text{B}_1}  \big\rangle \bigg\} 
\label{symbarestatereac}
\end{multline}
where $\Delta_b = 2 \, ( 1 + \delta_{b_{\text{AB}_1}, \, b_{\text{AB}_2} })$, 
$\eta = +1$ for symmetric states and $\eta = -1$ for anti-symmetric states.
The counting of these states have to be well ordered \cite{Green_JCP_62_2271_1975}. 
$b_{\text{AB}_1} = 1,2, ..., (2 i_\text {A}+1) \times  (2 i_\text {B}+1) \times (2 n_1+1)$ 
and 
$b_{\text{AB}_2} = 1,2, ..., (2 i_\text {A}+1) \times  (2 i_\text {B}+1) \times (2 n_2+1)$ 
are indices to count the states, for given numbers
$n_1$ and $n_2$ in the rotational basis.
To avoid double counting
the symmetrized states 
of the reactants, the condition $b_{\text{AB}_2} \ge b_{\text{AB}_1}$
or $b_{\text{AB}_2} \le b_{\text{AB}_1}$ has to be made.
Here we choose arbitrarily 
$b_{\text{AB}_2} \ge b_{\text{AB}_1}$.
Note that when the symmetrized anti-symmetric states ($\eta=-1$) are considered,
$b_{\text{AB}_2}$ cannot be equal to $b_{\text{AB}_1}$, which implies
$b_{\text{AB}_2} > b_{\text{AB}_1}$. These general arguments
hold whenever the well ordered states are mentioned hereafter.
\\

In the presence of a magnetic field, the symmetrized bare states transform into the symmetrized dressed states
\begin{multline}
\big| \, d_{\text {AB}_1}, d_{\text {AB}_2} ; \eta \big\rangle 
=\frac{1}{\sqrt{\Delta_d}} \, \bigg\{ \big| \, d_{\text{AB}_1} \big\rangle \otimes \big| \, d_{\text{AB}_2} \big\rangle \\
\shoveright{ + \eta \, \big| \, d_{\text{AB}_2} \big\rangle \otimes \big| \, d_{\text{AB}_1} \big\rangle \bigg\} } \\
= \frac{1}{\sqrt{\Delta_d}} \,
\sum_{b_{\text {AB}_1}  }  \, \sum_{ b_{\text {AB}_2}  } \, 
\bigg\{ \, \big| \, b_{\text {AB}_1} \big\rangle  \otimes \big| \, b_{\text {AB}_2} \big\rangle 
+ \eta \ \big| \, b_{\text {AB}_2} \big\rangle \otimes \big| \, b_{\text {AB}_1} \big\rangle   \bigg\}
\\
\shoveright{  \big\langle b_{\text {AB}_1} \, \big| \, d_{\text {AB}_1} \big\rangle \, \big\langle b_{\text {AB}_2} \, \big| \, d_{\text {AB}_2} \big\rangle  } \\
\shoveleft{ =  \sum_{m_{n_1}} \,  \sum_{m_{n_2}} \,
\sum_{ m_{\text{A}_1} = -i_{\text{A}}}^{+i_{\text{A}}} \, \sum_{ m_{\text{B}_1} = -i_{\text{B}}}^{+i_{\text{B}}}  \, \sum_{ m_{\text{A}_2} = -i_{\text{A}}}^{+i_{\text{A}}} \, \sum_{ m_{\text{B}_2} = -i_{\text{B}}}^{+i_{\text{B}}}  } \\ 
\big\langle n_{1}  \, m_{n_1}  \, m_{\text{A}_1} \, m_{\text{B}_1}  \, \big| \, d_{\text {AB}_1} \big\rangle 
\, \big\langle n_{2}  \, m_{n_2}  \, m_{\text{A}_2} \, m_{\text{B}_2}  \, \big| \, d_{\text {AB}_2} \big\rangle  \\
\ \delta_{m_{n_1} + m_{\text{A}_1}+m_{\text{B}_1}, M_1} 
\ \delta_{m_{n_2} + m_{\text{A}_2}+m_{\text{B}_2}, M_2}  \\
\frac{1}{\sqrt{\Delta_d}} \, \bigg\{ \big| \, n_{1}  \, m_{n_1}  \, m_{\text{A}_1} \, m_{\text{B}_1} \big\rangle \otimes \big| \, n_{2}  \, m_{n_2}  \, m_{\text{A}_2} \, m_{\text{B}_2}   \big\rangle \\
+ \eta \, \big| \, n_{2}  \, m_{n_2}  \, m_{\text{A}_2} \, m_{\text{B}_2}  \big\rangle 
\otimes \big| \, n_{1}  \, m_{n_1}  \, m_{\text{A}_1} \, m_{\text{B}_1}  \big\rangle \bigg\} 
\label{symdresstatereac0}
\end{multline}
where now $\Delta_d = 2 \, ( 1 + \delta_{ d_{\text{AB}_1}, \, d_{\text{AB}_2} })$
and the well ordering of the states is chosen to be $d_{\text{AB}_2} \ge d_{\text{AB}_1}$. 
It is important to note that here there is no well ordering of the $b_{\text{AB}_1}$ and $b_{\text{AB}_2}$ states, as they span all values that come from
the dressing of the bare states Eq.~\eqref{dresstatereac1} and Eq.~\eqref{dresstatereac2}.
The Kronecker delta terms insure the conservation of the 
total projection quantum number of each molecule AB$_1$ and AB$_2$, as it is the case in a magnetic field.

\subsubsection{For the products}

In contrast with the two identical reactants AB, 
the products AA and BB are different molecules.
Therefore there is no symmetrization of the wavefunction of the combined products that is required.
However, as the A atoms (B atoms) are identical, we have to symmetrize the wavefunction of 
the individual molecule AA (BB).
The bare symmetrized states are
\begin{eqnarray}
\big| \, b_{\text {AA}}, b_{\text {BB}} ; \eta_\text{A}, \eta_\text{B} \big\rangle 
= \big| \, b_{\text {AA}} ; \eta_\text{A} \big\rangle \otimes \big| \,  b_{\text {BB}} ; \eta_\text{B} \big\rangle
\label{symbarestateprod}
\end{eqnarray}
with
\begin{eqnarray}
\big| \, b_{\text {AA}} ; \eta_\text{A} \big\rangle &=& 
\big|  \, n_{\text{A}} \, m_{n_\text{A}}  \big\rangle \nonumber \\
&\times&
\frac{1}{\sqrt{\Delta_\text{A}}} \, \bigg\{  \big|  \, m_{\text{A}_1} \,  m_{\text{A}_2}  \big\rangle + \eta_\text{A} \, \big|  \, m_{\text{A}_2} \, m_{\text{A}_1}  \big\rangle  \bigg\} \nonumber \\
& = & \frac{1}{\sqrt{\Delta_\text{A}}} \, 
\bigg\{ \big| \, n_{\text{A}} \, m_{n_\text{A}} \, m_{\text{A}_1} \,  m_{\text{A}_2}  \big\rangle \nonumber \\
& & \qquad  \qquad \qquad + \eta_\text{A} \, \big| \, n_{\text{A}} \, m_{n_\text{A}} \, m_{\text{A}_2} \, m_{\text{A}_1}  \big\rangle  \bigg\} \nonumber \\
& \equiv &  \big|  \, n_{\text{A}} \, m_{n_\text{A}} \, m_{\text{A}_1} \, m_{\text{A}_2}; \eta_{\text {A}}  \big\rangle \nonumber \\
\, \big| \,  b_{\text {BB}} ; \eta_\text{B} \big\rangle &=&
\big|  \, n_{\text{B}} \, m_{n_\text{B}}  \big\rangle \nonumber \\
&\times&
\frac{1}{\sqrt{\Delta_\text{B}}} \, \bigg\{ \big|  \, m_{\text{B}_1} \,  m_{\text{B}_2}  \big\rangle + \eta_\text{B} \, \big|  \, m_{\text{B}_2} \, m_{\text{B}_1}  \big\rangle  \bigg\} \nonumber \\
& = & \frac{1}{\sqrt{\Delta_\text{B}}} \, 
\bigg\{ \big| \, n_{\text{B}} \, m_{n_\text{B}} \, m_{\text{B}_1} \,  m_{\text{B}_2}  \big\rangle \nonumber \\
& & \qquad  \qquad \qquad + \eta_\text{B} \, \big| \, n_{\text{B}} \, m_{n_\text{B}} \, m_{\text{B}_2} \, m_{\text{B}_1}  \big\rangle  \bigg\} \nonumber \\
& \equiv &  \big|  \, n_{\text{B}} \, m_{n_\text{B}} \, m_{\text{B}_1} \, m_{\text{B}_2}; \eta_{\text {B}}  \big\rangle. \nonumber \\ 
\label{symbarestateprod2}
\end{eqnarray}
$\Delta_\text{A} = 2 \, ( 1 + \delta_{m_{\text{A}_1}, \, m_{\text{A}_2}})$, $\Delta_B = 2 \, ( 1 + \delta_{m_{\text{B}_1}, \, m_{\text{B}_2}})$
and $\eta_\text{A} = \pm 1$, $\eta_\text{B} = \pm 1$ for symmetric or anti-symmetric states of AA and BB.
Again, these states have to be well ordered. 
The nuclear spin indices
span $m_{\text{A}_{1,2}} = - i_\text{A}, + i_\text{A}$
and $m_{\text{B}_{1,2}} = - i_\text{B}, + i_\text{B}$. 
We choose $m_{\text{A}_2}  \ge m_{\text{A}_1}$ as well as
$m_{\text{B}_2} \le m_{\text{B}_1}$.
This is arbitrary but convenient
for the case when the molecules have $m_1 = m_2$.
In such a case, the conservation of 
$m_{\text{A}_1} + m_{\text{B}_1} = m_{\text{A}_2} + m_{\text{B}_2}$ 
implies that if $m_{\text{A}_2} \ge m_{\text{A}_1}$, then $m_{\text{B}_2} \le m_{\text{B}_1}$.\\

In the presence of a magnetic field, the symmetrized bare states transform 
into the symmetrized dressed states for each individual molecule
\begin{multline}
\big| \, d_{\text {AA}} ; \eta_{\text {A}} \big\rangle 
= \sum_{b_{\text {AA}} } \, \big| \, b_{\text {AA}} ; \eta_{\text {A}} \big\rangle \, \big\langle b_{\text {AA}} ; \eta_{\text {A}} \, \big| \, d_{\text {AA}} ; \eta_{\text {A}} \big\rangle  \\
\shoveleft{ = \sum_{ m_{n_\text{A}} } \,
\sum_{m_{\text{A}_1}} \sum_{m_{\text{A}_2} \ge m_{\text{A}_1} } 
\big|  \, n_{\text{A}} \, m_{n_\text{A}} \, m_{\text{A}_1} \, m_{\text{A}_2}; \eta_{\text {A}}  \big\rangle } \\
\shoveright{ \big\langle n_{\text{A}} \, m_{n_\text{A}} \, m_{\text{A}_1} \, m_{\text{A}_2}; \eta_{\text {A}}  \, \big| \, d_{\text {AA}} ; \eta_{\text {A}} \big\rangle  } \\
\shoveleft{ \big| \, d_{\text {BB}} ; \eta_{\text {B}} \big\rangle 
= \sum_{b_{\text {BB}} } \, \big| \, b_{\text {BB}} ; \eta_{\text {B}} \big\rangle \, \big\langle b_{\text {BB}} ; \eta_{\text {B}} \, \big| \, d_{\text {BB}} ; \eta_{\text {B}} \big\rangle  }  \\
\shoveleft{ =\sum_{  m_{n_\text{B}} } \,
\sum_{m_{\text{B}_1}} \sum_{m_{\text{B}_2} \le m_{\text{B}_1} } 
\big|  \, n_{\text{B}} \, m_{n_\text{B}} \, m_{\text{B}_1} \, m_{\text{B}_2}; \eta_{\text {B}}  \big\rangle } \\
\big\langle n_{\text{B}} \, m_{n_\text{B}} \, m_{\text{B}_1} \, m_{\text{B}_2}; \eta_{\text {B}}  \, \big| \, d_{\text {BB}} ; \eta_{\text {B}} \big\rangle .
\label{symdresstateprodAABB}
\end{multline}
Then we get the following symmetrized dressed state for the products 
\begin{multline}
\big| \, d_{\text {AA}}, d_{\text {BB}} ; \eta_{\text {A}}, \eta_{\text {B}} \big\rangle 
= \big| \, d_{\text {AA}} ; \eta_{\text {A}} \big\rangle \otimes 
\big| \, d_{\text {BB}} ; \eta_{\text {B}} \big\rangle \\ 
\shoveleft{ = \sum_{b_{\text {AA}} }   \, \sum_{b_{\text {BB}} }  \, 
\big| \, b_{\text {AA}}, b_{\text {BB}} ; \eta_\text{A}, \eta_\text{B}  \big\rangle }
 \\ 
\shoveright{ \big\langle b_{\text {AA}} ; \eta_{\text {A}} \, \big| \, d_{\text {AA}} ; \eta_{\text {A}} \big\rangle  \, \big\langle b_{\text {BB}} ; \eta_{\text {B}} \, \big| \, d_{\text {BB}} ; \eta_{\text {B}} \big\rangle } \\
\shoveleft{ = \sum_{ m_{n_\text{A}} } \, \sum_{ m_{n_\text{B}} } \,
\sum_{ m_{\text{A}_1} = -i_{\text{A}}}^{+i_{\text{A}}} \sum_{m_{\text{A}_2} \ge m_{\text{A}_1} }   \, \sum_{ m_{\text{B}_1} = -i_{\text{B}}}^{+i_{\text{B}}}
 \sum_{m_{\text{B}_2} \le m_{\text{B}_1} }  } \\
  \, \big\langle n_{\text{A}} \, m_{n_\text{A}} \, m_{\text{A}_1} \, m_{\text{A}_2}; \eta_{\text {A}}  \, \big| \, d_{\text {AA}} ; \eta_{\text {A}} \big\rangle \\
  \big\langle n_{\text{B}} \, m_{n_\text{B}} \, m_{\text{B}_1} \, m_{\text{B}_2}; \eta_{\text {B}}  \, \big| \, d_{\text {BB}} ; \eta_{\text {B}} \big\rangle \\
 \ \delta_{m_{\text{A}}+m_{\text{B}},m_{{1}} + m_{{2}}} \\ 
  \delta_{m_{n_\text{A}} + m_{\text{A}_1} + m_{\text{A}_2} + m_{n_\text{B}} + m_{\text{B}_1} + m_{\text{B}_2} + m_{l_p}, M}  \\
  \frac{1}{\sqrt{\Delta_\text{A}}} \, \bigg\{  \big| \, n_{\text{A}} \, m_{n_\text{A}} \, m_{\text{A}_1} \,  m_{\text{A}_2}  \rangle + \, \eta_\text{A} \, \big| \, n_{\text{A}} \, m_{n_\text{A}} \,  m_{\text{A}_2} \, m_{\text{A}_1} \big\rangle \bigg\} \\
\otimes  \frac{1}{\sqrt{\Delta_\text{B}}} \, \bigg\{ \big| \, n_{\text{B}} \, m_{n_\text{B}} \,   m_{\text{B}_1} \,  m_{\text{B}_2}  \big\rangle + \, \eta_\text{B} \, \big| \, n_{\text{B}} \, m_{n_\text{B}} \,  m_{\text{B}_2} \, m_{\text{B}_1} \big\rangle  \bigg\}  . \\
\label{symdresstateprod}
\end{multline}
Among the multiple possible combined nuclear spin states of the molecular products AA and BB, 
only those which satisfy the conservation of 
$m_{{\text{A}}} + m_{{\text{B}}} = m_{\text{A}_1}+m_{\text{A}_2}+m_{\text{B}_1}+m_{\text{B}_2} = m_{{1}} + m_{{2}} $ 
are retained, as imposed by the first Kronecker delta term.
This is a consequence of the first assumption.
Also, only those which satisfy the conservation of total $M$ are retained,
as imposed by the second Kronecker delta term.
Note that because $M = m_{n_A} + m_{\text{A}_1} + m_{\text{B}_1} + m_{n_2} + m_{\text{A}_2} + m_{\text{B}_2} + m_{l_r}$, then the second Kronecker symbol can be recast as
$  \delta_{m_{n_\text{A}} + m_{n_\text{B}}, m_{n_1} + m_{n_2} + m_{l_r} - m_{l_p}} $
and this is what we will employ in the following.

\subsection{State-to-state probabilities}

Our model will consider that the state-to-state probability $P_{i \to j}$, 
from the combined dressed state of the indistinguishable reactants
$\big|  \, d_{\text {AB}_1}, d_{\text {AB}_2} ; \eta \big\rangle $ denoted by $|  i \rangle $,
to the combined dressed state of the products
$\big|  d_{\text {AA}}, d_{\text {BB}} ; \eta_{\text {A}}, \eta_{\text {B}} \big\rangle$
denoted by $| j \rangle $, is simply the modulus square of 
the probability amplitude, the overlap between those 
two states
\begin{eqnarray}
P_{i \to j} = \big| \langle j \big| i \rangle \big|^2 = \bigg| \langle  \, d_{\text {AA}}, d_{\text {BB}} ; \eta_{\text {A}}, \eta_{\text {B}}  \big|  \, d_{\text {AB}_1}, d_{\text {AB}_2} ; \eta \big\rangle \bigg|^2 . \nonumber \\
\label{proba-sts}
\end{eqnarray}
It reflects the amount of the molecular product wavefunction obtained from 
the molecular reactant wavefunction, both dressed by the magnetic field,
taking into account the appropriate permutation
symmetry considerations
characterized by $\eta_{\text {A}}, \eta_{\text {B}}$ and $\eta $.
This probability is for a transition from initial quantum numbers
$n_1, m_{n_1}, n_2, m_{n_2}, l_r, m_{l_r}$ 
to final ones
$n_{\text {A}}, m_{n_{\text {A}}}, n_{\text {B}}, m_{n_{\text {B}}}, l_p, m_{l_p}$.
The conservation of the total angular momentum $M$
as mentioned above, is implicit here
and leads to the selection rule
$m_{n_1} + m_{n_2} + m_{l_r} = m_{n_\text{A}} + m_{n_\text{B}} + m_{l_p}$.
From the assumptions made in the model and as shown in Appendix A, the probability amplitude 
related to Eq.~\eqref{proba-sts} can be recast into the product of two parts,
namely 
$\langle  \, d_{\text {AA}}, d_{\text {BB}} ; \eta_{\text {A}}, \eta_{\text {B}}  \big|  \, d_{\text {AB}_1}, d_{\text {AB}_2} ; \eta \big\rangle^\text{ns}$
for the nuclear spin degree of freedom,
and $\big\langle {n_\text{A}} \, {n_\text{B}} \, \big| \, n_{1}   \,  n_{2}  \big\rangle $
for the rotational degree of freedom (or all the other degrees of freedom but the nuclear spin
if we consider for example vibration etc ...).
This implies that Eq.~\eqref{proba-sts} becomes a product
of two probabilities
\begin{eqnarray}
P_{i \to j} = P^\text{rot} \times P_{i \to j}^\text{ns} 
\label{proba-sts-indep}
\end{eqnarray}
with a global one related to a given rotational transition
$P^\text{rot} = \big| \big\langle {n_\text{A}} \, {n_\text{B}} \, \big| \, n_{1}   \,  n_{2}  \big\rangle \big|^2$,
and a state-to-state specific one related to a nuclear spin transition (of 
a given rotational transition)
$P_{i \to j}^\text{ns}  = \bigg| \langle  \, d_{\text {AA}}, d_{\text {BB}} ; \eta_{\text {A}}, \eta_{\text {B}}  \big|  \, d_{\text {AB}_1}, d_{\text {AB}_2} ; \eta \big\rangle^\text{ns} \bigg|^2. $
The probability 
$P^\text{rot}$ 
highly depends on the complicated short-range dynamics of the tetramer complex
and the different angular momenta couplings. It is 
not treated in our study as mentionned above
but can represented by the statistical expressions
developped in Refs \cite{Gonzalez-Martinez_PRA_90_052716_2014,Bonnet_JCP_152_084117_2020,Liu_N_593_379_2021}.
The probability $P_{i \to j}^\text{ns}$ (and the amplitude) is treated in our study and 
depending on the initial preparation of the reactants,
different simplifications occur as derived in Appendices A and B.
This is on what we focus now.

\onecolumngrid

\subsubsection{General case of reactants in non-zero rotational states}

For the general case where the reactants are in a non-zero rotational states,
the probability amplitude for the nuclear spin is given by the general expression

\begin{multline}
\langle  \, d_{\text {AA}}, d_{\text {BB}} ; \eta_{\text {A}}, \eta_{\text {B}}  \big|  \, d_{\text {AB}_1}, d_{\text {AB}_2} ; \eta \big\rangle^\text{ns}  =  
\delta_{\eta_{\text {A}} \eta_{\text {B}}, \eta} \times \\ 
\sum_{ m_{n_\text{A}}} \sum_{m_{n_\text{B}}} 
\sum_{ m_{\text{A}_1} = -i_{\text{A}}}^{+i_{\text{A}}} \sum_{m_{\text{A}_2} \ge m_{\text{A}_1} }   \, \sum_{ m_{\text{B}_1} = -i_{\text{B}}}^{+i_{\text{B}}}
 \sum_{m_{\text{B}_2} \le m_{\text{B}_1} }
  \, \big\langle d_{\text {AA}} ; \eta_{\text {A}} \, \big| \, n_{\text{A}} \, m_{n_\text{A}} \, m_{\text{A}_1} \, m_{\text{A}_2}; \eta_{\text {A}}  \big\rangle  \, \big\langle d_{\text {BB}} ; \eta_{\text {B}} \, \big| \, n_{\text{B}} \, m_{n_\text{B}} \, m_{\text{B}_1} \, m_{\text{B}_2} ; \eta_{\text {B}}   \big\rangle  \\ 
  \frac{1}{\sqrt{\Delta_\text{A}}} \, \frac{1}{\sqrt{\Delta_\text{B}}} \, \frac{1}{\sqrt{\Delta_d}} 
  \, \sum_{ m_{n_1}} \sum_{m_{n_2} } 
   \delta_{m_{n_\text{A}} + m_{n_\text{B}}, m_{n_1} + m_{n_2} + m_{l_r} - m_{l_p}} \\
   \bigg\{  
  \big\langle n_{1}  \, m_{n_1}  \, m_{\text{A}_1} \, m_{\text{B}_1}  \, \big| \, d_{\text {AB}_1} \big\rangle \, \big\langle n_{2}  \, m_{n_2}  \, m_{\text{A}_2} \, m_{\text{B}_2} \, \big| \, d_{\text {AB}_2} \big\rangle 
   \bigg[ 1 + \eta_\text{A} \, \eta_\text{B} \, \eta  \bigg] 
\ \delta_{ m_{n_1} + m_{\text{A}_1}+m_{\text{B}_1}, M_1} 
\ \delta_{ m_{n_2} + m_{\text{A}_2}+m_{\text{B}_2}, M_2} \\
\shoveright{    \big\langle n_{1}  \, m_{n_1}  \, m_{\text{A}_2} \, m_{\text{B}_2}  \, \big| \, d_{\text {AB}_1} \big\rangle \, \big\langle n_{2}  \, m_{n_2}  \, m_{\text{A}_1} \, m_{\text{B}_1} \, \big| \, d_{\text {AB}_2} \big\rangle  \bigg[ \eta_\text{A} \, \eta_\text{B} + \eta \bigg] 
\ \delta_{m_{n_1} + m_{\text{A}_2}+m_{\text{B}_2}, M_1} 
\ \delta_{m_{n_2} + m_{\text{A}_1}+m_{\text{B}_1}, M_2} } \\
\shoveright{ \big\langle n_{1}  \, m_{n_1}  \, m_{\text{A}_1} \, m_{\text{B}_2}  \, \big| \, d_{\text {AB}_1} \big\rangle \, \big\langle n_{2}  \, m_{n_2}  \, m_{\text{A}_2} \, m_{\text{B}_1} \, \big| \, d_{\text {AB}_2} \big\rangle   \bigg[ \eta_\text{B} + \eta_\text{A} \, \eta \bigg] 
\ \delta_{m_{n_1} + m_{\text{A}_1}+m_{\text{B}_2}, M_1} 
\ \delta_{m_{n_2} + m_{\text{A}_2}+m_{\text{B}_1}, M_2} } \\
 \big\langle n_{1}  \, m_{n_1}  \, m_{\text{A}_2} \, m_{\text{B}_1}  \, \big| \, d_{\text {AB}_1} \big\rangle \, \big\langle n_{2}  \, m_{n_2}  \, m_{\text{A}_1} \, m_{\text{B}_2} \, \big| \, d_{\text {AB}_2} \big\rangle  \bigg[  \eta_\text{A} + \eta_\text{B} \, \eta \bigg]  
\ \delta_{m_{n_1} + m_{\text{A}_2}+m_{\text{B}_1}, M_1} 
\ \delta_{m_{n_2} + m_{\text{A}_1}+m_{\text{B}_2}, M_2}  \bigg\} .
   \label{ampproba-sts-general-nonzerorot}
\end{multline}
\noindent Important to note is the fact that
the Kronecker delta term $\delta_{\eta_{\text {A}} \eta_{\text {B}}, \eta} $
imposes a selection rule for the symmetry of the AA and BB molecular products.
If $\eta = +1$, only the cases
$\eta_{\text {A}} = \eta_{\text {B}} = +1$ or $\eta_{\text {A}} = \eta_{\text {B}} = -1$
are allowed.
If $\eta = -1$, only the cases
$\eta_{\text {A}} = +1$, $\eta_{\text {B}} = -1$ or $\eta_{\text {A}} = -1$, $\eta_{\text {B}} = +1$
are allowed.

\subsubsection{Specific case of reactants in zero rotational states}

Now if the reactants are in a zero rotational state, 
we have $n_1 = n_2 = 0$ and $m_{n_1} = m_{n_2} = 0$,
which implies $M_1 \equiv m_1$, $M_2 \equiv m_2$. 
Eq.~\eqref{ampproba-sts-general-nonzerorot}
simplifies to
%
\begin{multline}
\langle  \, d_{\text {AA}}, d_{\text {BB}} ; \eta_{\text {A}}, \eta_{\text {B}}  \big|  \, d_{\text {AB}_1}, d_{\text {AB}_2} ; \eta \big\rangle^\text{ns}  =  \delta_{\eta_{\text {A}} \eta_{\text {B}}, \eta} \times 
\sum_{ m_{n_\text{A}}} \sum_{m_{n_\text{B}}}    \delta_{m_{n_\text{A}} + m_{n_\text{B}}, m_{l_r} - m_{l_p}} \\
\sum_{ m_{\text{A}_1} = -i_{\text{A}}}^{+i_{\text{A}}} \sum_{m_{\text{A}_2} \ge m_{\text{A}_1} }   \, \sum_{ m_{\text{B}_1} = -i_{\text{B}}}^{+i_{\text{B}}}
 \sum_{m_{\text{B}_2} \le m_{\text{B}_1} }
 \, \big\langle d_{\text {AA}} ; \eta_{\text {A}} \, \big| \, n_{\text{A}} \, m_{n_\text{A}} \, m_{\text{A}_1} \, m_{\text{A}_2}; \eta_{\text {A}}  \big\rangle  \, \big\langle d_{\text {BB}} ; \eta_{\text {B}} \, \big| \, n_{\text{B}} \, m_{n_\text{B}} \, m_{\text{B}_1} \, m_{\text{B}_2} ; \eta_{\text {B}}   \big\rangle   \\ 
  \frac{1}{\sqrt{\Delta_\text{A}}} \, \frac{1}{\sqrt{\Delta_\text{B}}} \, \frac{1}{\sqrt{\Delta_d}} \,
   \bigg\{  
   \big\langle m_{\text{A}_1} \, m_{\text{B}_1}  \, \big| \, d_{\text {AB}_1} \big\rangle \, \big\langle m_{\text{A}_2} \, m_{\text{B}_2} \, \big| \, d_{\text {AB}_2} \big\rangle   \bigg[ 1 + \eta_\text{A} \, \eta_\text{B} \, \eta  \bigg] 
\ \delta_{m_{\text{A}_1}+m_{\text{B}_1}, m_1} 
\ \delta_{m_{\text{A}_2}+m_{\text{B}_2}, m_2} \\
\shoveright{     +    \big\langle m_{\text{A}_2} \, m_{\text{B}_2}  \, \big| \, d_{\text {AB}_1} \big\rangle \, \big\langle m_{\text{A}_1} \, m_{\text{B}_1} \, \big| \, d_{\text {AB}_2} \big\rangle   \bigg[\eta_\text{A} \, \eta_\text{B} +  \eta \bigg] 
\ \delta_{m_{\text{A}_2}+m_{\text{B}_2}, m_1} 
\ \delta_{m_{\text{A}_1}+m_{\text{B}_1}, m_2} } \\
\shoveright{  +     \big\langle m_{\text{A}_1} \, m_{\text{B}_2}  \, \big| \, d_{\text {AB}_1} \big\rangle \, \big\langle m_{\text{A}_2} \, m_{\text{B}_1} \, \big| \, d_{\text {AB}_2} \big\rangle   \bigg[  \eta_\text{B} 
+ \eta_\text{A} \, \eta \bigg] 
\ \delta_{m_{\text{A}_1}+m_{\text{B}_2}, m_1} 
\ \delta_{m_{\text{A}_2}+m_{\text{B}_1}, m_2} } \\
 \shoveright{   +    \big\langle m_{\text{A}_2} \, m_{\text{B}_1}  \, \big| \, d_{\text {AB}_1} \big\rangle \, \big\langle m_{\text{A}_1} \, m_{\text{B}_2} \, \big| \, d_{\text {AB}_2} \big\rangle  \bigg[ \eta_\text{A} + \eta_\text{B} \, \eta \bigg]  
\ \delta_{m_{\text{A}_2}+m_{\text{B}_1}, m_1} 
\ \delta_{m_{\text{A}_1}+m_{\text{B}_2}, m_2}  \bigg\}    . }  \\
   \label{ampproba-sts-general}
\end{multline}

\subsubsection{Specific case of reactants in zero rotational with $m_1 = m_2$}

When the molecules (indistinguishable or not) have the same values $m_1 = m_2$, 
Eq.~\eqref{ampproba-sts-general} simplifies to 
%
\begin{multline}
\langle  \, d_{\text {AA}}, d_{\text {BB}} ; \eta_{\text {A}}, \eta_{\text {B}}  \big|  \, d_{\text {AB}_1}, d_{\text {AB}_2} ; \eta \big\rangle^\text{ns} = \delta_{\eta_{\text {A}} \eta_{\text {B}}, \eta} \times 
  \sum_{ m_{n_\text{A}}} \sum_{m_{n_\text{B}}}  \delta_{m_{n_\text{A}} + m_{n_\text{B}}, m_{l_r} - m_{l_p}} \\
\frac{1}{\sqrt{\Delta_d}} \,
 \sum_{ m_{\text{A}_1} = -i_{\text{A}}}^{+i_{\text{A}}} \sum_{m_{\text{A}_2} \ge m_{\text{A}_1} }   \, \sum_{ m_{\text{B}_1} = -i_{\text{B}}}^{+i_{\text{B}}}
 \sum_{m_{\text{B}_2} \le m_{\text{B}_1} }
  \, \big\langle d_{\text {AA}} ; \eta_{\text {A}} \, \big| \, n_{\text{A}} \, m_{n_\text{A}} \, m_{\text{A}_1} \, m_{\text{A}_2}; \eta_{\text {A}}  \big\rangle  \, \big\langle d_{\text {BB}} ; \eta_{\text {B}} \, \big| \, n_{\text{B}} \, m_{n_\text{B}} \, m_{\text{B}_1} \, m_{\text{B}_2} ; \eta_{\text {B}}   \big\rangle \\ 
     \bigg\{  
   \big\langle m_{\text{A}_1} \, m_{\text{B}_1}  \, \big| \, d_{\text {AB}_1} \big\rangle \, \big\langle m_{\text{A}_2} \, m_{\text{B}_2} \, \big| \, d_{\text {AB}_2} \big\rangle   
   + \eta \, \big\langle m_{\text{A}_2} \, m_{\text{B}_2}  \, \big| \, d_{\text {AB}_1} \big\rangle \, \big\langle m_{\text{A}_1} \, m_{\text{B}_1} \, \big| \, d_{\text {AB}_2} \big\rangle 
     \bigg\}  \\
 \    \delta_{m_{\text{A}_1}+m_{\text{B}_1}, m_1} 
\ \delta_{m_{\text{A}_2}+m_{\text{B}_2}, m_1}  .
\label{ampproba-sts-indist-dist}
\end{multline}
\twocolumngrid
\noindent The Kronecker delta terms $ \delta_{m_{\text{A}_1}+m_{\text{B}_1}, m_1}$ and $\delta_{m_{\text{A}_2}+m_{\text{B}_2}, m_1} $ impose some restrictions for the quadruple sum of the nuclear spins. 
If $m_{\text{A}_2} = m_{\text{A}_1}$, this implies automatically $m_{\text{B}_2} = m_{\text{B}_1}$.
There is no case with $m_{\text{A}_2} = m_{\text{A}_1}$ and $m_{\text{B}_2} \ne m_{\text{B}_1}$ or the reverse, $m_{\text{B}_2} = m_{\text{B}_1}$ and $m_{\text{A}_2} \ne m_{\text{A}_1}$, for the same reasons. 
Therefore, only the cases
$m_{\text{A}_2} = m_{\text{A}_1}$,  $m_{\text{B}_2} = m_{\text{B}_1}$
and $m_{\text{A}_2} > m_{\text{A}_1}$, $m_{\text{B}_2} < m_{\text{B}_1}$
have to be considered for $\eta = +1$, and only 
$m_{\text{A}_2} > m_{\text{A}_1}$, $m_{\text{B}_2} < m_{\text{B}_1}$
for $\eta = -1$ (see Appendix B). \\

\paragraph{Indistinguishable case}

In addition, if the reactants are indistinguishable, 
$d_{\text {AB}_2}  \equiv d_{\text {AB}_1} \equiv d_{\text {AB}}$
and only the component $\eta=+1$ has to be computed.
The selection rule necessarily implies
$\eta_{\text {A}} = \eta_{\text {B}} = +1$ or $\eta_{\text {A}} = \eta_{\text {B}} = -1$.
Eq.~\eqref{ampproba-sts-indist-dist} further simplifies to 
\begin{multline}
\langle  \, d_{\text {AA}}, d_{\text {BB}} ; \eta_{\text {A}}, \eta_{\text {B}}  \big|  \, d_{\text {AB}}, d_{\text {AB}} ; +1 \big\rangle^\text{ns} =  
\delta_{\eta_{\text {A}} \eta_{\text {B}}, +1} \times \\
\sum_{ m_{n_\text{A}}} \sum_{m_{n_\text{B}}}    \delta_{m_{n_\text{A}} + m_{n_\text{B}}, m_{l_r} - m_{l_p}} \\
 \sum_{ m_{\text{A}_1} = -i_{\text{A}}}^{+i_{\text{A}}} \sum_{m_{\text{A}_2} \ge m_{\text{A}_1} }   \, \sum_{ m_{\text{B}_1} = -i_{\text{B}}}^{+i_{\text{B}}}
 \sum_{m_{\text{B}_2} \le m_{\text{B}_1} }   \\
  \, \big\langle d_{\text {AA}} ; \eta_{\text {A}} \, \big| \, n_{\text{A}} \, m_{n_\text{A}} \, m_{\text{A}_1} \, m_{\text{A}_2}; \eta_{\text {A}}  \big\rangle  \\ \big\langle d_{\text {BB}} ; \eta_{\text {B}} \, \big| \, n_{\text{B}} \, m_{n_\text{B}} \, m_{\text{B}_1} \, m_{\text{B}_2} ; \eta_{\text {B}}   \big\rangle \\ 
     \big\langle m_{\text{A}_1} \, m_{\text{B}_1}  \, \big| \, d_{\text {AB}} \big\rangle \, \big\langle m_{\text{A}_2} \, m_{\text{B}_2} \, \big| \, d_{\text {AB}} \big\rangle  \\
 \    \delta_{m_{\text{A}_1}+m_{\text{B}_1}, m_1} 
\ \delta_{m_{\text{A}_2}+m_{\text{B}_2}, m_1}  .
\label{ampproba-sts-indist}
\end{multline}
This is the case treated in Sec.~\ref{Application}.
In that section, we will also define
a symmetric state probability
\begin{eqnarray}
P_{i \to j}^{S \, \text{ns}} = \bigg| \langle  \, d_{\text {AA}}, d_{\text {BB}} ; +1, +1 \big|  \, d_{\text {AB}}, d_{\text {AB}} ; +1 \big\rangle^\text{ns} \bigg|^2, \nonumber \\
\label{proba-sts-S}
\end{eqnarray}
the probability that corresponds
to the case of a nuclear spin symmetric wavefunction of the AA and BB products
$\eta_{\text {A}} = \eta_{\text {B}} = +1$,
and an anti-symmetric state probability
\begin{eqnarray}
P_{i \to j}^{A \, \text{ns}} = \bigg| \langle  \, d_{\text {AA}}, d_{\text {BB}} ; -1, -1 \big|  \, d_{\text {AB}}, d_{\text {AB}} ; +1 \big\rangle^\text{ns} \bigg|^2, \nonumber \\
\label{proba-sts-A}
\end{eqnarray}
the one that corresponds
to the case of a nuclear spin anti-symmetric wavefunction of the AA and BB products
$\eta_{\text {A}} = \eta_{\text {B}} = -1$.
As it will be discussed in Sec.~\ref{Application}, the symmetric or anti-symmetric state probabilities
imply a specific parity of the rotational quantum numbers of the molecular products, depending on the bosonic or fermionic character of the involved atoms.  \\

\paragraph{Distinguishable case}

If the reactants are distinguishable,
both components $\eta=\pm1$ have to be computed and
Eq.~\eqref{ampproba-sts-indist-dist} 
becomes
\begin{multline}
\langle  \, d_{\text {AA}}, d_{\text {BB}} ; \eta_{\text {A}}, \eta_{\text {B}}  \big|  \, d_{\text {AB}_1}, d_{\text {AB}_2} ; \eta \big\rangle^\text{ns} =  
\delta_{\eta_{\text {A}} \eta_{\text {B}}, \eta} \times \\
\sum_{ m_{n_\text{A}}} \sum_{m_{n_\text{B}}}  \delta_{m_{n_\text{A}} + m_{n_\text{B}}, m_{l_r} - m_{l_p}} \\
\frac{1}{\sqrt{2}} \,
  \sum_{ m_{\text{A}_1} = -i_{\text{A}}}^{+i_{\text{A}}} \sum_{m_{\text{A}_2} \ge m_{\text{A}_1} }   \, \sum_{ m_{\text{B}_1} = -i_{\text{B}}}^{+i_{\text{B}}}
 \sum_{m_{\text{B}_2} \le m_{\text{B}_1} }   \\
  \, \big\langle d_{\text {AA}} ; \eta_{\text {A}} \, \big| \, n_{\text{A}} \, m_{n_\text{A}} \, m_{\text{A}_1} \, m_{\text{A}_2}; \eta_{\text {A}}  \big\rangle  \\ \big\langle d_{\text {BB}} ; \eta_{\text {B}} \, \big| \, n_{\text{B}} \, m_{n_\text{B}} \, m_{\text{B}_1} \, m_{\text{B}_2} ; \eta_{\text {B}}   \big\rangle \\ 
     \bigg\{  
   \big\langle m_{\text{A}_1} \, m_{\text{B}_1}  \, \big| \, d_{\text {AB}_1} \big\rangle \, \big\langle m_{\text{A}_2} \, m_{\text{B}_2} \, \big| \, d_{\text {AB}_2} \big\rangle   \\
\shoveright{   + \eta \, \big\langle m_{\text{A}_2} \, m_{\text{B}_2}  \, \big| \, d_{\text {AB}_1} \big\rangle \, \big\langle m_{\text{A}_1} \, m_{\text{B}_1} \, \big| \, d_{\text {AB}_2} \big\rangle 
     \bigg\}  } \\
  \delta_{m_{\text{A}_1}+m_{\text{B}_1}, m_1} 
\ \delta_{m_{\text{A}_2}+m_{\text{B}_2}, m_1} .
\label{ampproba-sts-dist}
\end{multline}

\subsection{Sum over the state-to-state probabilities}

For a given rotational transition, 
the probability summed over all the 
final combined dressed states of the products $| j \rangle $
simplifies due to appropriate closure relations as derived in Appendix C,
and is given by
\begin{eqnarray}
\sum_j P_{i \to j} =  P^\text{rot} \times \sum_j P_{i \to j}^\text{ns}  .
\end{eqnarray}
We quote only the expression for the indistinguishable case here
\begin{multline}
\sum_j P_{i \to j}^\text{ns}  = 
\delta_{\eta_{\text {A}} \eta_{\text {B}}, +1} \times \\ 
 \sum_{ m_{\text{A}_1} = -i_{\text{A}}}^{+i_{\text{A}}} \sum_{m_{\text{A}_2} \ge m_{\text{A}_1} }   \, \sum_{ m_{\text{B}_1} = -i_{\text{B}}}^{+i_{\text{B}}}
 \sum_{m_{\text{B}_2} \le m_{\text{B}_1} }   \\
    \bigg|     \big\langle m_{\text{A}_1} \, m_{\text{B}_1}  \, \big| \, d_{\text {AB}} \big\rangle \, \big\langle m_{\text{A}_2} \, m_{\text{B}_2} \, \big| \, d_{\text {AB}} \big\rangle   \bigg|^2 \\  \delta_{m_{\text{A}_1}+m_{\text{B}_1}, m_1} 
\ \delta_{m_{\text{A}_2}+m_{\text{B}_2}, m_1}   .
\label{proba-sts-tot}
\end{multline}
This is the generalized form of the expression used 
in Ref.~\cite{Hu_NC_13_435_2021} to compare with experimental data.
Similar to Eq.~\eqref{proba-sts-S}
and Eq.~\eqref{proba-sts-A},
two types of sums can be computed,
$\sum_j P_{i \to j}^{S \, \text{ns}}$ 
and $\sum_j P_{i \to j}^{A \, \text{ns}}$.
From the conservation of the total probability, the sum of them has to be unity: 
$\sum_j P_{i \to j}^{S \, \text{ns}} + 
\sum_j P_{i \to j}^{A \, \text{ns}}  = 1$.
We show here that this summed probability only requires the eigenfunctions
of the dressed reactants in the magnetic field, not the ones of the dressed products.
Similar simplifications are found for the other cases, as discussed in Appendix C.

\section{Application to state-to-state chemical reactions}
\label{Application}

To illustrate our model, we present results
of nuclear spin state-to-state distributions in a magnetic field $B$
for reactants in the ground rotational states
$n_1 = n_2 = 0$ starting with bi-alkali molecules
in indistinguishable states (so that $m_{n_1} = m_{n_2} = 0$, $\eta = +1$ and $m_1 = m_2$).
We assume the molecules being prepared in their electronic $X^1\Sigma^+$ and vibrational $v=0$
ground state.
To simplify the calculation
of the interaction of the molecules
with the magnetic field, we also consider
products in the ground rotational states 
$n_{\text {A}} = 0$ and $n_{\text {B}} = 0$
(so that $m_{n_\text{A}} = m_{n_\text{B}} = 0$, $(\eta_\text{A}, \eta_\text{B}) = (+1, +1)$ or $(\eta_\text{A}, \eta_\text{B})= (-1, -1)$, and $m_1 = m_2$).
Due to symmetry reasons, having both type of products in 
the zero rotational state implies 
necessarily both molecular products 
with all the atoms involved of either bosonic character
of fermionic character.
When the atoms are all fermionic, 
the only possible reaction with bi-alkali molecules is
$^6$Li$^{40}$K + $^6$Li$^{40}$K $\to$ $^6$Li$_2$ + $^{40}$K$_2$,
and the nuclear spin state distribution
is symmetric $(\eta_\text{A}, \eta_\text{B})= (+1, +1)$.
When the atoms are all bosonic, 
for example in the reaction
$^{41}$K$^{87}$Rb + $^{41}$K$^{87}$Rb $\to$ $^{41}$K$_2$ + $^{87}$Rb$_2$, 
the nuclear spin state distribution
is anti-symmetric $(\eta_\text{A}, \eta_\text{B})= (-1, -1)$.
We discuss these two types of distribution in the following, applying the present model to these two systems 
taken as examples.

\subsection{$^6$Li$^{40}$K + $^6$Li$^{40}$K $\to$ $^6$Li$_2$ + $^{40}$K$_2$ chemical reaction}

\begin{figure}[h]
\begin{center}
\includegraphics*[width=8.cm, trim=0cm 0cm 0cm 0cm]{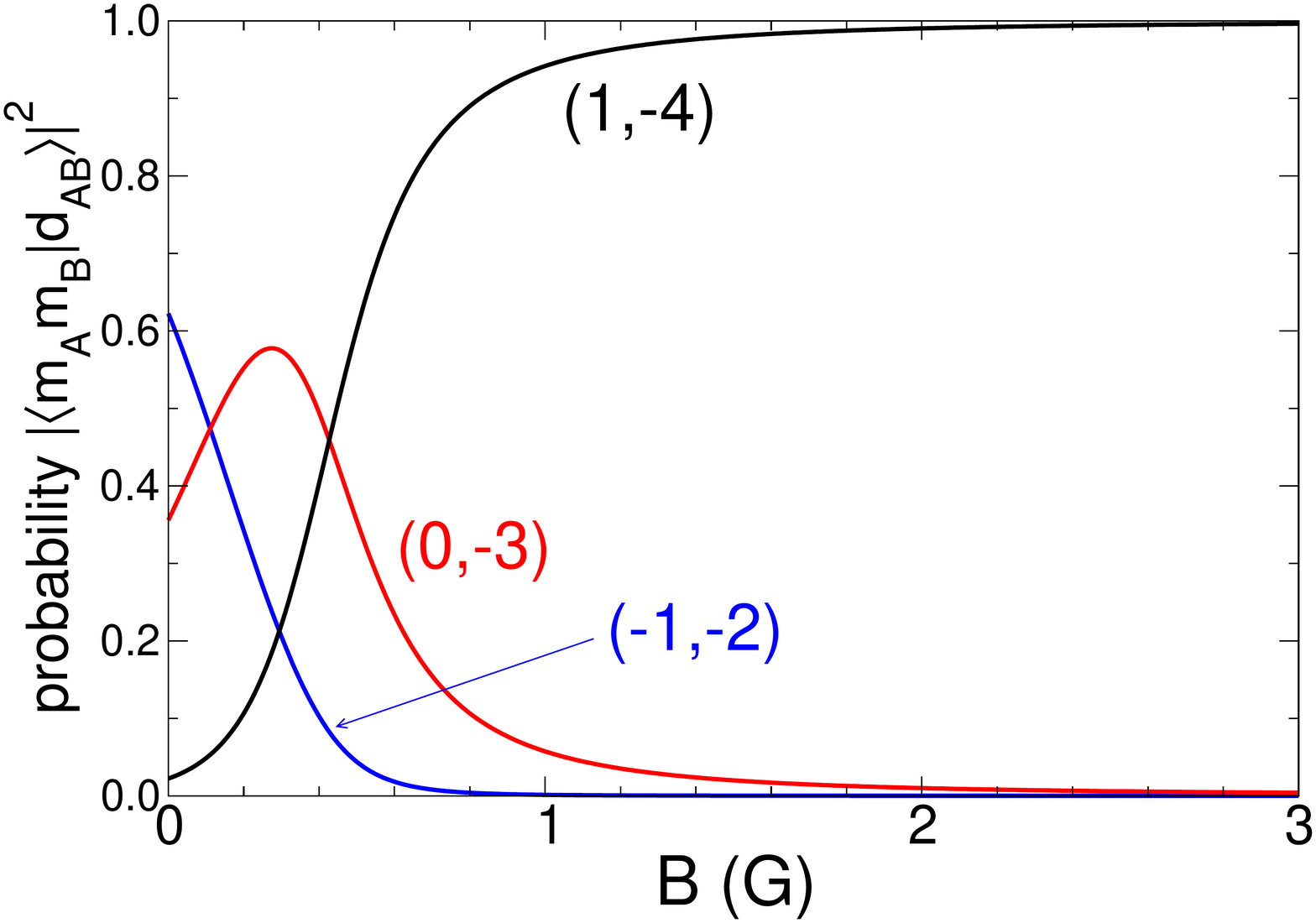}
\caption{The quantity 
$\big| \big\langle m_{\text{A}} \, m_{\text{B}} \, \big| \, d_{\text{AB}} \big\rangle \big|^2$ 
gives the probability of a component of the wavefunction of a $^6$Li$^{40}$K molecule in the zero rotational state
as a function of the magnetic field $B$, for its first dressed state $\big| d_{\text{AB}} \big\rangle = 1$
of the $m_1=m_2=-3$ manifold. 
The components of the wavefunction are denoted ($m_{\text{A}}, m_{\text{B}}$). The black, red, blue curves correspond respectively to the components ($1, -4$), ($0, -3$), ($-1, -2$).}
\label{FIG-EIGENVEC-LIK-SYM}
\end{center}
\end{figure}

Among the list of the ten bi-alkali possible as reactants, 
only the bosonic $^6$Li$^{40}$K molecule can illustrate
the example of a distribution with
symmetric nuclear spin states
with both products in the ground rotational state.
For symmetry reasons, this is indeed the only heteronuclear bi-alkali molecule
that can be made with two different fermionic atoms.
We consider that they are all prepared in the lowest state of the $m_1 = m_2 = -3$ manifold.
At large magnetic fields,
the Zeeman interaction starts to dominate over the other hyperfine interactions 
($B > 1$ G for this system) and the nuclear spin projections become 
good quantum numbers.
Therefore, the dressed states correspond to a nearly pure character 
of ${m}_{\text{Li}}$, ${m}_{\text{K}}$,
the quantum numbers associated with 
projections of the nuclear spins of the corresponding isotopes
${i}_{\text{Li}} = 1$ and ${i}_{\text{K}} = 4$ 
onto the magnetic field axis.
This can be seen in Fig.~\ref{FIG-EIGENVEC-LIK-SYM} as a black line, where
the probability of a component of the wavefunction of a $^6$Li$^{40}$K molecule
is plotted as a function of the magnetic field $B$ in its first dressed state $\big| d_{\text{AB}} \big\rangle = 1$ for $m_1=m_2=-3$. 
At lower magnetic fields, $\big| d_{\text{AB}} \big\rangle = 1$ can gain other characters, 
as other components of the LiK wavefunction,
plotted as the red and blue lines are also present due to the fact that other hyperfine interactions
cannot be neglected.
As the formalism assumes that the nuclear spins are spectators during the chemical reaction,
we expect that the chemical reaction favors at large magnetic fields 
a simple re-arrangement of the original nuclear spins of the initial reactants $(1, -4) + (1,-4)$
into the products, namely $(1, 1) + (-4, -4)$ by simply swapping them.
In the following, the short-hand notation ($m_{\text{A}_{1}}, m_{\text{B}_{1}}$) + 
($m_{\text{A}_{2}}, m_{\text{B}_{2}}$) will be sometimes used to express the main characters 
of the reactants,
and ($m_{\text{A}_{1}}, m_{\text{A}_{2}}$) +  ($m_{\text{B}_{1}}, m_{\text{B}_{2}}$) 
of the products. \\

To obtain the eigenfunctions and eigenenergies of the LiK, Li$_2$ and K$_2$ 
molecules in a magnetic field,
we diagonalize the corresponding molecular Hamiltonians 
using the basis sets of the nuclear spin states Eq.~\eqref{barestatereac}
and Eq.~\eqref{barestateprod}.
In the following we will take A = $^{6}$Li, B = $^{40}$K.
It is then obvious that the notations AA and BB used for the theoretical 
developments in the previous section stand for the notations of the molecules Li$_2$ and K$_2$
in the following.
As we consider only molecules in zero rotational states (see discussion about the second assumption), 
the form of the molecular Hamiltonian simplifies and reduces to
\cite{Aldegunde_PRA_78_033434_2008,Aldegunde_PRA_79_013401_2009,Aldegunde_PRA_96_042506_2017}
\begin{eqnarray} \label{Hamiltonian}
H = H_{hf} + H_{Z}
\end{eqnarray} 
with the hyperfine and Zeeman Hamiltonian for the AB molecule given respectively  by
\begin{eqnarray} 
H_{hf} &=& c_\text{AB} \ \vec{i}_\text{A}.\vec{i}_{\text {B}}  \nonumber \\  
H_{Z} &=& 
- g_\text{A} \, \mu_N \, \vec{i}_\text{A}.\vec{B} \, (1 - \sigma_{\text{A of AB}})  \nonumber \\  
& & - g_\text{B} \, \mu_N \, \vec{i}_\text{B}.\vec{B} \, (1 - \sigma_{\text{B of AB}}) ,
\end{eqnarray} 
for the AA molecule given respectively by
\begin{eqnarray} 
H_{hf} &=& c_{\text{AA}} \ \vec{i}_{\text{A}_1}.\vec{i}_{\text{A}_2} \nonumber \\  
H_{Z} &=& 
- g_\text{A} \, \mu_N \, \vec{i}_{\text{A}_1}.\vec{B} \, (1 - \sigma_{\text{A of AA}})  \nonumber \\  
& & - g_\text{A} \, \mu_N \, \vec{i}_{\text{A}_2}.\vec{B} \, (1 - \sigma_{\text{A of AA}}) ,
\end{eqnarray} 
and for the BB molecule given respectively by
\begin{eqnarray} 
H_{hf} &=& c_{\text{BB}} \ \vec{i}_{\text{B}_1}.\vec{i}_{\text{B}_2} \nonumber \\  
H_{Z} &=& 
- g_\text{B} \, \mu_N \, \vec{i}_{\text{B}_1}.\vec{B} \, (1 - \sigma_{\text{B of BB}})  \nonumber \\  
& & - g_\text{B} \, \mu_N \, \vec{i}_{\text{B}_2}.\vec{B} \, (1 - \sigma_{\text{B of BB}}) .
\end{eqnarray} 
We took $c_\text{AB} = 48.2$~Hz,  $g_\text{A} = 0.822$, $g_\text{B} = -0.324$, 
$\sigma_{\text{A of AB}} = 104.1$ ppm, and $\sigma_{\text{B of AB}} = 1296.8 $ ppm
from  Ref.~\cite{Aldegunde_PRA_96_042506_2017}.
We took $c_\text{AA} = 161$~Hz, $c_\text{BB} = -42$~Hz, 
$\sigma_{\text{A of AA}} = 102$ ppm, and $\sigma_{\text{B of BB}} = 1313 $ ppm
from Ref.~\cite{Aldegunde_PRA_79_013401_2009}.
For two molecules, we have $m_1 + m_2 = -6$
which is conserved during the collision.
The products of the chemical reaction $^{6}$Li$_2$ and $^{40}$K$_2$ are
formed in their ground electronic state $X^1\Sigma_g^+$, ground vibrational state $v=0$, 
and in the ground rotational state $n_{\text {A}} = n_{\text {B}} = 0$. \\

\begin{figure}[h]
\begin{center}
\includegraphics*[width=8.cm, trim=0cm 0cm 0cm 0cm]{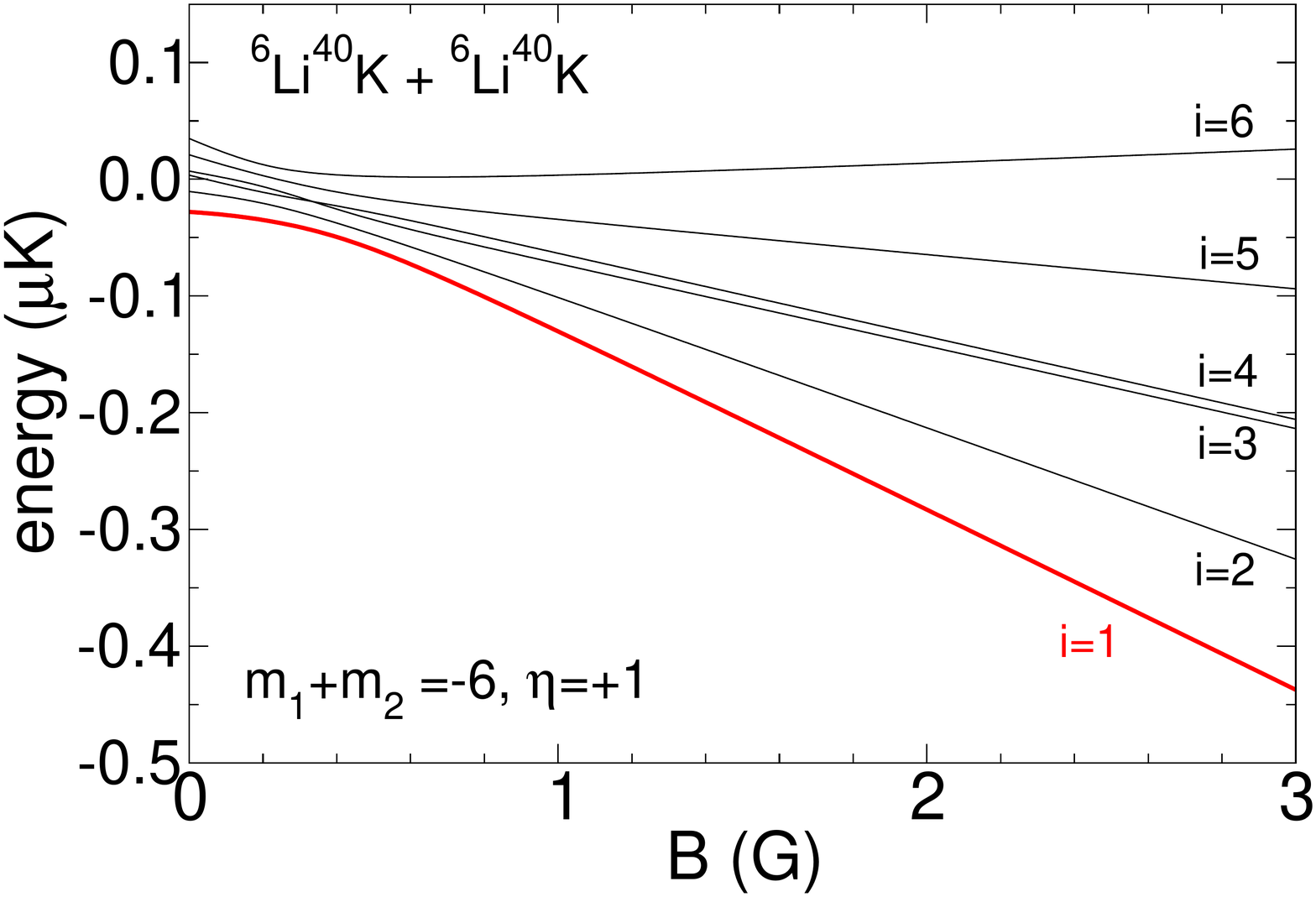}
\caption{Energies of the combined dressed states of the reactants 
$^{6}$Li$^{40}$K + $^{6}$Li$^{40}$K in the zero rotational states
as a function of the magnetic field $B$,
for $m_1+m_2=-6$ and $\eta = +1$.
There are 6 possible  $| i \rangle$ states as labeled on the figure. 
The $| i \rangle = 1$ state 
corresponds to two molecules in the state
$\big| d_{\text{AB}_1} \big\rangle = \big| d_{\text{AB}_2} \big\rangle = 1$.}
\label{FIG-NRG2PLE-LIK-LIK-SYM}
\end{center}
\end{figure}

\begin{table}[h]
\setlength{\extrarowheight}{4pt}
\begin{center}
\begin{tabular}{|ccc|cc|cc|}
\hline
                               $| i \rangle $ &  $\big| d_{\text{AB}_1} \big\rangle $ &  $\big| d_{\text{AB}_2} \big\rangle $ & ${m}_{\text{A}_1}$ & ${m}_{\text{B}_1}$ & ${m}_{\text{A}_2}$ & ${m}_{\text{B}_2}$ \\ \hline 
        1 & 1 & 1 & 1 & -4 & 1 & -4 \\ \hline
           2 & 1 & 2 & 1 & -4  & 0 & -3 \\ \hline
              3 & 1 & 3 & 1 & -4 & -1 & -2 \\ \hline
                 4 & 2 & 2 & 0 & -3 & 0 & -3 \\ \hline
                    5 & 2 & 3 & 0 & -3 & -1 & -2 \\ \hline
                       6 & 3 & 3 & -1 & -2  & -1 & -2  \\ 
\hline
\end{tabular}
\caption{Nomenclature for the combined dressed states of the reactants 
AB + AB = $^{6}$Li$^{40}$K + $^{6}$Li$^{40}$K
in the zero rotational states
for $\eta = +1$ and $m_1+m_2 = -6$, with A = $^{6}$Li, B = $^{40}$K.
The combined dressed states are denoted $| i \rangle$,
corresponding  to the combination of the dressed states 
$\big| d_{\text{AB}_1} \big\rangle $
and 
$\big| d_{\text{AB}_2} \big\rangle $.
In this study, the initial state will be $| i \rangle =1$,
corresponding to the case where 
$\big| d_{\text{AB}_2} \big\rangle \equiv \big| d_{\text{AB}_1} \big\rangle$.
The second and third columns display the main character of 
$\big| d_{\text{AB}_1} \big\rangle $
and 
$\big| d_{\text{AB}_2} \big\rangle $.
This is given at a large magnetic field, typically $B > 1$ G for this system,
as the dressed states tend to be nearly pure bare states.
}
\label{TableLIKLIKs}
\end{center}
\end{table}

Fig.~\ref{FIG-NRG2PLE-LIK-LIK-SYM} presents the energies of the
combined dressed states of the reactants 
AB + AB = $^{6}$Li$^{40}$K + $^{6}$Li$^{40}$K 
in the zero rotational states
as a function of the magnetic field,
for $m_1+m_2=-6$ and $\eta = +1$ for the case of indistinguishable states.
There are 6 dressed states denoted by $\big| i \big\rangle = 1, ... , 6$,
which correspond to different combinations of the individual dressed states
$\big| d_{\text{AB}} \big\rangle $
in the $m_1+m_2 = -6$ manifold.
Tab.~\ref{TableLIKLIKs} provides the nomenclature for these states.
We take $| i \big\rangle = 1$ as an example of initial state that could be prepared in experiments, 
presented as a red bold line in the figure.
The $| i \rangle = 1$ state corresponds to two molecules in 
the first dressed state $\big| d_{\text{AB}} \big\rangle = 1$.
At large magnetic fields, 
the initial state $| i \rangle = 1$ 
has a main character of ${m}_{\text{A}_1}=1$, ${m}_{\text{B}_1}=-4$, ${m}_{\text{A}_2}=1$, ${m}_{\text{B}_2}=-4$
and at lower fields it can gain other characters, 
as discussed above for the individual reactants. 
 \\

\begin{figure}[h]
\begin{center}
\includegraphics*[width=8.cm, trim=0cm 0cm 0cm 0cm]{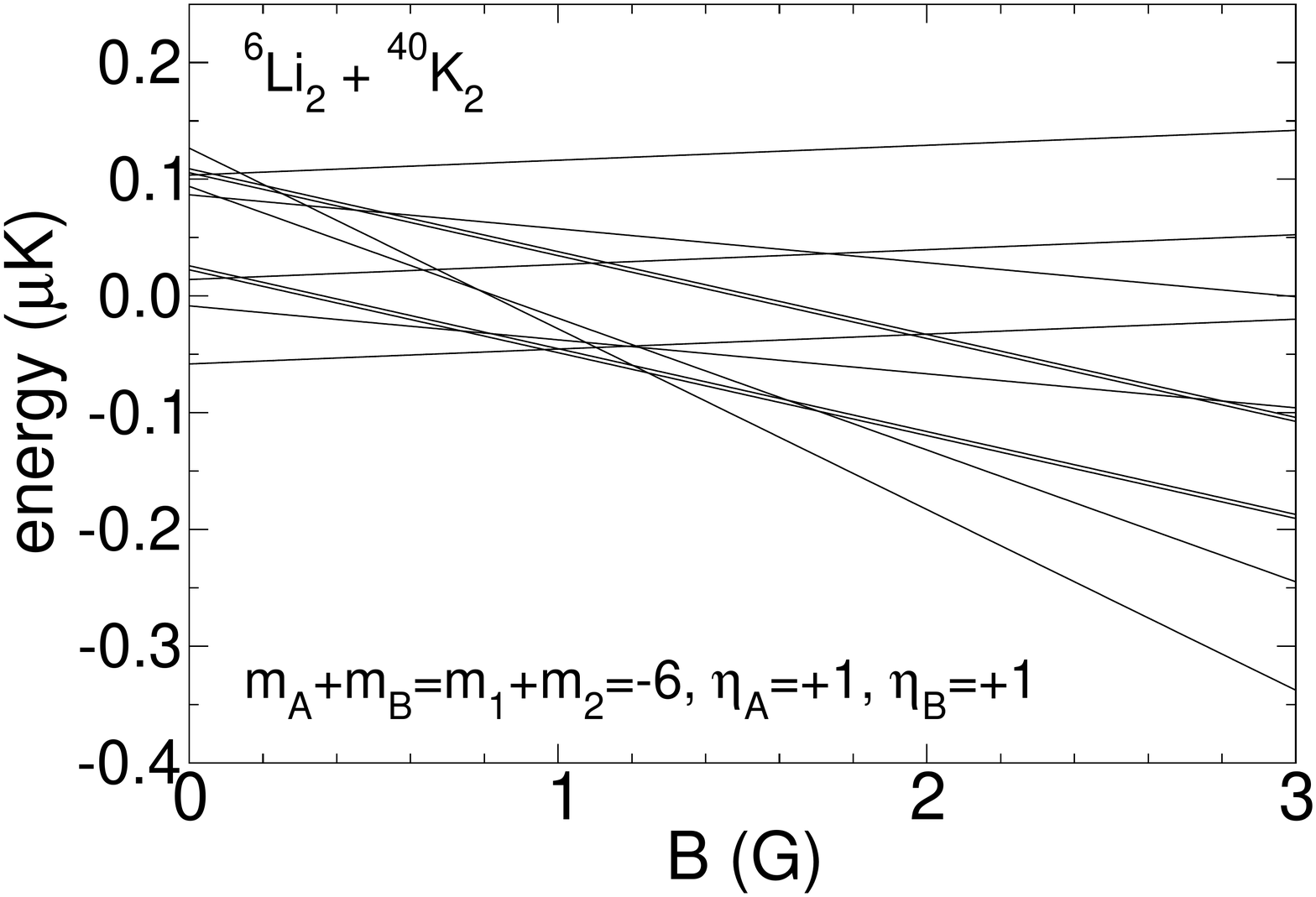}
\caption{Energies of the combined dressed states 
of the products $^{6}$Li$_2$ + $^{40}$K$_2$ in the zero rotational states
for $m_{\text {A}}+m_{\text {B}}=m_1+m_2=-6$ and $\eta_\text{A} = +1$, $\eta_\text{B} = +1$ (nuclear spin symmetric states).}
\label{FIG-NRG2PLE-LI2K2-SYM}
\end{center}
\end{figure}

Similarly, Fig.~\ref{FIG-NRG2PLE-LI2K2-SYM} 
presents the energies of the combined dressed states of the products 
AA + BB = $^{6}$Li$_2$ + $^{40}$K$_2$
in the zero rotational states
as a function of the magnetic field,
for $m_{\text {A}}+m_{\text {B}}=m_1+m_2=-6$ and $\eta_\text{A} = +1$, $\eta_\text{B} = +1$ (nuclear spin symmetric states).
$^{6}$Li and $^{40}$K nuclei are composite bosons. The permutation of two identical bosonic nuclei
in the $^{6}$Li$_2$ molecule and in the $^{40}$K$_2$ molecule 
should obey the symmetrization principle and lead to an overall symmetric wavefunction for each molecules.
As the rotational wavefunction of the $^{6}$Li$_2$ and $^{40}$K$_2$ molecules in the zero rotational state
are symmetric under the interchange of the nuclei, their nuclear spin wavefunctions 
have to be symmetric as well, and hence described by 
values of $\eta_\text{A} = +1$ and $\eta_\text{B} = +1$.
There are 11 dressed states denoted by $| j \rangle = 1, ... , 11$,
which correspond to different combinations of the individual dressed states
$\big| d_{\text{AA}} ; +1 \big\rangle $
and 
$\big| d_{\text{BB}} ; +1 \big\rangle $.
Tab.~\ref{TableLI2K2s} provides the nomenclature for the states.
These are all the possible final product states of the chemical reaction,
for the nuclear spin symmetric states of the molecules. \\

\begin{figure}[h]
\begin{center}
\includegraphics[width=8cm]{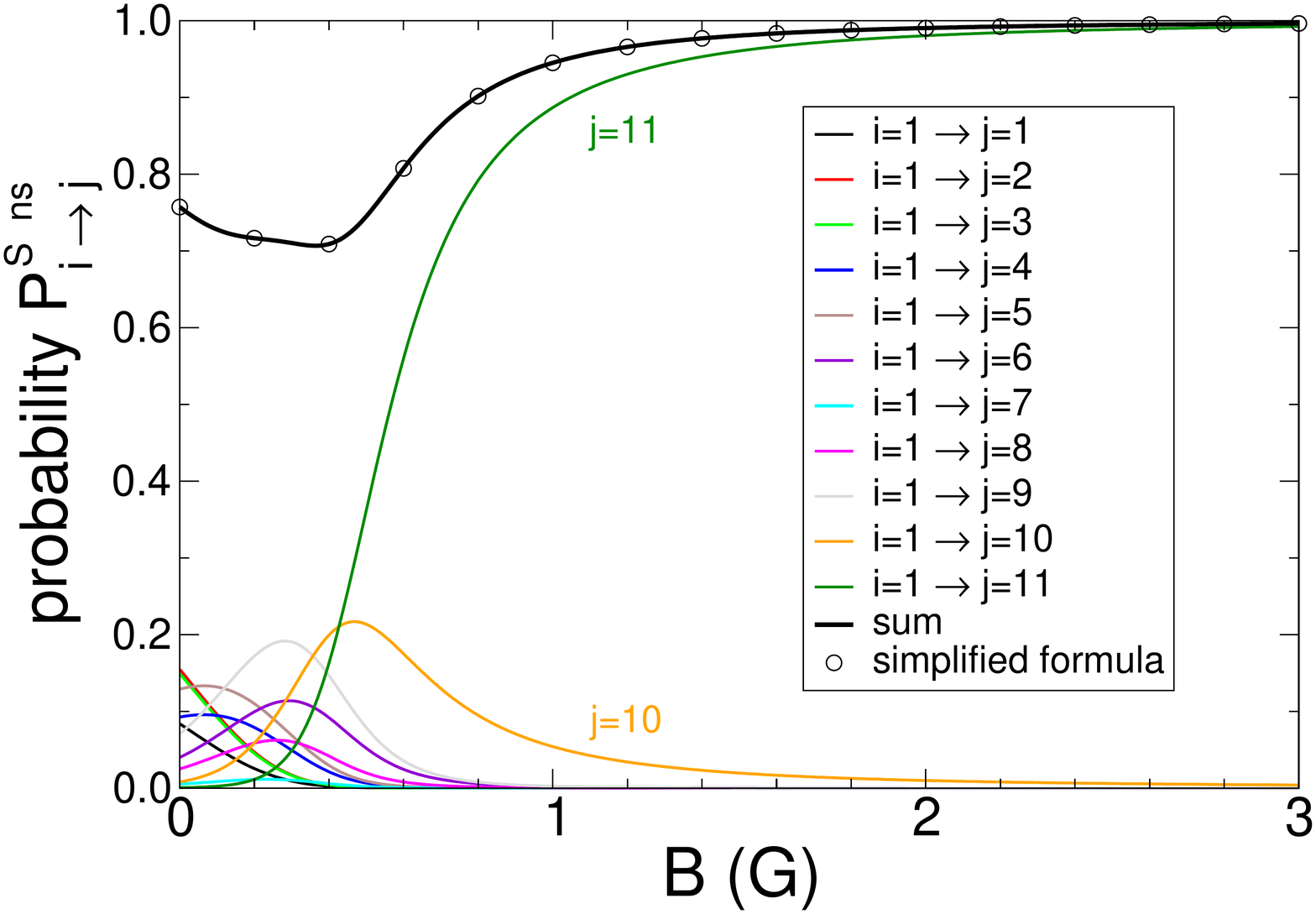} 
\includegraphics[width=8cm]{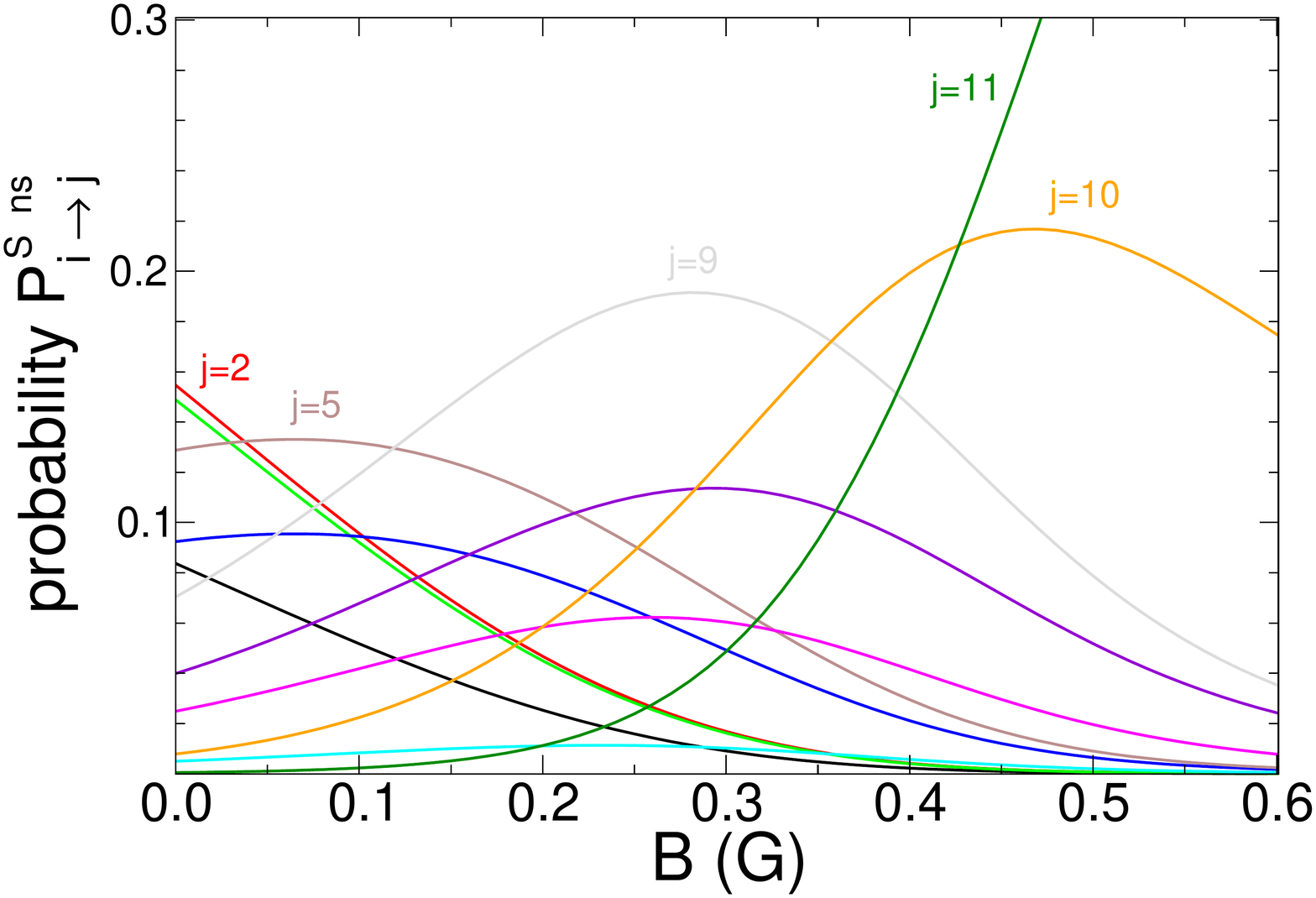} \\ 
\includegraphics[width=4cm]{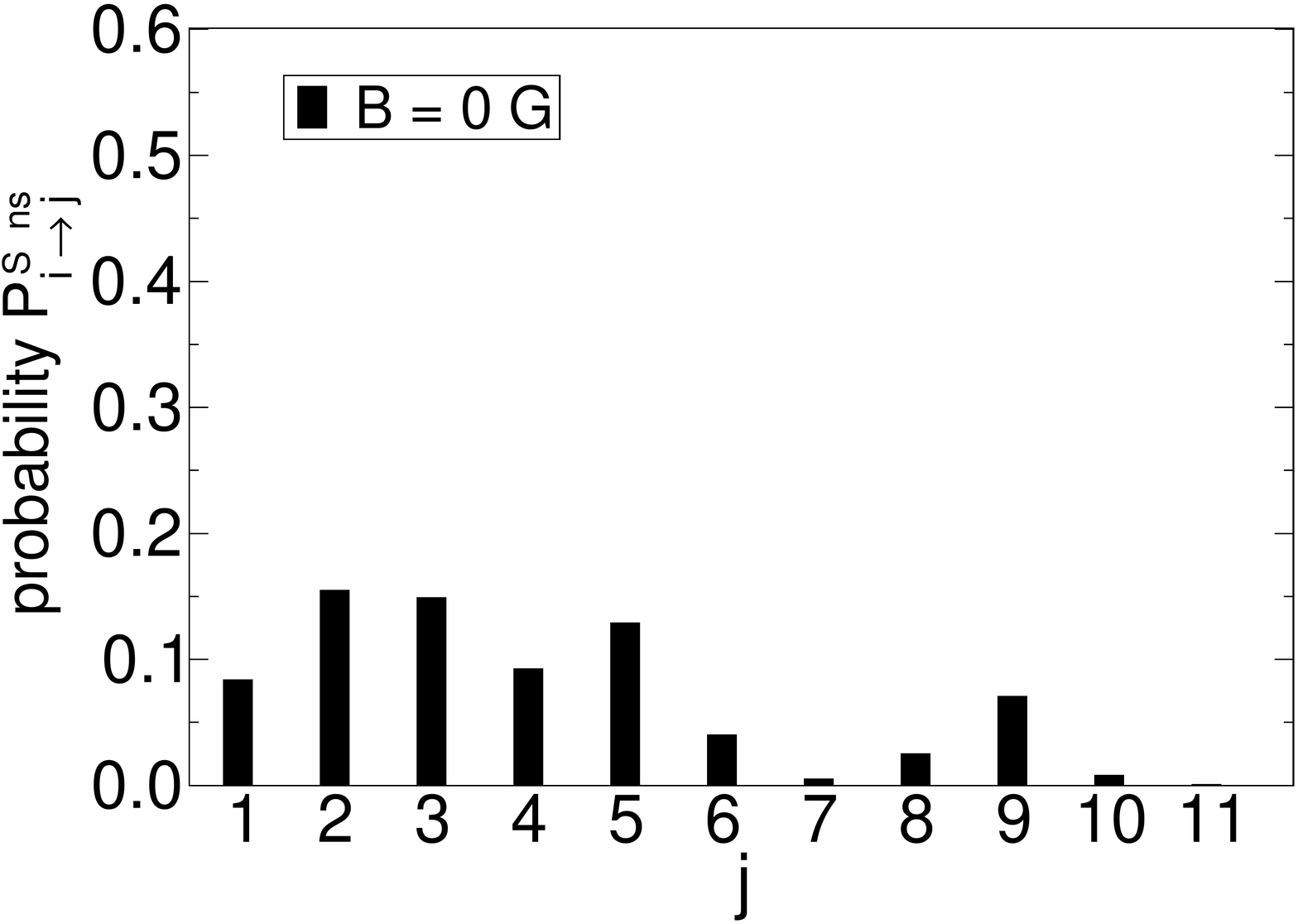} 
\includegraphics[width=4cm]{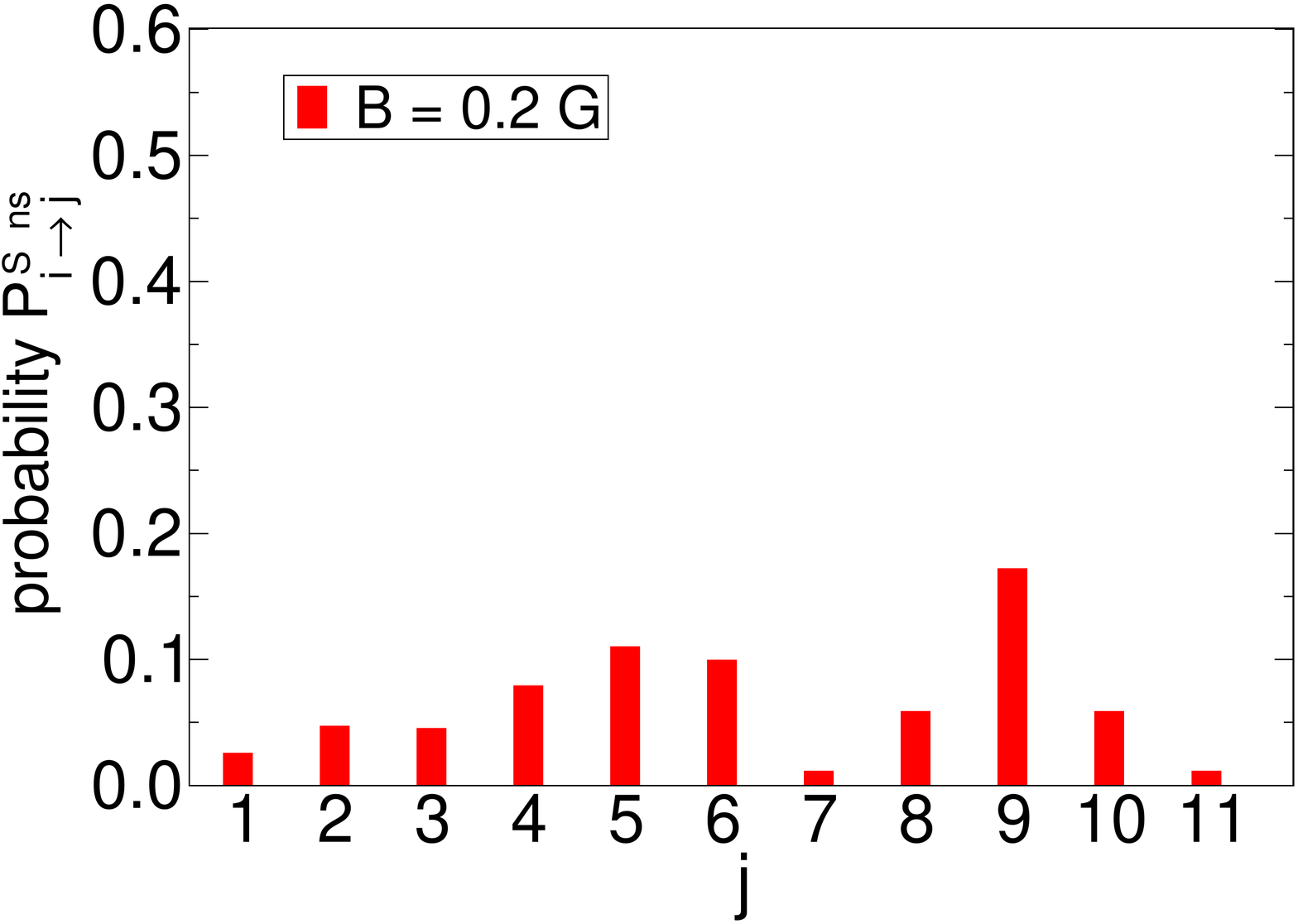} 
\includegraphics[width=4cm]{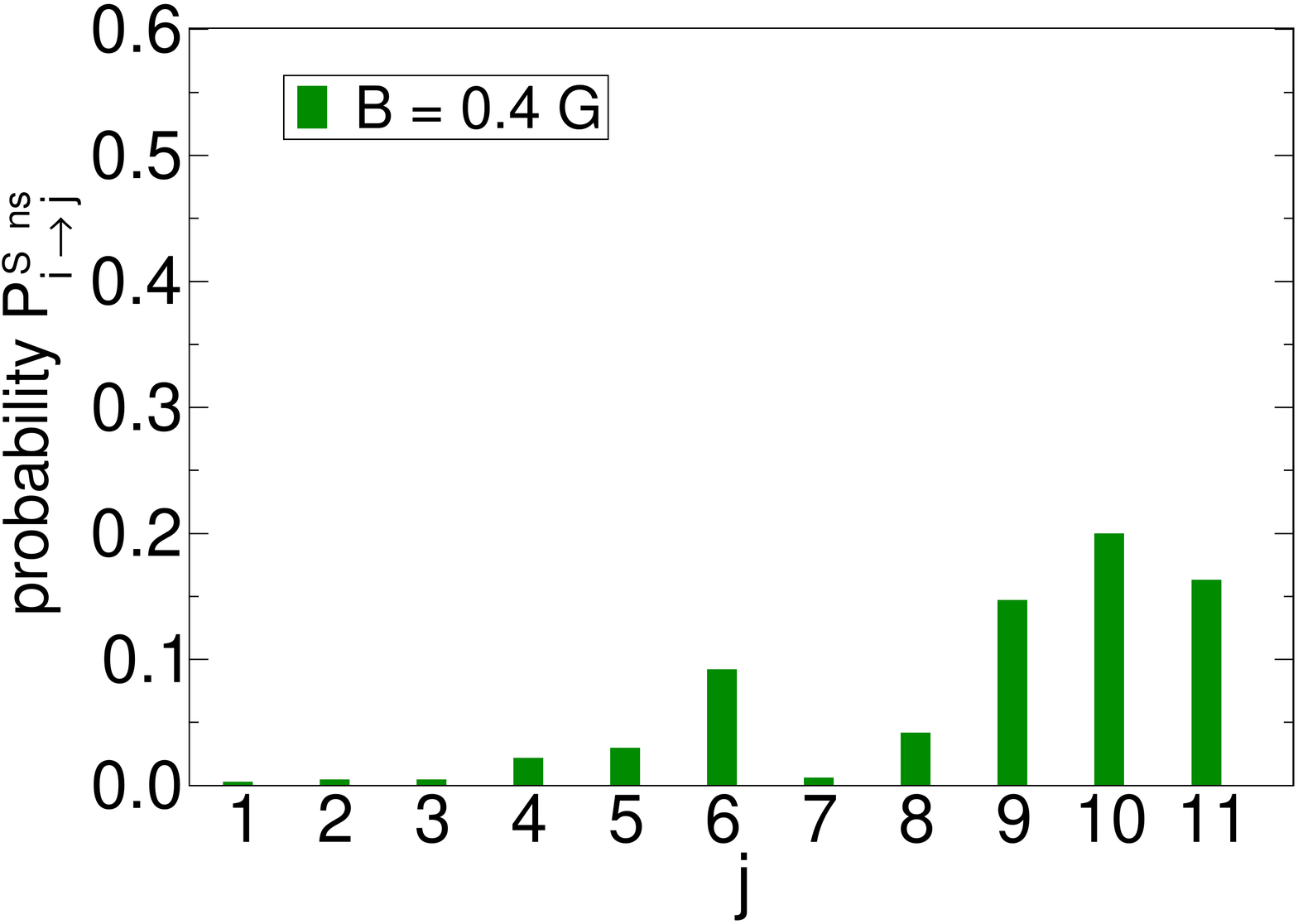} 
\includegraphics[width=4cm]{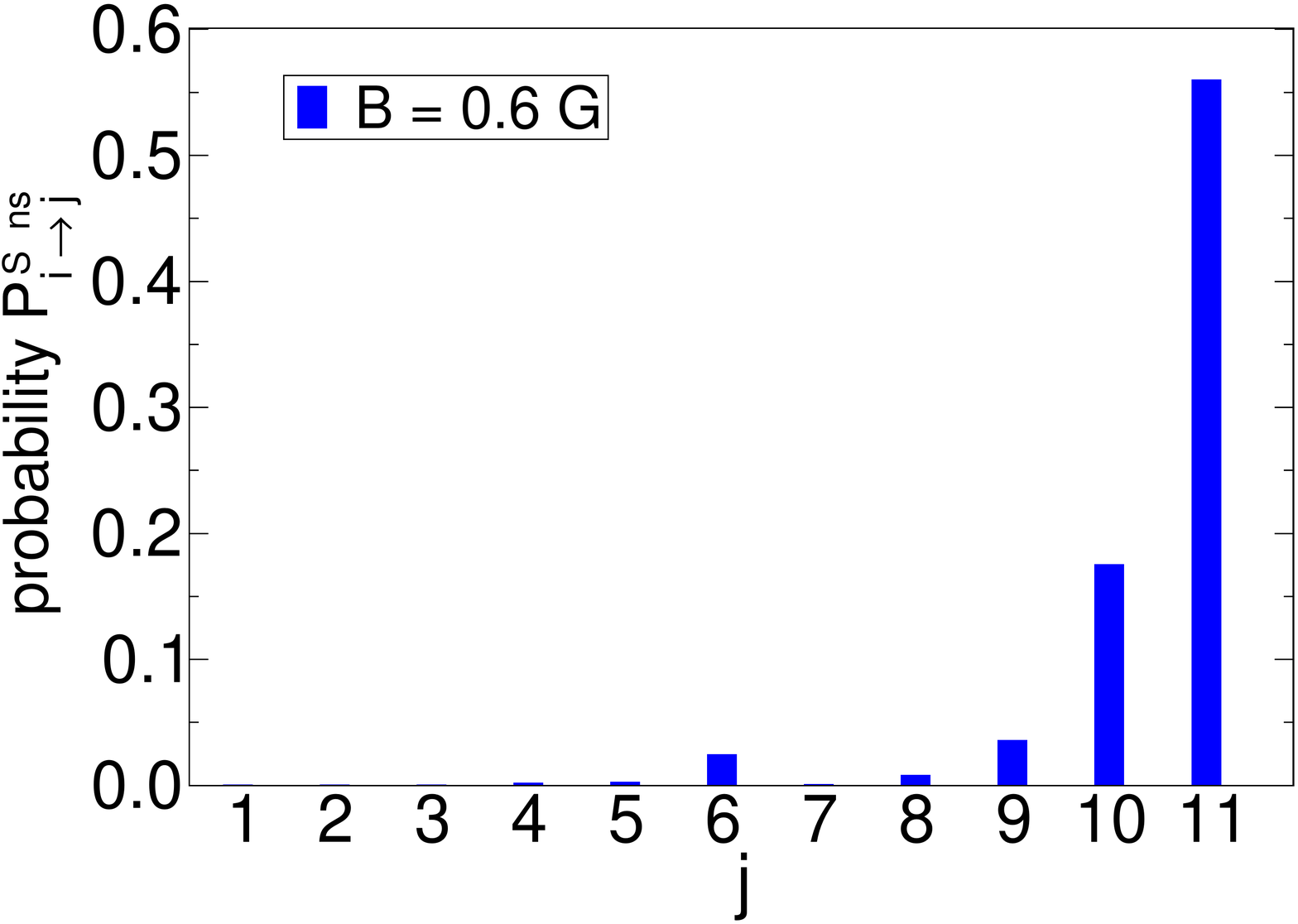} 
\caption{Top-left panel: Probability $P^{S \, \text{ns}}_{i \to j}$ 
for $^{6}$Li$_2$ + $^{40}$K$_2$
in zero rotational states 
as a function of the magnetic field $B$. 
The total sum is displayed as a thick black line.
The simplified formula (open circles) corresponds to Eq.~\eqref{proba-sts-tot}.
Top-right panel: zoom of the top-left one.
Bottom panels: Probability $P^{S \, \text{ns}}_{i \to j}$ 
as a function of the final states $| j \rangle = 1, ..., 11$ 
with $| i \rangle = 1$, for different magnetic fields.}
\label{FIG-PROBA-STS-SYM}
\end{center}
\end{figure}

From the eigenvectors of the combined dressed states expressed in the nuclear spin state basis set, namely 
$\big\langle m_{\text{A}_{1}} \, m_{\text{B}_{1}} \, \big| \, d_{\text {AB}} \big\rangle$ 
and  $\big\langle m_{\text{A}_{2}} \, m_{\text{B}_{2}} \, \big| \, d_{\text {AB}} \big\rangle$
for the reactants, 
$\big\langle m_{\text{A}_1} \, m_{\text{A}_2}; \eta_{\text {A}}   \, \big| \,  d_{\text {AA}} ; \eta_{\text {A}}\big\rangle $ and $\big\langle m_{\text{B}_1} \, m_{\text{B}_2}; \eta_{\text {B}}   \, \big| \,  d_{\text {BB}} ; \eta_{\text {B}}\big\rangle $ for the products (we omit the numbers
$n_{\text {A}}$, $m_{n_\text{A}}$, $n_{\text {B}}$, $m_{n_\text{B}}$), one can plot the 
nuclear spin state-to-state probability $P^{S \, \text{ns}}_{i \to j}$
expressed by Eq.~\eqref{proba-sts-S} and Eq.~\eqref{ampproba-sts-indist}
from the initial state $| i \rangle = 1 $ to all final states $| j \rangle$.
This is presented in the top panels of Fig.~\ref{FIG-PROBA-STS-SYM} as a function of the magnetic field.
The behavior of the probabilities as functions of 
$B$ really depends on the many admixtures of the bare states 
for the dressed states of the two reactants and the two products. 
The probabilities $P^{S \, \text{ns}}_{i \to j}$ are also plotted as functions of 
the final states $| j \rangle$ for different magnetic fields in the bottom panels of
Fig.~\ref{FIG-PROBA-STS-SYM}. 
It can be seen that the state-to-state distribution drastically changes 
with $B$ populating different final states. 
For $P^{S \, \text{ns}}_{i \to j}$, the $| j \rangle = 11$ final state (green curve)
is favored at high field while at low field it vanishes and other values 
of $|j \rangle$ are now more probable.
From Tab.~\ref{TableLI2K2s},
the $| j \rangle = 11$ state corresponds to a main character of $(1, 1) + (-4, -4)$,
entailing $^{6}$Li$_2$ and $^{40}$K$_2$ molecules with the same components 
of their atomic nuclear spins.
This state corresponds exactly to the original one $(1, -4) + (1, -4)$ of the reactants 
in the $| i \rangle = 1$ state, but with just a swap.
For lower magnetic fields (see the close-up figure), 
the chemical reaction ends up dominantly in other final product states,
respectively $| j \rangle = 10$ (orange curve), $| j \rangle = 9$ (grey curve), 
$| j \rangle = 5$ (brown curve), and $| j \rangle = 2$ (red curve)
for decreasing $B$, with a main character in $(0, 1) + (-3, -4)$, 
$(0, 0) + (-3, -3)$, $(-1, 0) + (-2, -3)$, and $(-1, -1) + (0, -4)$, respectively.
Then depending on the magnetic field that is applied, different types of states of
the molecular products can be predominantly composed.
Those with different nuclear spins, for example $(0, 1) + (-3, -4)$ for $| j \rangle = 10$
in the range $B = [0.35-0.4]$~G,
take the form of an entangled state, namely
$ \{  | 0, 1 \big\rangle + | 1, 0 \big\rangle  \} / \sqrt{2}$ for the $^{6}$Li nuclei, and 
$ \{  | -3, -4 \big\rangle + | -4, -3 \big\rangle  \} / \sqrt{2}$ 
for the $^{40}$K nuclei.
Those with same nuclear spins, for example $(1, 1) + (-4, -4)$ for $| j \rangle = 11$
at $B > 0.5$~G,
takes the form of a separable state, namely 
$| 1  \big\rangle \, | 1 \rangle $ for the $^{6}$Li nuclei,
and $| -4  \big\rangle \, | -4 \rangle $ for the $^{40}$K nuclei.
\\

The total probability $\sum_j P^{S \, \text{ns}}_{i \to j}$,
summed over all the 
final combined dressed states of the products $|  j \rangle $,
is reported as a black thick line in Fig.~\ref{FIG-PROBA-STS-SYM}
and consists of the sum of all individual curves given by
Eq.~\eqref{proba-sts-S}.
The  black thick line is compared with the simplified expression
given by Eq.~\eqref{proba-sts-tot}, 
presented as open circles for each figure.
We can see that the curve and the circle data are identical, 
so that this confirms that one can use the simplified formula Eq.~\eqref{proba-sts-tot} directly,
which requires only the knowledge of the eigenfunctions 
of the dressed states of the reactants.
There is no need to compute the dressed states of the products in a magnetic field as far as the 
total probability summed over all $|  j \rangle $ is concerned. 
This is what was done in \cite{Hu_NC_13_435_2021} for fermionic $^{40}$K$^{87}$Rb molecules, as the experimental data
involved a measurement of the total probability, summed over all possible nuclear spin states of the products. 
Eq.~\eqref{proba-sts-tot} was used to fit the y-axis of the experimental data,
which consisted of counting the product molecules 
(more precisely counting the ionized form of the product molecules).
A very good agreement was found in the trend of the data with the magnetic field,
confirming that the nuclear spin structure in the magnetic field is treated correctly 
and that the assumption of conserved nuclear spins is sufficient.
An overall fitting parameter was used to account for the overall magnitude of the counting
for each observed rotational state of the products, 
which is a characteristic of the intrinsic, complicated effect of the rotational structure of the molecules 
in the chemical reaction at short-range, something that is not taken into account in the present model. 
As the long-range physics where interactions between the magnetic field and nuclear spins occur is fully taken into account in our model, this overall fitting parameter 
depends only on the short-range physics. 
A recent experiment \cite{Liu_N_593_379_2021} suggested
that this fitting parameter can be simply obtained by a state-counting model based on a statistical theory argument
\cite{Gonzalez-Martinez_PRA_90_052716_2014,Bonnet_JCP_152_084117_2020}, when the products have a release of final kinetic energy larger than the initial one.

\clearpage

\onecolumngrid

\begin{table*}[t]
\setlength{\extrarowheight}{4pt}
\begin{center}
\begin{tabular}{|ccc|cc|cc|cc|cc|}
\hline
        $| j \rangle $ &  $\big| d_{\text{AA}} ; +1 \big\rangle $ &  $\big| d_{\text{BB}} ; +1 \big\rangle $ & ${m}_{\text{A}_1}$ & ${m}_{\text{A}_2}$ & ${m}_{\text{B}_1}$ & ${m}_{\text{B}_2}$  & ${m}_{\text{A}_1}$ & ${m}_{\text{A}_2}$ & ${m}_{\text{B}_1}$ & ${m}_{\text{B}_2}$ \\ \hline 
        1 & 1 & 7 & -1 & -1 & -1 & -3 &  &  & 0 & -4 \\ \hline
        2 & 1 & 8 & -1 & -1 & 0 & -4 &  &  & -2 & -2 \\ \hline
        3 & 1 & 9 & -1 & -1 & -1 & -3 &  &  & -2 & -2 \\ \hline
        4 & 2 & 5 & -1 & 0 &  -1 & -4  &  &  & -2 & -3 \\ \hline
        5 & 2 & 6 & -1 & 0 & -2 & -3 &  &  & -1 & -4 \\ \hline
        6 & 3 & 3 & -1 & 1 & -2 & -4  & 0 & 0 & -3 & -3 \\ \hline
        7 & 3 & 4 & -1 & 1 & -3 & -3  & 0 & 0  & -2 & -4 \\ \hline
        8 & 4 & 3 & 0 & 0 & -2 & -4  & -1 & 1  & -3 & -3  \\ \hline
        9 & 4 & 4 & 0 & 0 & -3 & -3  & -1 & 1 & -2 & -4 \\ \hline
      10 & 5 & 2 & 0 & 1 & -3 & -4 &  &  &  &  \\ \hline
      11 & 6 & 1 & 1 & 1 & -4 & -4 &  &  &  &  \\ 
\hline
\end{tabular}
\caption{Nomenclature for the combined dressed states of the products AA + BB
= $^{6}$Li$_2$ + $^{40}$K$_2$
for $\eta_\text{A} = +1$, $\eta_\text{B} = +1$ (nuclear spin symmetric states), 
with A = $^{6}$Li, B = $^{40}$K.
The combined dressed states are denoted by $| j \rangle$,
corresponding  to the combination of the dressed states 
$\big| d_{\text{AA}} ; \eta_{\text{A}} \big\rangle $
and 
$\big| d_{\text{BB}} ; \eta_{\text{B}} \big\rangle $.
The second and third columns display the main character of 
$\big| d_{\text{AA}} ; \eta_{\text{A}} \big\rangle $
and 
$\big| d_{\text{BB}} ; \eta_{\text{B}} \big\rangle $, while the fourth and fifth
display the second main character, if any.
This is given for large magnetic fields, typically $B > 1$ G for this system.
}
\label{TableLI2K2s}
\end{center}
\end{table*}

\twocolumngrid

\subsection{$^{41}$K$^{87}$Rb + $^{41}$K$^{87}$Rb $\to$ $^{41}$K$_2$ + $^{87}$Rb$_2$ chemical reaction}

The previous system is the only one involving
a state-to-state distribution of nuclear spin symmetric states $(\eta_\text{A}, \eta_\text{B}) = (+1, +1)$
if both products are in their ground rotational states.
To illustrate now 
a state-to-state distribution of nuclear spin anti-symmetric states $(\eta_\text{A}, \eta_\text{B}) = (-1, -1)$,
we take the example of bosonic $^{41}$K$^{87}$Rb molecules.
We consider that they are all prepared in the lowest state of the $m_1 = m_2 = -1$ manifold. \\

\begin{figure}[h]
\begin{center}
\includegraphics*[width=8.cm, trim=0cm 0cm 0cm 0cm]{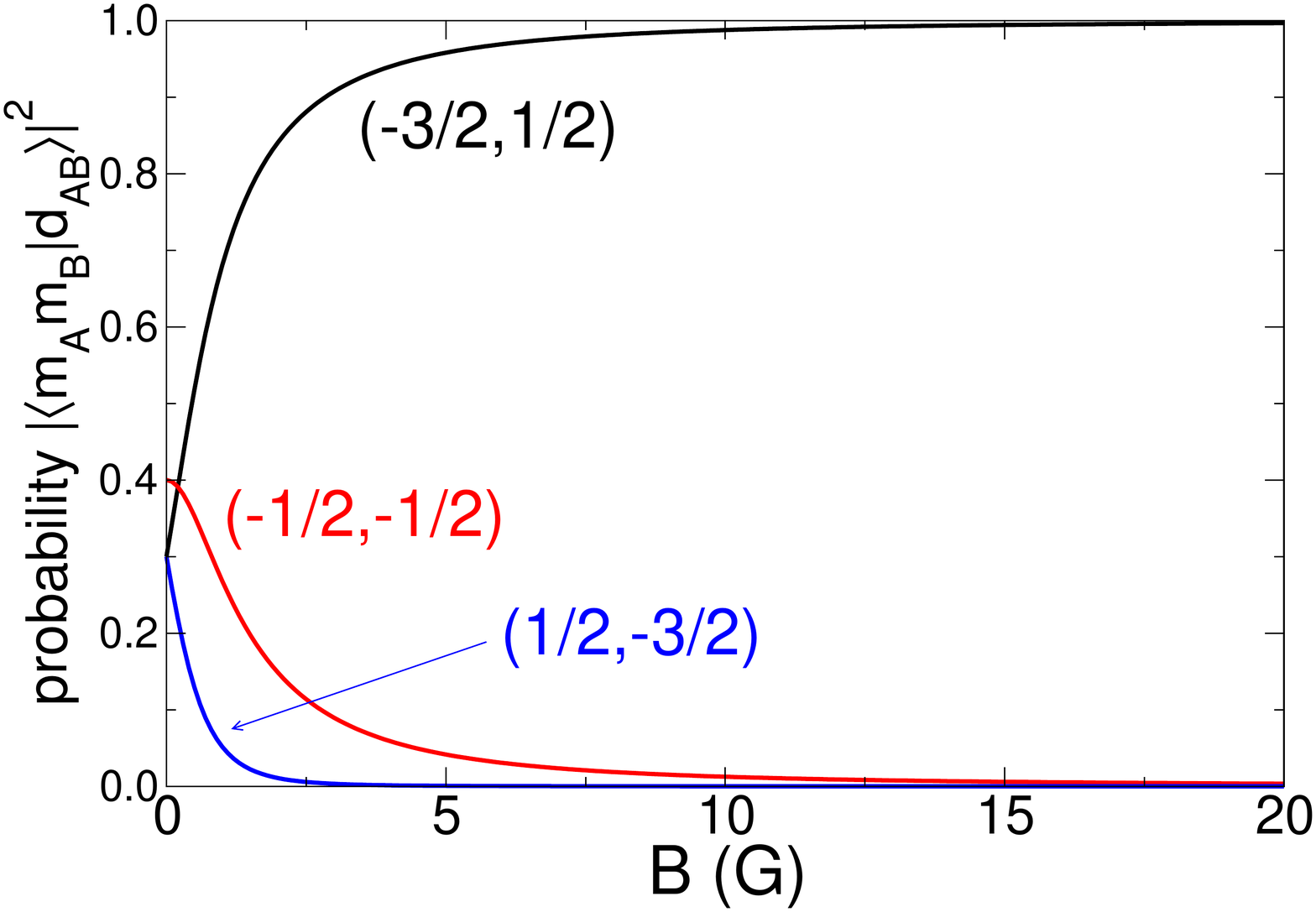}
\caption{Same as Fig.~\ref{FIG-EIGENVEC-LIK-SYM}
but for the $^{41}$K$^{87}$Rb molecule in the $m_1=-1$ manifold. 
The black, red, blue curves correspond respectively to the components 
($-3/2, 1/2$), ($-1/2, -1/2$), ($1/2, -3/2$).}
\label{FIG-EIGENVEC-KRB-SYM}
\end{center}
\end{figure}

Fig.~\ref{FIG-EIGENVEC-KRB-SYM} shows the probability of a component of the wavefunction of a $^{41}$K$^{87}$Rb  
molecule is plotted as a function of the magnetic field $B$ in its first dressed state 
$\big| d_{\text{AB}} \big\rangle = 1$ for $m_1=m_2=-1$. 
As can be seen with the black line, 
the dressed state at large magnetic field ($B > 5$ G for this system)
corresponds to a nearly pure character 
with ${m}_{\text{K}} = -3/2$, ${m}_{\text{Rb}} = 1/2$,
the nuclear spins of the corresponding isotopes being ${i}_{\text{K}} = 3/2$ and ${i}_{\text{Rb}} = 3/2$.
In the following we will take A = $^{41}$K, B = $^{87}$Rb.
The products of the chemical reaction $^{41}$K$_2$ and $^{87}$Rb$_2$ are
formed in their ground electronic state $X^1\Sigma_g^+$, ground vibrational state $v=0$, 
and in the ground rotational state $n_{\text {A}} = n_{\text {B}} = 0$.
To obtain the eigenfunctions and eigenenergies of the KRb, K$_2$ and Rb$_2$ 
molecules in a magnetic field,
we diagonalize the corresponding molecular Hamiltonians in Eq.~\eqref{Hamiltonian}.
As hyperfine parameters, we took $c_\text{AB} = 896.2$~Hz,  $g_\text{A} = 0.143$, $g_\text{B} = 1.834$, 
$\sigma_{\text{A of AB}} = 1321$ ppm, and $\sigma_{\text{B of AB}} = 3469 $ ppm
from  Ref.~\cite{Aldegunde_PRA_78_033434_2008}.
We took $c_\text{AA} = 32$~Hz, $c_\text{BB} = 25021$~Hz, 
$\sigma_{\text{A of AA}} = 1313$ ppm, and $\sigma_{\text{B of BB}} = 3489 $ ppm
from Ref.~\cite{Aldegunde_PRA_79_013401_2009}. \\

\begin{table}[h]
\setlength{\extrarowheight}{4pt}
\begin{center}
\begin{tabular}{|ccc|cc|cc|}
\hline
                               $| i \rangle $ &  $\big| d_{\text{AB}_1} \big\rangle $ &  $\big| d_{\text{AB}_2} \big\rangle $ & ${m}_{\text{A}_1}$ & ${m}_{\text{B}_1}$ & ${m}_{\text{A}_2}$ & ${m}_{\text{B}_2}$ \\ \hline 
        1 & 1 & 1 & -3/2 & 1/2 & -3/2 & 1/2 \\ \hline
           2 & 1 & 2 & -3/2 & 1/2 & -1/2 & -1/2 \\ \hline
              3 & 1 & 3 & -3/2 & 1/2 & 1/2 & -3/2 \\ \hline
                 4 & 2 & 2 & -1/2 & -1/2 & -1/2 & -1/2 \\ \hline
                    5 & 2 & 3 & -1/2 & -1/2 & 1/2 & -3/2 \\ \hline
                       6 & 3 & 3 & 1/2 & -3/2 & 1/2 &-3/2 \\ 
\hline
\end{tabular}
\caption{Same as Tab.~\ref{TableLIKLIKs}
but for AB + AB = $^{41}$K$^{87}$Rb + $^{41}$K$^{87}$Rb
with $m_1+m_2= -1$ and A = $^{41}$K, B = $^{87}$Rb.
}
\label{TableKRBKRBs}
\end{center}
\end{table}

\begin{figure}[h]
\begin{center}
\includegraphics*[width=8.cm, trim=0cm 0cm 0cm 0cm]{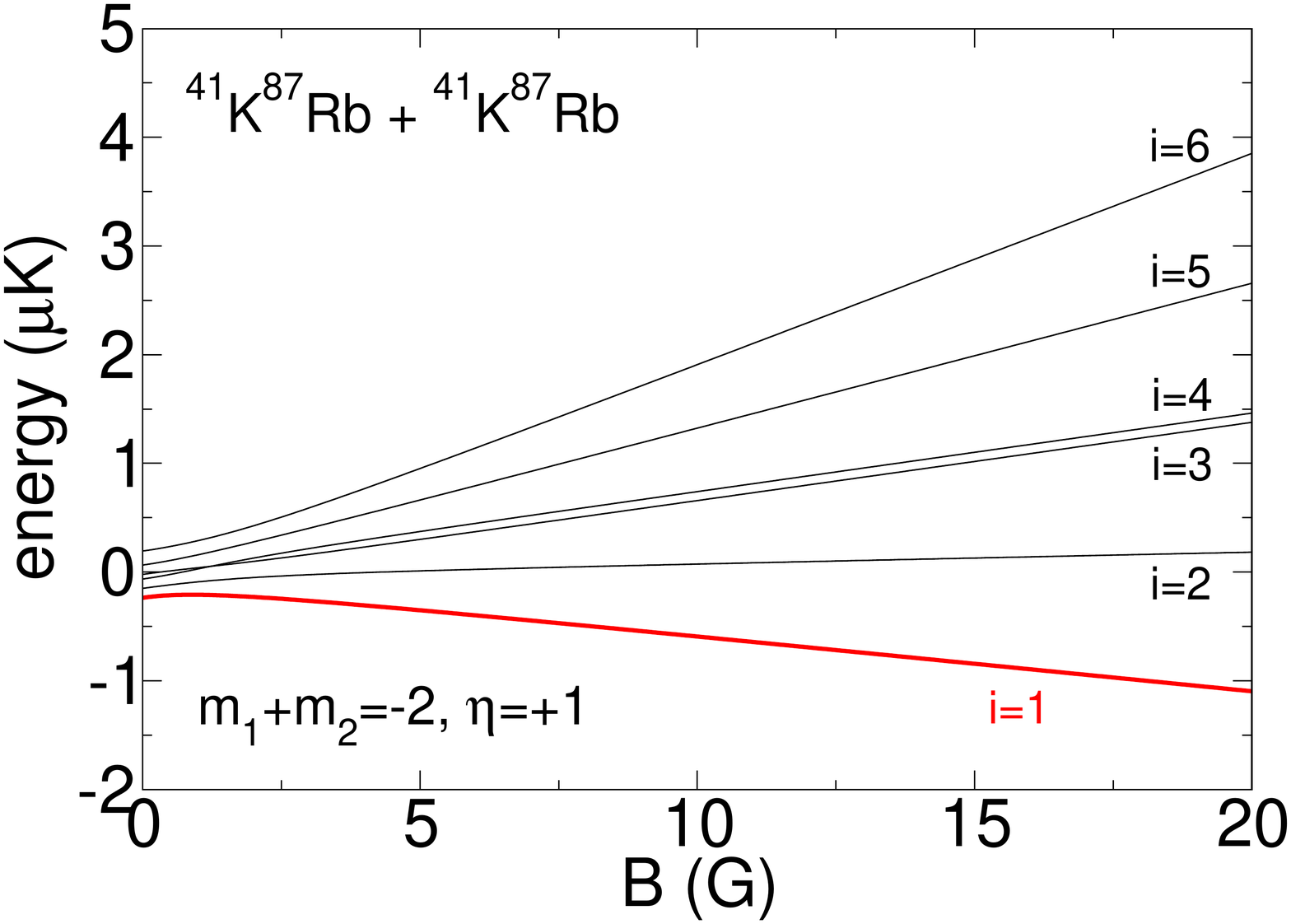}
\caption{Same as Fig.~\ref{FIG-NRG2PLE-LIK-LIK-SYM} 
but for $^{41}$K$^{87}$Rb + $^{41}$K$^{87}$Rb
with $m_1+m_2 = -1$.}
\label{FIG-NRG2PLE-KRB-KRB-SYM}
\end{center}
\end{figure}

Fig.~\ref{FIG-NRG2PLE-KRB-KRB-SYM} presents the energies of the
combined dressed states of the reactants 
AB + AB = $^{41}$K$^{87}$Rb + $^{41}$K$^{87}$Rb
in the zero rotational states
as a function of the magnetic field,
for $m_1+m_2=-2$ and $\eta = +1$ for the case of indistinguishable states.
There are 6 dressed states denoted by $\big| i \big\rangle = 1, ... , 6$.
Tab.~\ref{TableKRBKRBs} provides the nomenclature for these states.
We take $| i \big\rangle = 1$ as an example of initial state (red bold line in the figure).
At large magnetic fields, 
the initial state $| i \rangle = 1$ 
has a main character of ${m}_{\text{A}_1}=-3/2$, ${m}_{\text{B}_1}=1/2$, ${m}_{\text{A}_2}=-3/2$, ${m}_{\text{B}_2}=1/2$, as can be seen with the black line in Fig.~\ref{FIG-EIGENVEC-KRB-SYM} for the individual reactants,
but at lower fields $| i \rangle = 1$ gain other characters (red and blue lines in Fig.~\ref{FIG-EIGENVEC-LIK-SYM}). \\

\begin{figure}[h]
\begin{center}
\includegraphics*[width=8.cm, trim=0cm 0cm 0cm 0cm]{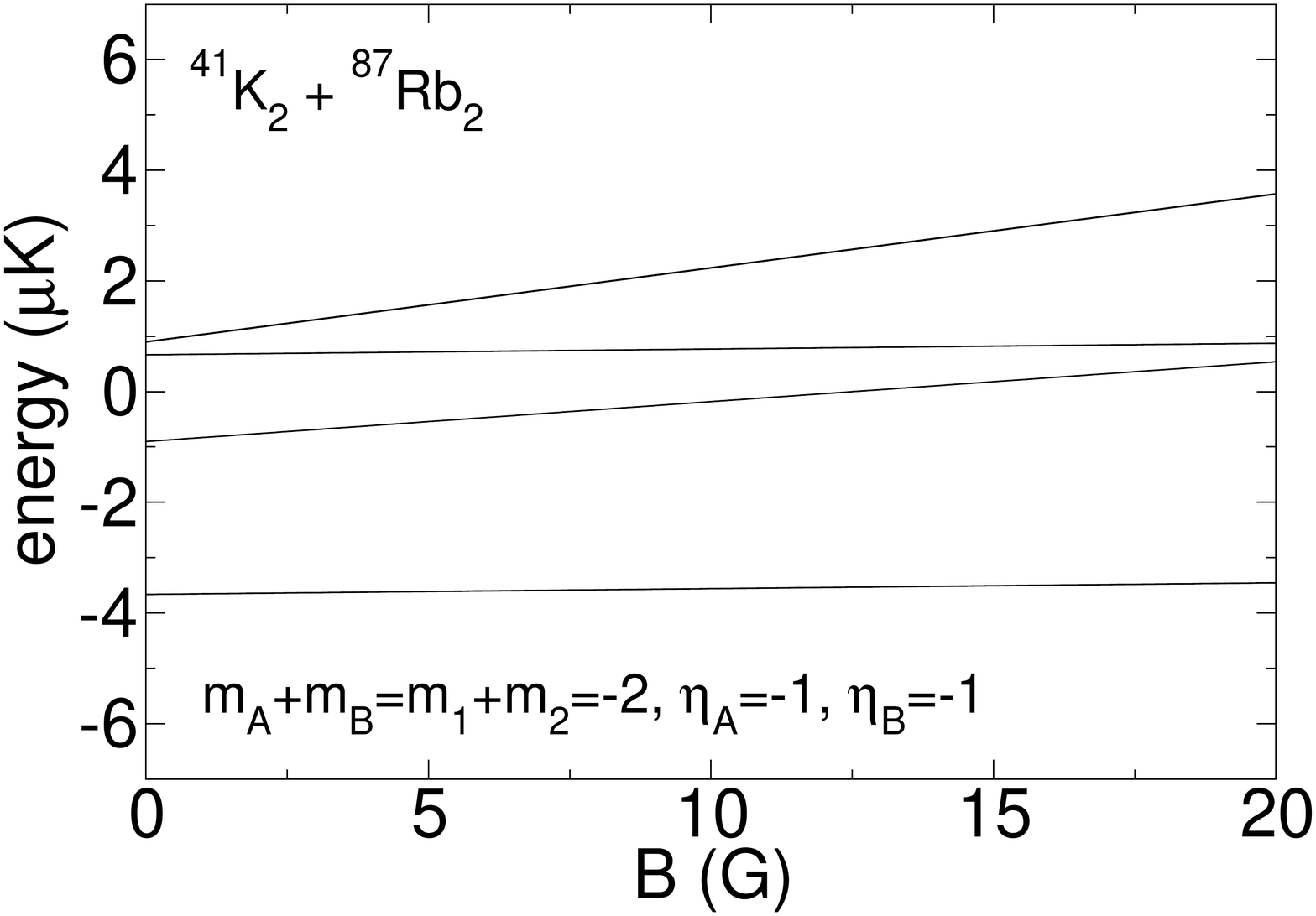}
\caption{Same as Fig.~\ref{FIG-NRG2PLE-LI2K2-SYM} but for 
 $^{41}$K$_2$ + $^{87}$Rb$_2$
for $m_\text{A} + m_\text{B} = m_1+m_2=-2$ and $\eta_\text{A} = -1$, $\eta_\text{B} = -1$ 
(nuclear spin anti-symmetric states).}
\label{FIG-NRG2PLE-K2RB2-ASYM}
\end{center}
\end{figure}

Fig.~\ref{FIG-NRG2PLE-K2RB2-ASYM} 
presents the energies of the combined dressed states of the products 
AA + BB = $^{41}$K$_2$ + $^{87}$Rb$_2$
in the zero rotational states
as a function of the magnetic field,
for $m_{\text {A}}+m_{\text {B}}=m_1+m_2=-2$ and $\eta_\text{A} = -1$, $\eta_\text{B} = -1$ (nuclear spin anti-symmetric states).
$^{41}$K and $^{87}$Rb nuclei are composite fermions. The permutation of two identical fermionic nuclei
in the $^{41}$K$_2$ molecule and in the $^{87}$Rb$_2$ molecule 
should lead to an overall anti-symmetric wavefunction for each molecules.
As the rotational wavefunction of the $^{41}$K$_2$ and $^{87}$Rb$_2$ molecules in the zero rotational state
are symmetric under the interchange of the nuclei, then their nuclear spin wavefunctions 
have to be anti-symmetric, and hence described by 
values of $\eta_\text{A} = -1$ and $\eta_\text{B} = -1$.
There are now 5 dressed states denoted by $| j \rangle = 1, ... , 5$,
which correspond to different combinations of the individual dressed states
$\big| d_{\text{AA}} ; -1 \big\rangle $
and 
$\big| d_{\text{BB}} ; -1 \big\rangle $.
Tab.~\ref{TableK2RB2a} provides the nomenclature for the states.
These are all the possible final product states of the chemical reaction,
for the nuclear spin anti-symmetric states of the molecules. \\

\begin{figure}[h]
\begin{center}
\includegraphics[width=8cm]{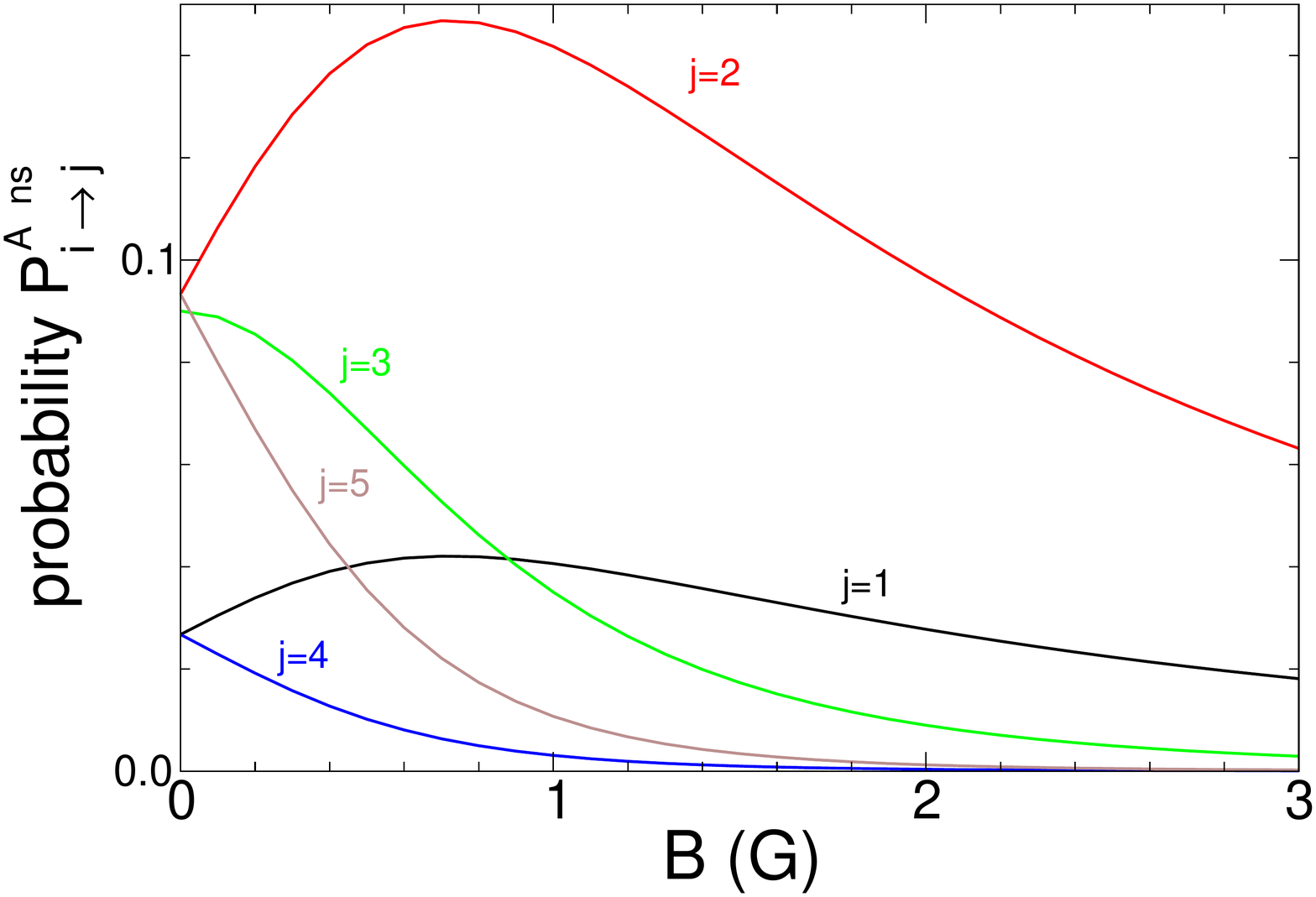} 
\includegraphics[width=8cm]{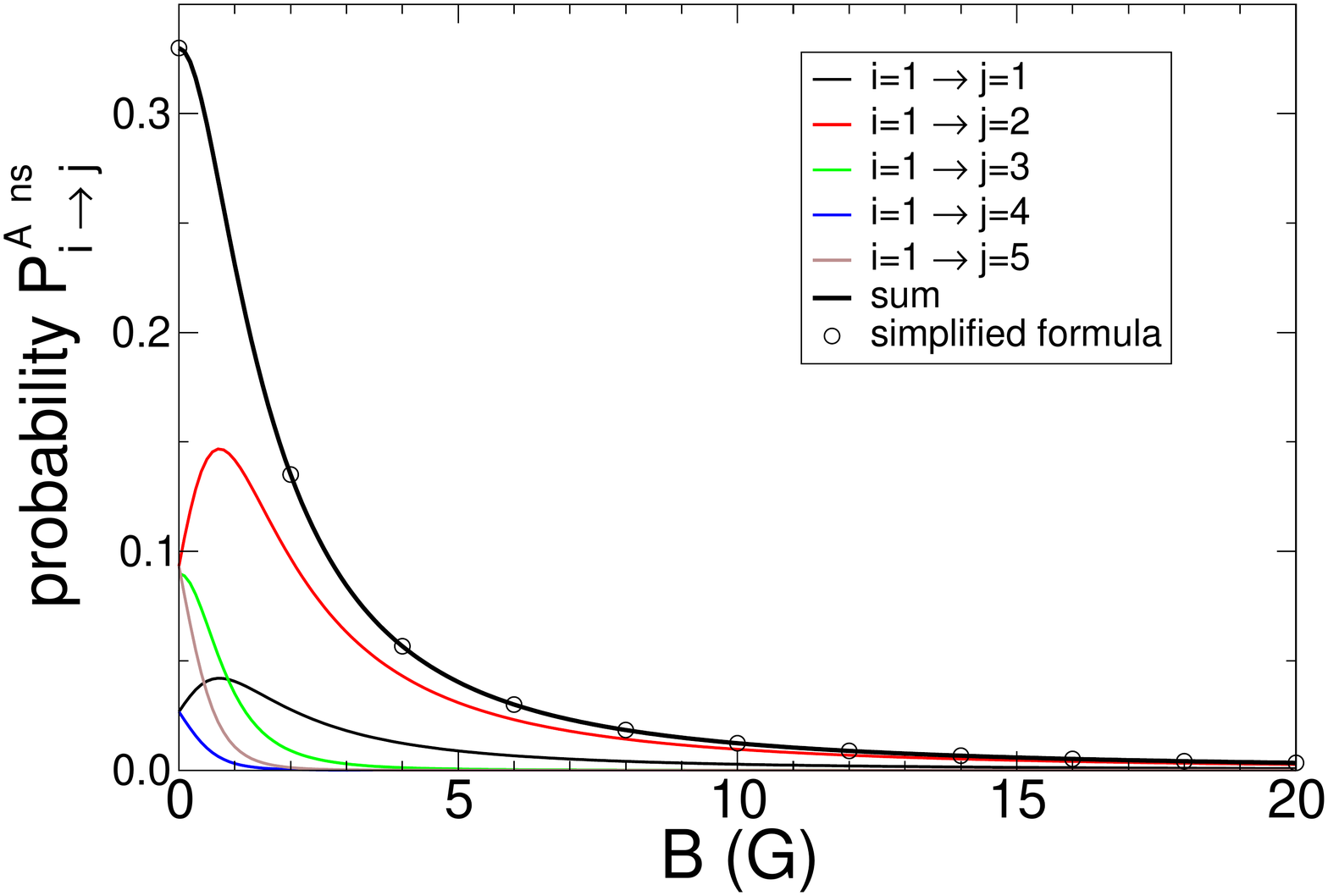} \\ 
\includegraphics[width=4cm]{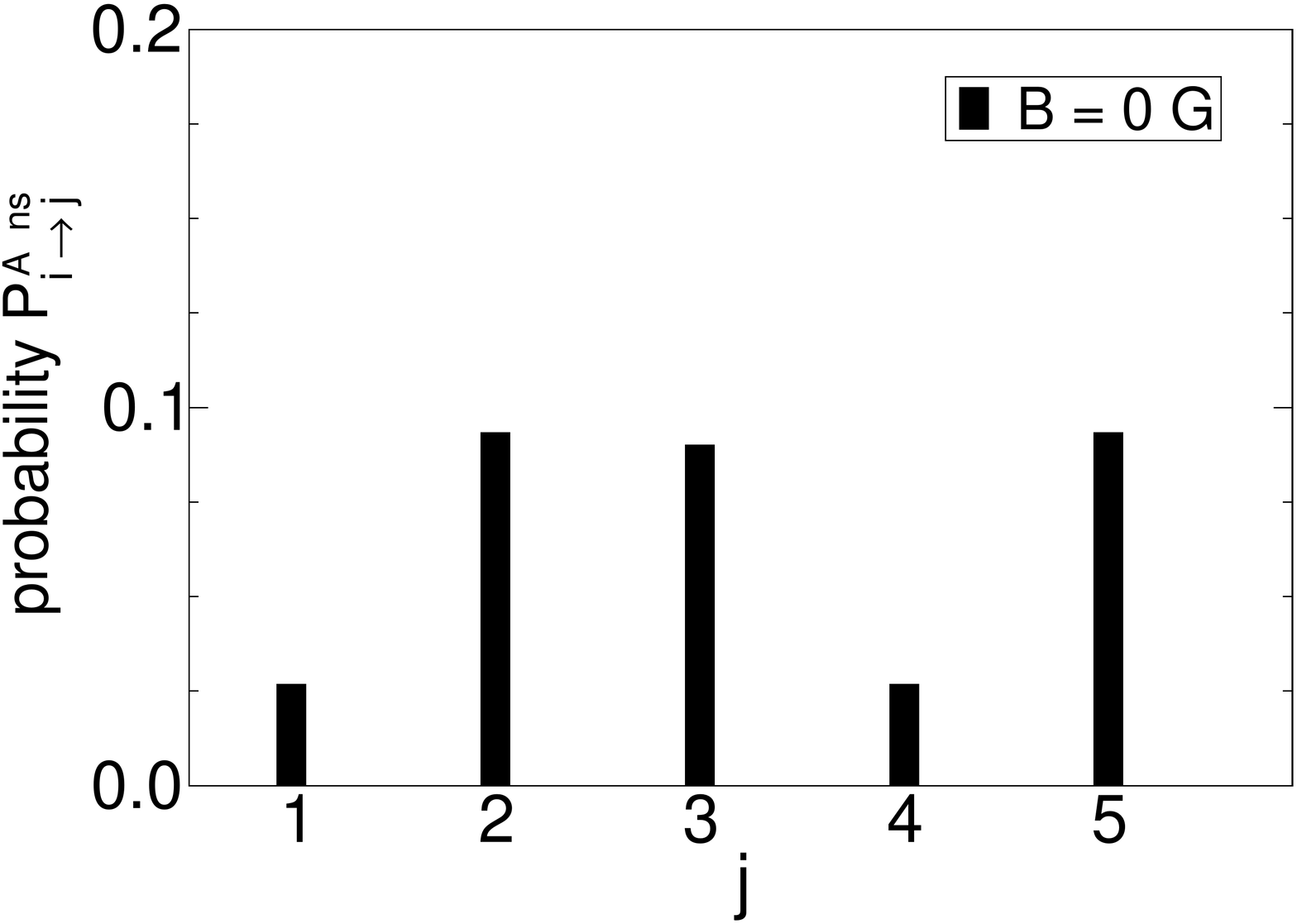} 
\includegraphics[width=4cm]{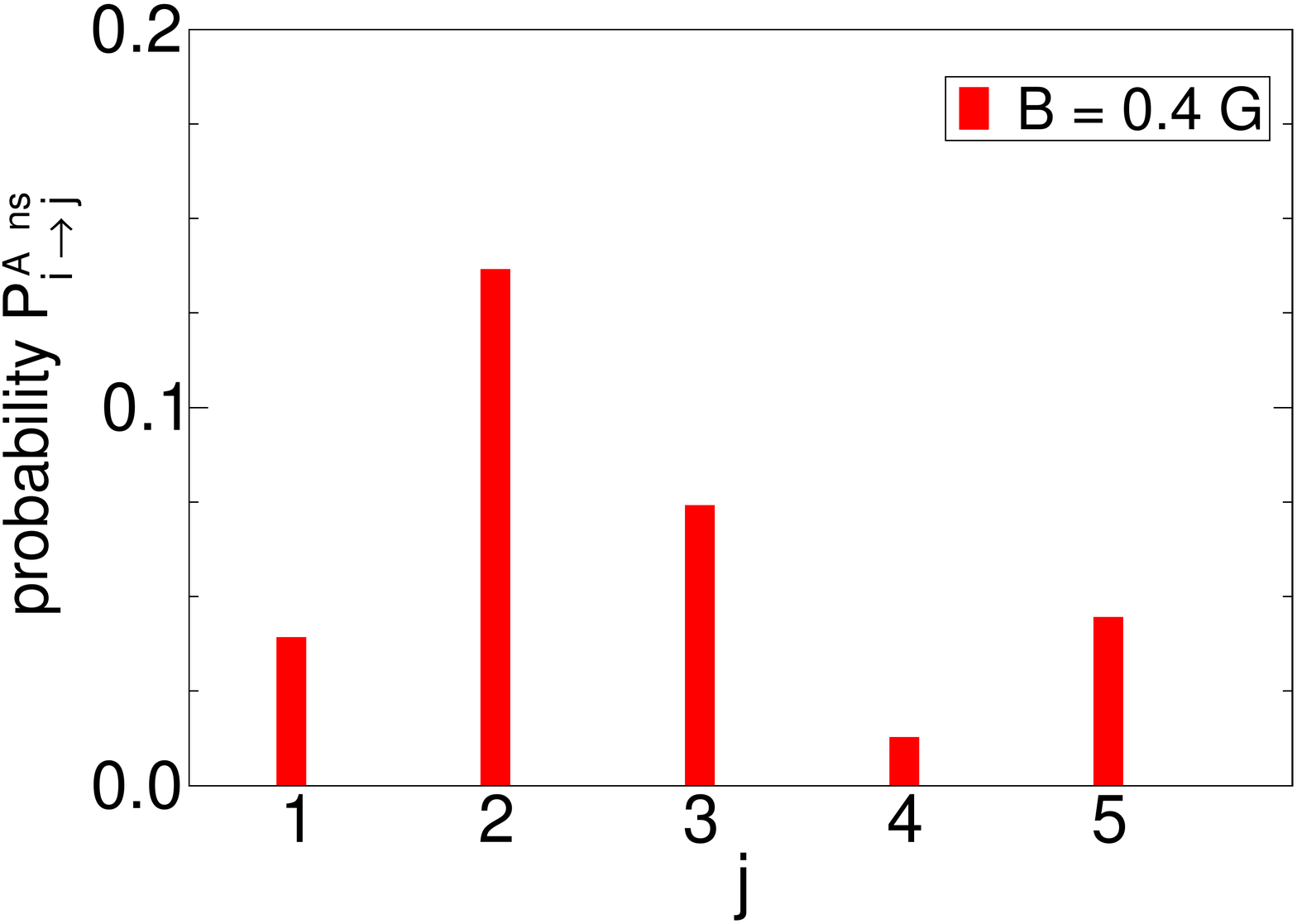} 
\includegraphics[width=4cm]{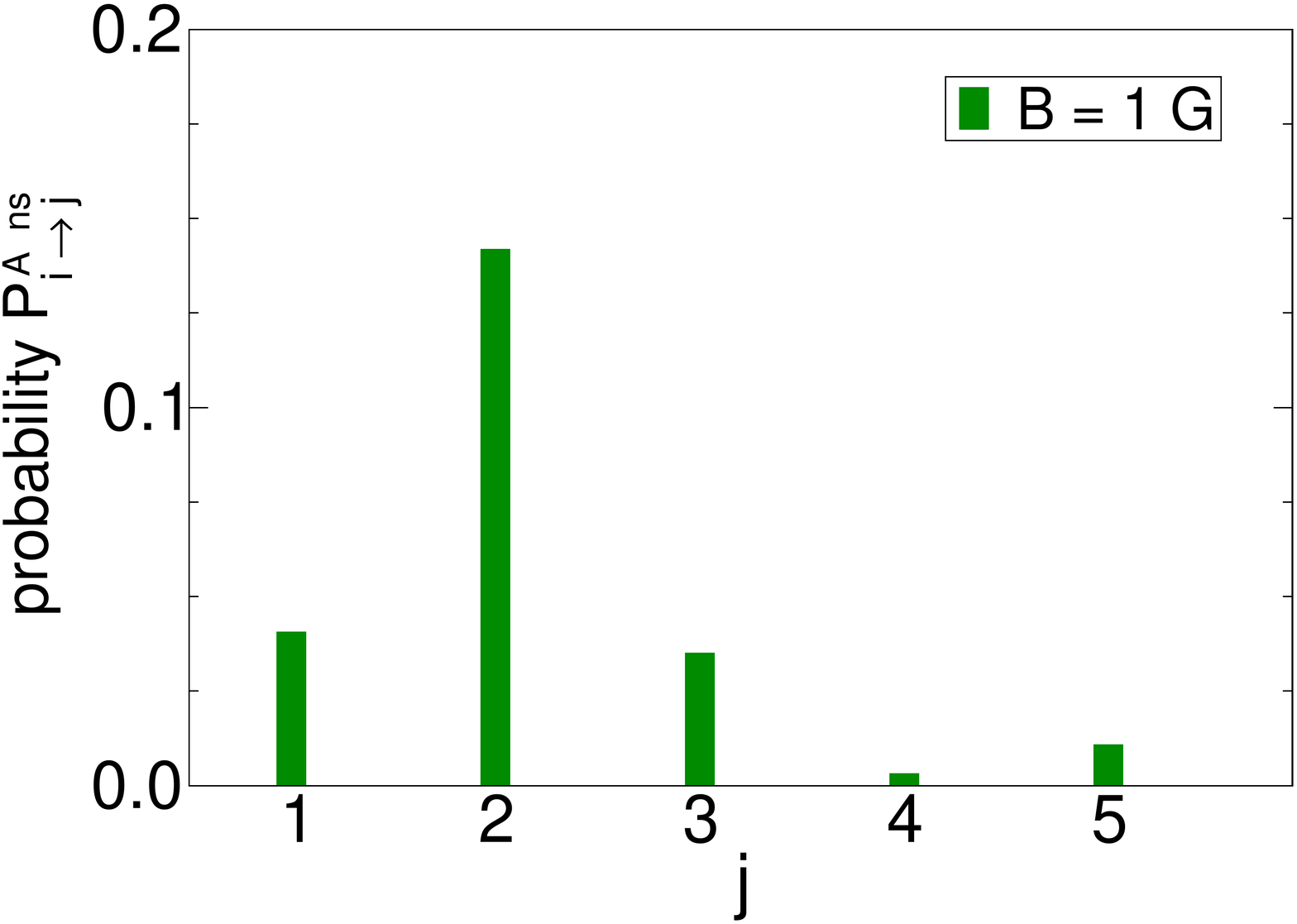} 
\includegraphics[width=4cm]{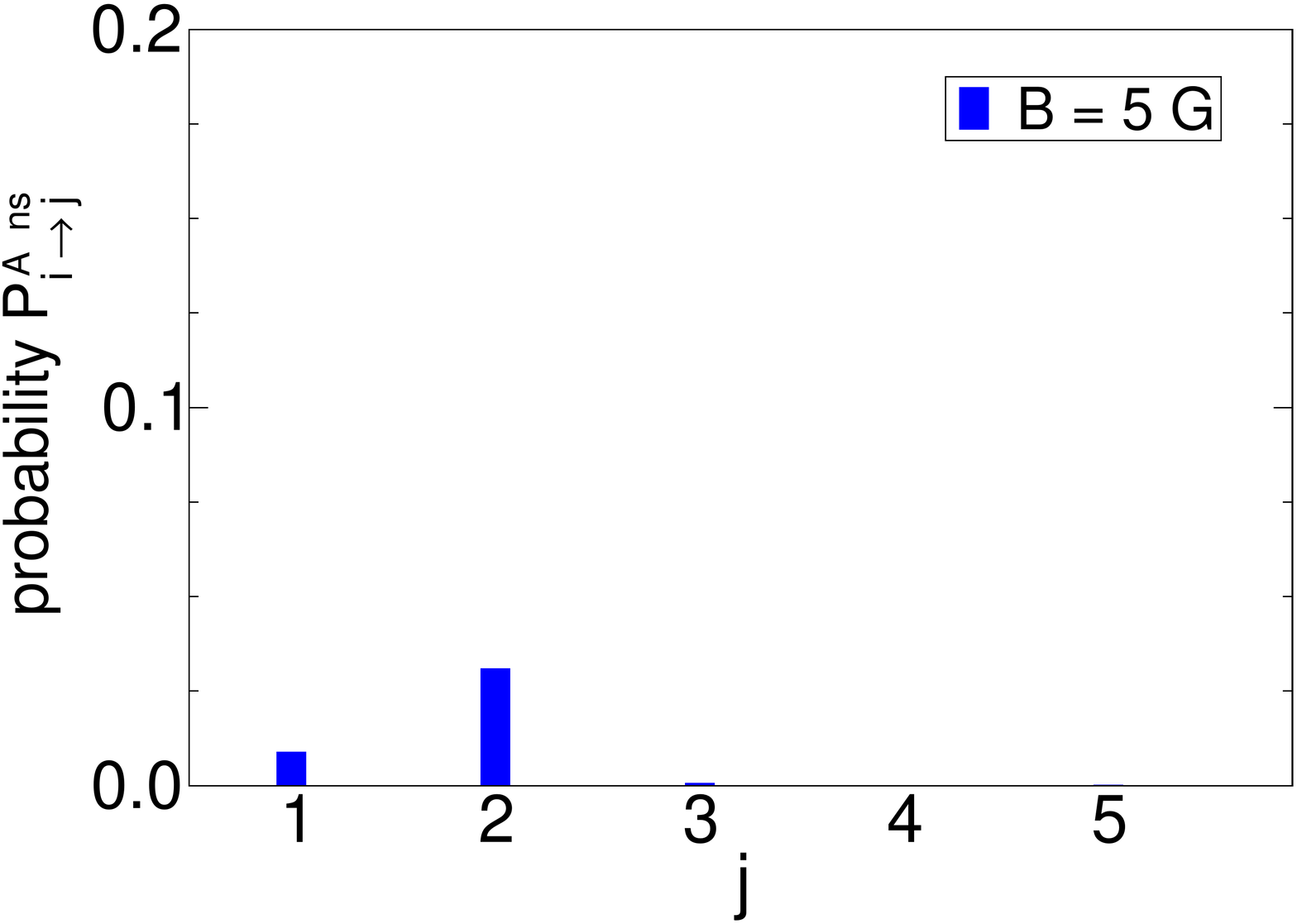} 
\caption{Same as Fig.~\ref{FIG-PROBA-STS-SYM} but for 
$P^{A \, \text{ns}}_{i \to j}$ for $^{40}$K$_2$ + $^{87}$Rb$_2$.}
\label{FIG-PROBA-STS-ASYM}
\end{center}
\end{figure}

Finally, the probability $P^{A \, \text{ns}}_{i \to j}$ expressed by Eq.~\eqref{proba-sts-A}
for the nuclear spin anti-symmetric states 
is presented in Fig.~\ref{FIG-PROBA-STS-ASYM}
as a function of the magnetic field (top panels)
and as a function of the final states $| j \rangle$ for different magnetic fields
(bottom panels).
We find a similar conclusion than for the case of the nuclear spin symmetric states
that the behavior of the probabilities depends on the magnetic field and on the admixtures of the bare states 
for the dressed states of the reactants and products. 
For all values of $B$, the  $| j \rangle = 2$ (red curve) dominates.
From Tab.~\ref{TableK2RB2a},
this state corresponds to a main character of $(-3/2, -1/2) + (1/2, -1/2)$.
Around $B \sim 0$~G,
the  $| j \rangle = 5$ (brown curve) with a main character of $(-1/2, 1/2) + (-1/2, -3/2)$,
and $| j \rangle = 3$ (green curve) with a main character of $(-3/2, 1/2) + (1/2, -3/2)$,
become as important as $| j \rangle = 2$.
All of these states take the form of an entangled state for the nuclei.
For example, for $| j \rangle = 2$ the wavefunction is
$ \{  | -3/2, -1/2 \big\rangle - | -1/2, -3/2 \big\rangle  \} / \sqrt{2}$ for the $^{41}$K nuclei, and 
 $\{  | 1/2, -1/2 \big\rangle - | -1/2, 1/2 \big\rangle  \} / \sqrt{2}$ for the $^{87}$Rb nuclei.
These entangled states carry now a minus sign in contrast with the entangled states of the products
discussed above for the case of nuclear spin symmetric states.
Therefore, symmetry considerations also impose the type of the entangled state (plus or minus).
At large magnetic fields, because both molecular states are anti-symmetric, there can't be
a character in the symmetric states $(-3/2, -3/2) + (1/2, 1/2)$, the character coming from the swap of the original nuclear spins in the reactants, for the present example.
The probabilities from the reactant state $| i \rangle = 1$ to all these anti-symmetric states
must then vanish. 
Therefore, identical reactants in indistinguishable states at large magnetic fields
will always populate nuclear spin symmetric states of the molecular products.
In the present case, this will populate the first excited rotational state
$n_\text{A} = n_\text{B} = 1$ of the products.
This explains the selected values of the rotation parities of the molecular products
observed in \cite{Hu_NC_13_435_2021}.

\section{Conclusion}
\label{Conclu}

We presented in this study a theoretical model to compute 
nuclear spin state-to-state distributions of ultracold chemical reactions,
from reactants to products. 
This distribution can be modified by an applied magnetic field.
The formalism is based on the fact that atomic nuclear spins of the products of a chemical reaction
ineherit the ones of the reactants in the magnetic field. The mechanism is then 
driven by long-range physics only and not by short-range interactions 
when the atoms are close together.
A simple atomic re-arrangement in the 
chemical reaction and symmetry considerations of identical particles 
are used to explain the symmetric
or anti-symmetric character of the nuclear spin wavefunctions, hence
the even/odd parity of the rotational wavefunction of the molecular products.
Depending on the magnetic field that is applied,
among the symmetric nuclear spin wavefunctions, 
the molecular products can end up in the form of a 
separable state or of an entangled state of the atomic nuclear spins.
Again depending on the magnetic field, 
for anti-symmetric states, the molecular products can end up only in the form of 
an entangled state. Otherwise the probability tends to zero, 
which is responsible for the selected values of the rotational parities
that have been observed in a recent experiment \cite{Hu_NC_13_435_2021}.
The state-to-state probabilities can be computed as functions of the magnetic field. 
It requires the knowledge of the eigenfunctions of the molecular reactants and products
in the magnetic field.
We showed that the final state-to-state distribution drastically changes with the magnetic field.
When the probability is summed over all the final dressed product states,
we showed that, after an analytic development, the probability 
requires only the knowledge of the eigenfunctions of the reactants.
This expression of the summed probabilities was used 
to explain and understand the magnetic field dependence
for the product distribution of a recent experiment
\cite{Hu_NC_13_435_2021}, validating the assumption that nuclear 
spins remain unchanged during an ultracold chemical reaction of bi-alkali molecules.

\section*{Acknowledgments}
G. Q. acknowledges funding from the FEW2MANY-SHIELD Project No. ANR-17-CE30-0015 from Agence Nationale de la Recherche.
M.-G. H., Y. L., M. A. N., L. Z., K.-K. N. acknowledge 
funding from the Department Of Energy (DE-SC0019020) and the David and Lucile Packard foundation (2016-65138).
M. A. N. is supported by the Arnold O. Beckman Postdoctoral Fellowship in Chemical Instrumentation.

\clearpage

\onecolumngrid

\begin{table*}[t]
\setlength{\extrarowheight}{4pt}
\begin{center}
\begin{tabular}{|ccc|cc|cc|cc|cc|}
\hline
        $| j \rangle $ &  $\big| d_{\text{AA}} ; -1 \big\rangle $ &  $\big| d_{\text{BB}} ; -1 \big\rangle $ & ${m}_{\text{A}_1}$ & ${m}_{\text{A}_2}$ & ${m}_{\text{B}_1}$ & ${m}_{\text{B}_2}$  & ${m}_{\text{A}_1}$ & ${m}_{\text{A}_2}$ & ${m}_{\text{B}_1}$ & ${m}_{\text{B}_2}$  \\ \hline 
        1 & 1 & 3 & -3/2 & -1/2 & 3/2 & -3/2 & &   & 1/2 & -1/2\\ \hline
        2 & 1 & 4 & -3/2 & -1/2 & 1/2 & -1/2 & &   & 3/2 & -3/2\\ \hline
        3 & 2 & 2 & -3/2 & 1/2 & 1/2 & -3/2  & &  &  &  \\ \hline
        4 & 3 & 1 & -3/2 & 3/2 & -1/2 & -3/2 & -1/2 & 1/2 &  &  \\ \hline
        5 & 4 & 1 & -1/2 & 1/2 & -1/2 & -3/2 & -3/2 & 3/2 &  &  \\ 
\hline
\end{tabular}
\caption{Same as Tab.~\ref{TableLI2K2s} 
but for AA + BB = $^{41}$K$_2$ + $^{87}$Rb$_2$ and 
$\eta_\text{A} = -1$, $\eta_\text{B} = -1$ (nuclear spin anti-symmetric states),
with A = $^{41}$K, B = $^{87}$Rb.
The main and second main characters are given for large magnetic fields, typically $B > 5$ G.
}
\label{TableK2RB2a}
\end{center}
\end{table*}

\twocolumngrid

\bibliography{../../../BIBLIOGRAPHY/bibliography}

\onecolumngrid
\appendix

\section*{Appendix A: State-to-state probability amplitudes}

We want to evaluate the probability amplitude
$ \langle  \, d_{\text {AA}}, d_{\text {BB}} ; \eta_{\text {A}}, \eta_{\text {B}}  \big|  \, d_{\text {AB}_1}, d_{\text {AB}_2} ; \eta \big\rangle $ in Eq.~\eqref{proba-sts},
using Eq.~\eqref{symdresstatereac0} and Eq.~\eqref{symdresstateprod}. 
We have
\begin{multline}
\langle  \, d_{\text {AA}}, d_{\text {BB}} ; \eta_{\text {A}}, \eta_{\text {B}}  \big|  \, d_{\text {AB}_1}, d_{\text {AB}_2} ; \eta \big\rangle = 
 \\
\shoveleft{ 
\sum_{  m_{n_\text{A}}' } \, \sum_{  m_{n_\text{B}}' } \, 
\sum_{ m_{\text{A}_1}' = -i_{\text{A}}}^{+i_{\text{A}}} \, \sum_{m_{\text{A}_2}' \ge m_{\text{A}_1}'}   \, \sum_{ m_{\text{B}_1}' = -i_{\text{B}}}^{+i_{\text{B}}}  \, 
 \sum_{m_{\text{B}_2}' \le m_{\text{B}_1}' }
  \, \big\langle d_{\text {AA}} ; \eta_{\text {A}} \, \big| \, {n_\text{A}}' \, m_{n_\text{A}}' \, m_{\text{A}_1}' \, m_{\text{A}_2}'; \eta_{\text {A}}  \big\rangle  \, \big\langle d_{\text {BB}} ; \eta_{\text {B}} \, \big|  \, {n_\text{B}}' \, m_{n_\text{B}}' \, m_{\text{B}_1}' \, m_{\text{B}_2}' ; \eta_{\text {B}} }  \big\rangle  \\  
\sum_{ m_{n_1}} \,  \sum_{ m_{n_2}} \, \sum_{ m_{\text{A}_1} = -i_{\text{A}}}^{+i_{\text{A}}} \, \sum_{ m_{\text{B}_1} = -i_{\text{B}}}^{+i_{\text{B}}}  \, \sum_{ m_{\text{A}_2} = -i_{\text{A}}}^{+i_{\text{A}}} \, \sum_{ m_{\text{B}_2} = -i_{\text{B}}}^{+i_{\text{B}}}  \, 
\big\langle  n_1 \, m_{n_1}  \, m_{\text{A}_1} \, m_{\text{B}_1}  \, \big| \, d_{\text {AB}_1} \big\rangle \, \big\langle  n_2 \, m_{n_2}  \, m_{\text{A}_2} \, m_{\text{B}_2} \, \big| \, d_{\text {AB}_2} \big\rangle \\
 \delta_{m_{n_\text{A}}' + m_{n_\text{B}}', m_{n_1} + m_{n_2} + m_{l_r} - m_{l_p}} 
\ \delta_{m_{n_1} + m_{\text{A}_1}+m_{\text{B}_1}, M_1} 
\ \delta_{m_{n_2} + m_{\text{A}_2}+m_{\text{B}_2}, M_2}  \\
\bigg(  \frac{1}{\sqrt{\Delta_\text{A}}} \, \bigg\{ \big\langle {n_\text{A}}' \, m_{n_\text{A}}' \, m_{\text{A}_1}' \,  m_{\text{A}_2}'  \big| + \, \eta_\text{A} \, \big\langle {n_\text{A}}' \, m_{n_\text{A}}' \, m_{\text{A}_2}' \, m_{\text{A}_1}' \big| \bigg\} 
 \otimes \frac{1}{\sqrt{\Delta_\text{B}}} \, \bigg\{ \big\langle {n_\text{B}}' \, m_{n_\text{B}}' \, m_{\text{B}_1}' \,  m_{\text{B}_2}'  \big| + \, \eta_\text{B} \, \big\langle  {n_\text{B}}' \, m_{n_\text{B}}' \, m_{\text{B}_2}' \, m_{\text{B}_1}' \big| \bigg\}  \bigg)  \\
\bigg( \frac{1}{\sqrt{\Delta_d}} \, \bigg\{ \big| \, n_{1}  \, m_{n_1}  \, m_{\text{A}_1} \, m_{\text{B}_1} \big\rangle \otimes \big| \, n_{2}  \, m_{n_2}  \, m_{\text{A}_2} \, m_{\text{B}_2}   \big\rangle 
+ \eta \, \big| \, n_{2}  \, m_{n_2}  \, m_{\text{A}_2} \, m_{\text{B}_2}  \big\rangle 
\otimes \big| \, n_{1}  \, m_{n_1}  \, m_{\text{A}_1} \, m_{\text{B}_1}  \big\rangle \bigg\} \bigg).
   \label{eqA1}
\end{multline}
We rewrite the last line of Eq.~\eqref{eqA1} using the first assumption
of the model that the nuclear spins remain spectators 
in the region of the tetra-atomic complex, so that after the reaction they remain the same
in the basis of the bare states. 
Then we can replace the kets 
$\big| \, n_{1}  \, m_{n_1}  \, m_{\text{A}_1} \, m_{\text{B}_1} \big\rangle \otimes \big| \, n_{2}  \, m_{n_2}  \, m_{\text{A}_2} \, m_{\text{B}_2}  \big\rangle $
and
$\big| \, n_{2}  \, m_{n_2}  \, m_{\text{A}_2} \, m_{\text{B}_2}  \big\rangle \otimes \big| \, n_{1}  \, m_{n_1}  \, m_{\text{A}_1} \, m_{\text{B}_1}  \big\rangle $
by
$\big| \, n_{1}  \, m_{n_1}  \, m_{\text{A}_1} \, m_{\text{A}_2} \big\rangle \otimes \big| \, n_{2}  \, m_{n_2}  \, m_{\text{B}_1} \, m_{\text{B}_2}   \big\rangle $
and
$\big| \, n_{2}  \, m_{n_2}  \, m_{\text{A}_2} \, m_{\text{A}_1}  \big\rangle \otimes \big| \, n_{1}  \, m_{n_1}  \, m_{\text{B}_2} \, m_{\text{B}_1}  \big\rangle $,
regardless of the order of the nuclear spins in the kets.
Note that the two atoms A and B are swapped at the same time, as the total wavefunction is invariant under the permutation of the molecules as a whole. 
One cannot therefore have a swap of atoms A without a swap of atoms B. 
We now develop the last two lines of the previous equation
\begin{multline}
 \bigg(  \frac{1}{\sqrt{\Delta_\text{A}}} \, \bigg\{ \big\langle {n_\text{A}}' \, m_{n_\text{A}}' \, m_{\text{A}_1}' \,  m_{\text{A}_2}'  \big| + \, \eta_\text{A} \, \big\langle {n_\text{A}}' \, m_{n_\text{A}}' \, m_{\text{A}_2}' \, m_{\text{A}_1}' \big| \bigg\} 
 \otimes \frac{1}{\sqrt{\Delta_\text{B}}} \, \bigg\{ \big\langle {n_\text{B}}' \, m_{n_\text{B}}' \, m_{\text{B}_1}' \,  m_{\text{B}_2}'  \big| + \, \eta_\text{B} \, \big\langle  {n_\text{B}}' \, m_{n_\text{B}}' \, m_{\text{B}_2}' \, m_{\text{B}_1}' \big| \bigg\}  \bigg) \\
\bigg( \frac{1}{\sqrt{\Delta_d}} \, \bigg\{ \big| \, n_{1}  \, m_{n_1}  \, m_{\text{A}_1} \, m_{\text{A}_2} \big\rangle \otimes \big| \, n_{2}  \, m_{n_2}  \, m_{\text{B}_1} \, m_{\text{B}_2}   \big\rangle 
+ \eta \, \big| \, n_{2}  \, m_{n_2}  \, m_{\text{A}_2} \, m_{\text{A}_1}  \big\rangle 
\otimes \big| \, n_{1}  \, m_{n_1}  \, m_{\text{B}_2} \, m_{\text{B}_1}  \big\rangle \bigg\} \bigg) \\
\shoveleft{ = \frac{1}{\sqrt{\Delta_\text{A}}} \, \frac{1}{\sqrt{\Delta_\text{B}}} \, \frac{1}{\sqrt{\Delta_d}} \,
   \bigg(   \big\langle {n_\text{A}}' \, m_{n_\text{A}}' \, m_{\text{A}_1}' \,  m_{\text{A}_2}'  \big| 
   \otimes \big\langle {n_\text{B}}' \, m_{n_\text{B}}' \, m_{\text{B}_1}' \,  m_{\text{B}_2}'  \big| 
   + \, \eta_\text{A} \, \big\langle {n_\text{A}}' \, m_{n_\text{A}}' \, m_{\text{A}_2}' \, m_{\text{A}_1}' \big|  
    \otimes  \big\langle {n_\text{B}}' \, m_{n_\text{B}}' \, m_{\text{B}_1}' \,  m_{\text{B}_2}'  \big| } \\
\shoveright{ + \, \eta_\text{B} \, \big\langle  {n_\text{A}}' \, m_{n_\text{A}}' \, m_{\text{A}_1}' \, m_{\text{A}_2}' \big|  
    \otimes  \big\langle {n_\text{B}}' \, m_{n_\text{B}}' \, m_{\text{B}_2}' \,  m_{\text{B}_1}'  \big| 
 + \, \eta_\text{A} \, \eta_\text{B} \, \big\langle {n_\text{A}}' \, m_{n_\text{A}}' \, m_{\text{A}_2}' \,  m_{\text{A}_1}'  \big| 
   \otimes  \big\langle {n_\text{B}}' \, m_{n_\text{B}}' \, m_{\text{B}_2}' \,  m_{\text{B}_1}'  \big| \bigg) } \\
\bigg(  \big| \, n_{1}  \, m_{n_1}  \, m_{\text{A}_1} \, m_{\text{A}_2} \big\rangle \otimes \big| \, n_{2}  \, m_{n_2}  \, m_{\text{B}_1} \, m_{\text{B}_2}   \big\rangle 
+ \eta \, \big| \, n_{2}  \, m_{n_2}  \, m_{\text{A}_2} \, m_{\text{A}_1}  \big\rangle 
\otimes \big| \, n_{1}  \, m_{n_1}  \, m_{\text{B}_2} \, m_{\text{B}_1}  \big\rangle  \bigg) \\
\shoveleft{  = \frac{1}{\sqrt{\Delta_\text{A}}} \, \frac{1}{\sqrt{\Delta_\text{B}}} \, \frac{1}{\sqrt{\Delta_d}} \,
   \bigg\{  } \\
 \delta_{m_{\text{A}_1},m_{\text{A}_1}'} \, \delta_{m_{\text{A}_2},m_{\text{A}_2}'} \, \delta_{m_{\text{B}_1},m_{\text{B}_1}'}  \, \delta_{m_{\text{B}_2},m_{\text{B}_2}'} 
   \bigg[ \big\langle {n_\text{A}}' \, m_{n_\text{A}}' \, {n_\text{B}}' \, m_{n_\text{B}}' \big| \, n_{1}  \, m_{n_1}  \,  n_{2}  \, m_{n_2}   \big\rangle 
  + \eta_\text{A} \, \eta_\text{B} \, \eta  \,   
  \big\langle {n_\text{A}}' \, m_{n_\text{A}}' \, {n_\text{B}}' \, m_{n_\text{B}}' \big| \, n_{2}  \, m_{n_2}  \,  n_{1}  \, m_{n_1}   \big\rangle    \bigg]  \\ 
   +   \delta_{m_{\text{A}_1},m_{\text{A}_2}'} \, \delta_{m_{\text{A}_2},m_{\text{A}_1}'} \, \delta_{m_{\text{B}_1},m_{\text{B}_2}'}  \, \delta_{m_{\text{B}_2},m_{\text{B}_1}'} 
   \bigg[  \eta_\text{A} \, \eta_\text{B} \, 
   \big\langle {n_\text{A}}' \, m_{n_\text{A}}' \, {n_\text{B}}' \, m_{n_\text{B}}' \big| \, n_{1}  \, m_{n_1}  \,  n_{2}  \, m_{n_2}   \big\rangle +
   \eta \, \big\langle {n_\text{A}}' \, m_{n_\text{A}}' \, {n_\text{B}}' \, m_{n_\text{B}}' \big| \, n_{2}  \, m_{n_2}  \,  n_{1}  \, m_{n_1}   \big\rangle   \bigg]  \\
  +   \delta_{m_{\text{A}_1},m_{\text{A}_1}'} \, \delta_{m_{\text{A}_2},m_{\text{A}_2}'} \, \delta_{m_{\text{B}_1},m_{\text{B}_2}'}  \, \delta_{m_{\text{B}_2},m_{\text{B}_1}'} 
  \bigg[  \eta_\text{B} \, \big\langle {n_\text{A}}' \, m_{n_\text{A}}' \, {n_\text{B}}' \, m_{n_\text{B}}' \big| \, n_{1}  \, m_{n_1}  \,  n_{2}  \, m_{n_2}   \big\rangle 
  + \eta_\text{A} \, \eta \, \big\langle {n_\text{A}}' \, m_{n_\text{A}}' \, {n_\text{B}}' \, m_{n_\text{B}}' \big| \, n_{2}  \, m_{n_2}  \,  n_{1}  \, m_{n_1}   \big\rangle  \bigg] \\
    +   \delta_{m_{\text{A}_1},m_{\text{A}_2}'} \, \delta_{m_{\text{A}_2},m_{\text{A}_1}'} \, \delta_{m_{\text{B}_1},m_{\text{B}_1}'}  \, \delta_{m_{\text{B}_2},m_{\text{B}_2}'} 
    \bigg[ \eta_\text{A} \, 
    \big\langle {n_\text{A}}' \, m_{n_\text{A}}' \, {n_\text{B}}' \, m_{n_\text{B}}' \big| \, n_{1}  \, m_{n_1}  \,  n_{2}  \, m_{n_2}   \big\rangle 
    + \eta_\text{B} \, \eta \, \big\langle {n_\text{A}}' \, m_{n_\text{A}}' \, {n_\text{B}}' \, m_{n_\text{B}}' \big| \, n_{2}  \, m_{n_2}  \,  n_{1}  \, m_{n_1}   \big\rangle  \bigg]  \bigg\} . \\
      \label{eqA2}
\end{multline}
From the third assumption, we assume that the terms
$\big\langle {n_\text{A}}' \, m_{n_\text{A}}' \, {n_\text{B}}' \, m_{n_\text{B}}' \big| \, n_{1}  \, m_{n_1}  \,  n_{2}  \, m_{n_2}   \big\rangle $
and 
$ \big\langle {n_\text{A}}' \, m_{n_\text{A}}' \, {n_\text{B}}' \, m_{n_\text{B}}' \big| \, n_{2}  \, m_{n_2}  \,  n_{1}  \, m_{n_1}   \big\rangle$
are independent of the projection quantum numbers
and are equal regardless of the order of the rotational quantum numbers.
We then rewrite these terms as
$\big\langle {n_\text{A}}' \, {n_\text{B}}' \, \big| \, n_{1}   \,  n_{2}  \big\rangle $.
If we get rid of the dummy prime variables, we get the general expression
\begin{eqnarray}
\langle  \, d_{\text {AA}}, d_{\text {BB}} ; \eta_{\text {A}}, \eta_{\text {B}}  \big|  \, d_{\text {AB}_1}, d_{\text {AB}_2} ; \eta \big\rangle =  
\big\langle {n_\text{A}} \, {n_\text{B}} \, \big| \, n_{1}   \,  n_{2}  \big\rangle \times 
\langle  \, d_{\text {AA}}, d_{\text {BB}} ; \eta_{\text {A}}, \eta_{\text {B}}  \big|  \, d_{\text {AB}_1}, d_{\text {AB}_2} ; \eta \big\rangle^\text{ns} 
\end{eqnarray}
with
\begin{multline}
\langle  \, d_{\text {AA}}, d_{\text {BB}} ; \eta_{\text {A}}, \eta_{\text {B}}  \big|  \, d_{\text {AB}_1}, d_{\text {AB}_2} ; \eta \big\rangle^\text{ns}  =  
\delta_{\eta_{\text {A}} \eta_{\text {B}}, \eta} \times \\ 
\sum_{ m_{n_\text{A}}} \sum_{m_{n_\text{B}}} 
\sum_{ m_{\text{A}_1} = -i_{\text{A}}}^{+i_{\text{A}}} \sum_{m_{\text{A}_2} \ge m_{\text{A}_1} }   \, \sum_{ m_{\text{B}_1} = -i_{\text{B}}}^{+i_{\text{B}}}
 \sum_{m_{\text{B}_2} \le m_{\text{B}_1} }
  \, \big\langle d_{\text {AA}} ; \eta_{\text {A}} \, \big| \, n_{\text{A}} \, m_{n_\text{A}} \, m_{\text{A}_1} \, m_{\text{A}_2}; \eta_{\text {A}}  \big\rangle  \, \big\langle d_{\text {BB}} ; \eta_{\text {B}} \, \big| \, n_{\text{B}} \, m_{n_\text{B}} \, m_{\text{B}_1} \, m_{\text{B}_2} ; \eta_{\text {B}}   \big\rangle  \\ 
  \frac{1}{\sqrt{\Delta_\text{A}}} \, \frac{1}{\sqrt{\Delta_\text{B}}} \, \frac{1}{\sqrt{\Delta_d}} 
  \, \sum_{ m_{n_1}} \sum_{m_{n_2} } 
   \delta_{m_{n_\text{A}} + m_{n_\text{B}}, m_{n_1} + m_{n_2} + m_{l_r} - m_{l_p}} \\
   \bigg\{  
  \big\langle n_{1}  \, m_{n_1}  \, m_{\text{A}_1} \, m_{\text{B}_1}  \, \big| \, d_{\text {AB}_1} \big\rangle \, \big\langle n_{2}  \, m_{n_2}  \, m_{\text{A}_2} \, m_{\text{B}_2} \, \big| \, d_{\text {AB}_2} \big\rangle 
   \bigg[ 1 + \eta_\text{A} \, \eta_\text{B} \, \eta  \bigg] 
\ \delta_{ m_{n_1} + m_{\text{A}_1}+m_{\text{B}_1}, M_1} 
\ \delta_{ m_{n_2} + m_{\text{A}_2}+m_{\text{B}_2}, M_2} \\
\shoveright{    \big\langle n_{1}  \, m_{n_1}  \, m_{\text{A}_2} \, m_{\text{B}_2}  \, \big| \, d_{\text {AB}_1} \big\rangle \, \big\langle n_{2}  \, m_{n_2}  \, m_{\text{A}_1} \, m_{\text{B}_1} \, \big| \, d_{\text {AB}_2} \big\rangle  \bigg[ \eta_\text{A} \, \eta_\text{B} + \eta \bigg] 
\ \delta_{m_{n_1} + m_{\text{A}_2}+m_{\text{B}_2}, M_1} 
\ \delta_{m_{n_2} + m_{\text{A}_1}+m_{\text{B}_1}, M_2} } \\
\shoveright{ \big\langle n_{1}  \, m_{n_1}  \, m_{\text{A}_1} \, m_{\text{B}_2}  \, \big| \, d_{\text {AB}_1} \big\rangle \, \big\langle n_{2}  \, m_{n_2}  \, m_{\text{A}_2} \, m_{\text{B}_1} \, \big| \, d_{\text {AB}_2} \big\rangle   \bigg[ \eta_\text{B} + \eta_\text{A} \, \eta \bigg] 
\ \delta_{m_{n_1} + m_{\text{A}_1}+m_{\text{B}_2}, M_1} 
\ \delta_{m_{n_2} + m_{\text{A}_2}+m_{\text{B}_1}, M_2} } \\
 \big\langle n_{1}  \, m_{n_1}  \, m_{\text{A}_2} \, m_{\text{B}_1}  \, \big| \, d_{\text {AB}_1} \big\rangle \, \big\langle n_{2}  \, m_{n_2}  \, m_{\text{A}_1} \, m_{\text{B}_2} \, \big| \, d_{\text {AB}_2} \big\rangle  \bigg[  \eta_\text{A} + \eta_\text{B} \, \eta \bigg]  
\ \delta_{m_{n_1} + m_{\text{A}_2}+m_{\text{B}_1}, M_1} 
\ \delta_{m_{n_2} + m_{\text{A}_1}+m_{\text{B}_2}, M_2}  \bigg\} .
 \label{eqA3} 
\end{multline}
The Kronecker symbol $\delta_{\eta_{\text {A}} \eta_{\text {B}}, \eta} $ 
has been added explicitely from the related selection rules in the square brackets.
The expression has been separated into two parts with
a part related to the rotational degree of freedom
$\big\langle {n_\text{A}} \, {n_\text{B}} \, \big| \, n_{1}   \,  n_{2}  \big\rangle $
and for a given transition from initial rotational quantum numbers 
$n_{1}, n_{2} $ to final ones
$n_{\text{A}}, n_{\text{B}}$,
a part related to the nuclear spin degree of freedom
$\langle  \, d_{\text {AA}}, d_{\text {BB}} ; \eta_{\text {A}}, \eta_{\text {B}}  \big|  \, d_{\text {AB}_1}, d_{\text {AB}_2} ; \eta \big\rangle^\text{ns}$.
The corresponding probability is the modulus square of the amplitude
\begin{eqnarray}
P_{i \to j}  = \bigg| \langle  \, d_{\text {AA}}, d_{\text {BB}} ; \eta_{\text {A}}, \eta_{\text {B}}  \big|  \, d_{\text {AB}_1}, d_{\text {AB}_2} ; \eta \big\rangle \bigg|^2 = P^\text{rot}  \times P_{i \to j}^\text{ns}  \end{eqnarray}
with
\begin{align}
P^\text{rot} &= \bigg| \big\langle {n_\text{A}} \, {n_\text{B}} \, \big| \, n_{1}   \,  n_{2}  \big\rangle \bigg|^2 &
P_{i \to j}^\text{ns}  &= \bigg| \langle  \, d_{\text {AA}}, d_{\text {BB}} ; \eta_{\text {A}}, \eta_{\text {B}}  \big|  \, d_{\text {AB}_1}, d_{\text {AB}_2} ; \eta \big\rangle^\text{ns}  \bigg|^2 .
\end{align}

\section*{Appendix B: Simplifications for reactants in zero rotational states}

If we consider reactants in $n_1 = n_2 = 0$, then 
$m_{n_1} = m_{n_2} = 0$ and 
$M_1 \equiv m_1$, $M_2 \equiv m_2$. 

\subsection*{Distinguishable case with $m_1 \ne m_2$}

We conisder first reactants with $m_1 \ne m_2$.
Eq.~\eqref{eqA3} can simplify and we get
\begin{multline}
\langle  \, d_{\text {AA}}, d_{\text {BB}} ; \eta_{\text {A}}, \eta_{\text {B}}  \big|  \, d_{\text {AB}_1}, d_{\text {AB}_2} ; \eta \big\rangle^\text{ns}  =  \delta_{\eta_{\text {A}} \eta_{\text {B}}, \eta} \times 
\sum_{ m_{n_\text{A}}} \sum_{m_{n_\text{B}}}    \delta_{m_{n_\text{A}} + m_{n_\text{B}}, m_{l_r} - m_{l_p}} \\
\sum_{ m_{\text{A}_1} = -i_{\text{A}}}^{+i_{\text{A}}} \sum_{m_{\text{A}_2} \ge m_{\text{A}_1} }   \, \sum_{ m_{\text{B}_1} = -i_{\text{B}}}^{+i_{\text{B}}}
 \sum_{m_{\text{B}_2} \le m_{\text{B}_1} }
 \, \big\langle d_{\text {AA}} ; \eta_{\text {A}} \, \big| \, n_{\text{A}} \, m_{n_\text{A}} \, m_{\text{A}_1} \, m_{\text{A}_2}; \eta_{\text {A}}  \big\rangle  \, \big\langle d_{\text {BB}} ; \eta_{\text {B}} \, \big| \, n_{\text{B}} \, m_{n_\text{B}} \, m_{\text{B}_1} \, m_{\text{B}_2} ; \eta_{\text {B}}   \big\rangle   \\ 
  \frac{1}{\sqrt{\Delta_\text{A}}} \, \frac{1}{\sqrt{\Delta_\text{B}}} \, \frac{1}{\sqrt{\Delta_d}} \,
   \bigg\{  
   \big\langle m_{\text{A}_1} \, m_{\text{B}_1}  \, \big| \, d_{\text {AB}_1} \big\rangle \, \big\langle m_{\text{A}_2} \, m_{\text{B}_2} \, \big| \, d_{\text {AB}_2} \big\rangle   \bigg[ 1 + \eta_\text{A} \, \eta_\text{B} \, \eta  \bigg] 
\ \delta_{m_{\text{A}_1}+m_{\text{B}_1}, m_1} 
\ \delta_{m_{\text{A}_2}+m_{\text{B}_2}, m_2} \\
\shoveright{     +    \big\langle m_{\text{A}_2} \, m_{\text{B}_2}  \, \big| \, d_{\text {AB}_1} \big\rangle \, \big\langle m_{\text{A}_1} \, m_{\text{B}_1} \, \big| \, d_{\text {AB}_2} \big\rangle   \bigg[\eta_\text{A} \, \eta_\text{B} +  \eta \bigg] 
\ \delta_{m_{\text{A}_2}+m_{\text{B}_2}, m_1} 
\ \delta_{m_{\text{A}_1}+m_{\text{B}_1}, m_2} } \\
\shoveright{  +     \big\langle m_{\text{A}_1} \, m_{\text{B}_2}  \, \big| \, d_{\text {AB}_1} \big\rangle \, \big\langle m_{\text{A}_2} \, m_{\text{B}_1} \, \big| \, d_{\text {AB}_2} \big\rangle   \bigg[  \eta_\text{B} 
+ \eta_\text{A} \, \eta \bigg] 
\ \delta_{m_{\text{A}_1}+m_{\text{B}_2}, m_1} 
\ \delta_{m_{\text{A}_2}+m_{\text{B}_1}, m_2} } \\
 \shoveright{   +    \big\langle m_{\text{A}_2} \, m_{\text{B}_1}  \, \big| \, d_{\text {AB}_1} \big\rangle \, \big\langle m_{\text{A}_1} \, m_{\text{B}_2} \, \big| \, d_{\text {AB}_2} \big\rangle  \bigg[ \eta_\text{A} + \eta_\text{B} \, \eta \bigg]  
\ \delta_{m_{\text{A}_2}+m_{\text{B}_1}, m_1} 
\ \delta_{m_{\text{A}_1}+m_{\text{B}_2}, m_2}  \bigg\}    . }  \\
    \label{eqA3bis}
\end{multline}

\subsection*{Indistinguishable case with $m_1 = m_2$}

We further develop the expression of Eq.~\eqref{eqA3bis}
when, in addition, the reactants are indistinguishable.
In this case $d_{\text {AB}_1} = d_{\text {AB}_2} = d_{\text {AB}}$, 
$\eta = +1$, $\Delta_d = 4$.
From the first Kronecker delta term, this implies necessarily 
$\eta_{\text {A}} = \eta_{\text {B}} = +1$ or $\eta_{\text {A}} = \eta_{\text {B}} = -1$.
Also, molecules have necessarily the same values of projection so that $m_1 = m_2$. 
Eq.~\eqref{eqA3bis} simplifies to
\begin{multline}
\langle  \, d_{\text {AA}}, d_{\text {BB}} ; \eta_{\text {A}}, \eta_{\text {B}}  \big|  \, d_{\text {AB}}, d_{\text {AB}} ; +1 \big\rangle^\text{ns} =  \delta_{\eta_{\text {A}} \eta_{\text {B}}, +1} \times 
\sum_{ m_{n_\text{A}}} \sum_{m_{n_\text{B}}}    \delta_{m_{n_\text{A}} + m_{n_\text{B}}, m_{l_r} - m_{l_p}} \\
 \sum_{ m_{\text{A}_1} = -i_{\text{A}}}^{+i_{\text{A}}} \sum_{m_{\text{A}_2} \ge m_{\text{A}_1} }   \, \sum_{ m_{\text{B}_1} = -i_{\text{B}}}^{+i_{\text{B}}}
 \sum_{m_{\text{B}_2} \le m_{\text{B}_1} }
  \,  \big\langle d_{\text {AA}} ; \eta_{\text {A}} \, \big| \, n_{\text{A}} \, m_{n_\text{A}} \, m_{\text{A}_1} \, m_{\text{A}_2}; \eta_{\text {A}}  \big\rangle  \, \big\langle d_{\text {BB}} ; \eta_{\text {B}} \, \big| \, n_{\text{B}} \, m_{n_\text{B}} \, m_{\text{B}_1} \, m_{\text{B}_2} ; \eta_{\text {B}}  \big\rangle  \\ 
  \frac{1}{\sqrt{\Delta_\text{A}}} \, \frac{1}{\sqrt{\Delta_\text{B}}} \
   \bigg\{  
  \big\langle  m_{\text{A}_1} \, m_{\text{B}_1}  \, \big| \, d_{\text {AB}} \big\rangle \, \big\langle  m_{\text{A}_2} \, m_{\text{B}_2} \, \big| \, d_{\text {AB}} \big\rangle 
   \bigg[ 1 + \eta_\text{A} \, \eta_\text{B}  \bigg] 
\ \delta_{  m_{\text{A}_1}+m_{\text{B}_1}, m_1} 
\ \delta_{  m_{\text{A}_2}+m_{\text{B}_2}, m_1}  \\
\big\langle  m_{\text{A}_1} \, m_{\text{B}_2}  \, \big| \, d_{\text {AB}} \big\rangle \, \big\langle m_{\text{A}_2} \, m_{\text{B}_1} \, \big| \, d_{\text {AB}} \big\rangle   \bigg[ \eta_\text{A} + \eta_\text{B} \bigg] 
\ \delta_{ m_{\text{A}_1}+m_{\text{B}_2}, m_1} 
\ \delta_{ m_{\text{A}_2}+m_{\text{B}_1}, m_1} \bigg\}    .    \\  
%
%
    \label{eqA4}
\end{multline}
Let's consider the case 
$m_{\text{A}_2} = m_{\text{A}_1}$ and $m_{\text{B}_2} = m_{\text{B}_1}$ in the quadruple sum of nuclear spins,
with necessarily $\eta_{\text {A}} = +1$, $\eta_{\text {B}} = +1$, 
$\Delta_\text{A} = \Delta_\text{B} = 4$. 
After simplifications, this gives a term
\begin{eqnarray}
 \big\langle d_{\text {AA}} ; +1 \, \big| \, n_{\text{A}} \, m_{n_\text{A}} \, m_{\text{A}_1} \, m_{\text{A}_1}; +1 \big\rangle  \, \big\langle d_{\text {BB}} ; +1 \, \big| \, n_{\text{B}} \, m_{n_\text{B}} \, m_{\text{B}_1} \, m_{\text{B}_1} ; +1  \big\rangle \,
 \big\langle m_{\text{A}_1} \, m_{\text{B}_1}  \, \big| \, d_{\text {AB}} \big\rangle \, \big\langle m_{\text{A}_1} \, m_{\text{B}_1} \, \big| \, d_{\text {AB}} \big\rangle .
    \label{eqA5}
\end{eqnarray}
The case $m_{\text{A}_2} > m_{\text{A}_1}$ and $m_{\text{B}_2} < m_{\text{B}_1}$ in the quadruple sum,
with $\Delta_\text{A} = \Delta_\text{B} = 2$,
$\eta_{\text {A}} = \eta_{\text {B}} = +1$ or $\eta_{\text {A}} = \eta_{\text {B}} = -1$,
gives
\begin{eqnarray}
 \big\langle d_{\text {AA}} ; \pm1 \, \big| \, n_{\text{A}} \, m_{n_\text{A}} \, m_{\text{A}_1} \, m_{\text{A}_1}; \pm1 \big\rangle  \, \big\langle d_{\text {BB}} ; \pm1 \, \big| \, n_{\text{B}} \, m_{n_\text{B}}  \, m_{\text{B}_1} \, m_{\text{B}_1} ; \pm1 \big\rangle \,
 \big\langle m_{\text{A}_1} \, m_{\text{B}_1}  \, \big| \, d_{\text {AB}} \big\rangle \, \big\langle m_{\text{A}_2} \, m_{\text{B}_2} \, \big| \, d_{\text {AB}} \big\rangle .
  \label{eqA6}
\end{eqnarray}
We used the fact that the coefficients $\big\langle m_{\text{A}_1} \, m_{\text{B}_2}  \, \big| \, d_{\text {AB}} \big\rangle$ and $\big\langle m_{\text{A}_2} \, m_{\text{B}_1} \, \big| \, d_{\text {AB}} \big\rangle $ in Eq.~\eqref{eqA4} both necessarily vanish when the A or B atoms are swapped,
because when $m_{\text{A}_2} > m_{\text{A}_1}$ ($m_{\text{B}_2} < m_{\text{B}_1}$) 
then $m_{\text{A}_2} + m_{\text{B}_1} > m_{\text{A}_1} + m_{\text{B}_1}$ ($m_{\text{A}_1} + m_{\text{B}_2} < m_{\text{A}_1} + m_{\text{B}_1}$),
so that $m_{\text{A}_2} + m_{\text{B}_1} > m_1$ ($m_{\text{A}_1} + m_{\text{B}_2} < m_1$).
For the same reasons, we 
cannot have a case $m_{\text{A}_2} = m_{\text{A}_1}$ and $m_{\text{B}_2} < m_{\text{B}_1}$,
or $m_{\text{A}_2} > m_{\text{A}_1}$ and $m_{\text{B}_2} = m_{\text{B}_1}$.
Therefore, one can write Eq.~\eqref{eqA4} in a compact form
\begin{multline}
\langle  \, d_{\text {AA}}, d_{\text {BB}} ; \eta_{\text {A}}, \eta_{\text {B}}  \big|  \, d_{\text {AB}}, d_{\text {AB}} ; +1 \big\rangle^\text{ns} = 
    \delta_{\eta_{\text {A}} \eta_{\text {B}}, +1} \times 
  \sum_{ m_{n_\text{A}}} \sum_{m_{n_\text{B}}}  \delta_{m_{n_\text{A}} + m_{n_\text{B}}, m_{l_r} - m_{l_p}} \\
  \sum_{ m_{\text{A}_1} = -i_{\text{A}}}^{+i_{\text{A}}} \sum_{m_{\text{A}_2} \ge m_{\text{A}_1} }   \, \sum_{ m_{\text{B}_1} = -i_{\text{B}}}^{+i_{\text{B}}}
 \sum_{m_{\text{B}_2} \le m_{\text{B}_1} }
  \,  \big\langle d_{\text {AA}} ; \eta_{\text {A}} \, \big| \, n_{\text{A}} \, m_{n_\text{A}} \, m_{\text{A}_1} \, m_{\text{A}_2}; \eta_{\text {A}}  \big\rangle  \, \big\langle d_{\text {BB}} ; \eta_{\text {B}} \, \big| \, n_{\text{B}} \, m_{n_\text{B}} \, m_{\text{B}_1} \, m_{\text{B}_2} ; \eta_{\text {B}}  \big\rangle  \\ 
     \big\langle m_{\text{A}_1} \, m_{\text{B}_1}  \, \big| \, d_{\text {AB}} \big\rangle \, \big\langle m_{\text{A}_2} \, m_{\text{B}_2} \, \big| \, d_{\text {AB}} \big\rangle 
 \    \delta_{m_{\text{A}_1}+m_{\text{B}_1}, m_1} 
\ \delta_{m_{\text{A}_2}+m_{\text{B}_2}, m_1}  .
 \label{eqA7}
\end{multline}

\subsection*{Distinguishable case with $m_1 = m_2$}

We further develop Eq.~\eqref{eqA3} when the reactants are distinguishable.
In this case $d_{\text {AB}_1} \ne d_{\text {AB}_2}$, $\Delta_d = 2$,
and both $\eta = \pm 1$ components have to be computed.  
From the first Kronecker delta term, this implies necessarily 
$\eta_{\text {A}} = \eta_{\text {B}} = +1$ or $\eta_{\text {A}} = \eta_{\text {B}} = -1$
when $\eta = + 1$,
and $\eta_{\text {A}} = +1$, $\eta_{\text {B}} = -1$ or $\eta_{\text {A}} = -1$, $\eta_{\text {B}} = +1$
when $\eta = - 1$.
We focus on the case where
molecules have the same values of projection so that $m_1 = m_2$.
Let's start with the component $\eta = +1$.
The case $m_{\text{A}_2} = m_{\text{A}_1}$ and $m_{\text{B}_2} = m_{\text{B}_1}$ in the quadruple sum
of nuclear spins,
with necessarily $\eta_{\text {A}} = +1$, $\eta_{\text {B}} = +1$, 
$\Delta_\text{A} = \Delta_\text{B} = 4$, $\eta = +1$, gives a term
\begin{eqnarray}
   \sqrt{2} \,  
    \big\langle d_{\text {AA}} ; +1 \, \big| \, n_{\text{A}} \, m_{n_\text{A}} \, m_{\text{A}_1} \, m_{\text{A}_1}; +1  \big\rangle  \, \big\langle d_{\text {BB}} ; +1 \, \big| \, n_{\text{B}} \, m_{n_\text{B}} \, m_{\text{B}_1} \, m_{\text{B}_1} ; +1   \big\rangle \,
     \big\langle m_{\text{A}_1} \, m_{\text{B}_1}  \, \big| \, d_{\text {AB}_1} \big\rangle \,
     \big\langle m_{\text{A}_1} \, m_{\text{B}_1}  \, \big| \, d_{\text {AB}_2} \big\rangle.
    \label{eqA9-0}
\end{eqnarray}
We prefer the following form for a later general notation convenience
\begin{multline}
   \frac{1}{\sqrt{2}} \,  
    \big\langle d_{\text {AA}} ; +1 \, \big| \, n_{\text{A}} \, m_{n_\text{A}} \, m_{\text{A}_1} \, m_{\text{A}_1}; +1  \big\rangle  \, \big\langle d_{\text {BB}} ; +1 \, \big| \, n_{\text{B}} \, m_{n_\text{B}} \, m_{\text{B}_1} \, m_{\text{B}_1} ; +1   \big\rangle  \\
  \bigg\{    \big\langle m_{\text{A}_1} \, m_{\text{B}_1}  \, \big| \, d_{\text {AB}_1} \big\rangle \, \big\langle m_{\text{A}_1} \, m_{\text{B}_1} \, \big| \, d_{\text {AB}_2} \big\rangle +
    \big\langle m_{\text{A}_1} \, m_{\text{B}_1}  \, \big| \, d_{\text {AB}_1} \big\rangle \, \big\langle m_{\text{A}_1} \, m_{\text{B}_1} \, \big| \, d_{\text {AB}_2} \big\rangle \bigg\} .
    \label{eqA9}
\end{multline}
The case $m_{\text{A}_2} > m_{\text{A}_1}$ and $m_{\text{B}_2} < m_{\text{B}_1}$ in the quadruple sum,
with $\Delta_\text{A} = \Delta_\text{B} = 2$,
$\eta_{\text {A}} = \eta_{\text {B}} = +1$ or $\eta_{\text {A}} = \eta_{\text {B}} = -1$,
gives
\begin{multline}
   \frac{1}{\sqrt{2}} \,  
    \big\langle d_{\text {AA}} ; \pm 1 \, \big| \, n_{\text{A}} \, m_{n_\text{A}} \, m_{\text{A}_1} \, m_{\text{A}_1}; \pm 1  \big\rangle  \, \big\langle d_{\text {BB}} ; \pm 1 \, \big| \, n_{\text{B}} \, m_{n_\text{B}} \, m_{\text{B}_1} \, m_{\text{B}_1} ; \pm 1   \big\rangle  \\
  \bigg\{    \big\langle m_{\text{A}_1} \, m_{\text{B}_1}  \, \big| \, d_{\text {AB}_1} \big\rangle \, \big\langle m_{\text{A}_2} \, m_{\text{B}_2} \, \big| \, d_{\text {AB}_2} \big\rangle +
    \big\langle m_{\text{A}_2} \, m_{\text{B}_2}  \, \big| \, d_{\text {AB}_1} \big\rangle \, \big\langle m_{\text{A}_1} \, m_{\text{B}_1} \, \big| \, d_{\text {AB}_2} \big\rangle \bigg\} .
    \label{eqA10}
\end{multline}
For the same reasons than above, 
the coefficients $\big\langle m_{\text{A}_1} \, m_{\text{B}_2}  \, \big| \, d_{\text {AB}_1} \big\rangle$, 
$\big\langle m_{\text{A}_2} \, m_{\text{B}_1} \, \big| \, d_{\text {AB}_1} \big\rangle $,
$\big\langle m_{\text{A}_1} \, m_{\text{B}_2}  \, \big| \, d_{\text {AB}_2} \big\rangle$, 
$\big\langle m_{\text{A}_2} \, m_{\text{B}_1} \, \big| \, d_{\text {AB}_2} \big\rangle $ all vanish.
Similarly, we cannot have a case $m_{\text{A}_2} = m_{\text{A}_1}$ and $m_{\text{B}_2} < m_{\text{B}_1}$,
or  $m_{\text{A}_2} > m_{\text{A}_1}$ and $m_{\text{B}_2} = m_{\text{B}_1}$.
For $\eta = -1$, similar arguments hold, in addition with the fact that the case
$m_{\text{A}_2} = m_{\text{A}_1}$ and $m_{\text{B}_2} = m_{\text{B}_1}$
cannot exist.
The case $m_{\text{A}_2} > m_{\text{A}_1}$ and $m_{\text{B}_2} < m_{\text{B}_1}$ in the quadruple sum,
with $\Delta_\text{A} = \Delta_\text{B} = 2$,
$\eta_{\text {A}} = +1$, $\eta_{\text {B}} = -1$ or $\eta_{\text {A}} = -1$, $\eta_{\text {B}} = +1$,
gives
\begin{multline}
   \frac{1}{\sqrt{2}} \, \big\langle d_{\text {AA}} ; \pm 1 \, \big| \, n_{\text{A}} \, m_{n_\text{A}} \, m_{\text{A}_1} \, m_{\text{A}_2}; \pm 1  \big\rangle  \, \big\langle d_{\text {BB}} ; \mp 1 \, \big| \, n_{\text{B}} \, m_{n_\text{B}} \, m_{\text{B}_1} \, m_{\text{B}_2} ; \mp 1   \big\rangle  \\ 
   \bigg\{  
   \big\langle m_{\text{A}_1} \, m_{\text{B}_1}  \, \big| \, d_{\text {AB}_1} \big\rangle \, \big\langle m_{\text{A}_2} \, m_{\text{B}_2} \, \big| \, d_{\text {AB}_2} \big\rangle   
   -    \big\langle m_{\text{A}_2} \, m_{\text{B}_2}  \, \big| \, d_{\text {AB}_1} \big\rangle \, \big\langle m_{\text{A}_1} \, m_{\text{B}_1} \, \big| \, d_{\text {AB}_2} \big\rangle   \bigg\}  
    \label{eqA11}
\end{multline}
with now a destructive term inside the bracket.
Gathering these expressions all together, Eq.~\eqref{eqA3} can then be written in a compact form
\begin{multline}
\langle  \, d_{\text {AA}}, d_{\text {BB}} ; \eta_{\text {A}}, \eta_{\text {B}}  \big|  \, d_{\text {AB}_1}, d_{\text {AB}_2} ; \eta \big\rangle^\text{ns} = \delta_{\eta_{\text {A}} \eta_{\text {B}}, \eta} \times 
  \sum_{ m_{n_\text{A}}} \sum_{m_{n_\text{B}}}  \delta_{m_{n_\text{A}} + m_{n_\text{B}}, m_{l_r} - m_{l_p}} \\
\frac{1}{\sqrt{2}} \,
 \sum_{ m_{\text{A}_1} = -i_{\text{A}}}^{+i_{\text{A}}} \sum_{m_{\text{A}_2} \ge m_{\text{A}_1} }   \, \sum_{ m_{\text{B}_1} = -i_{\text{B}}}^{+i_{\text{B}}}
 \sum_{m_{\text{B}_2} \le m_{\text{B}_1} }
  \, \big\langle d_{\text {AA}} ; \eta_{\text {A}} \, \big|  \, n_{\text{A}} \, m_{n_\text{A}} \, m_{\text{A}_1} \, m_{\text{A}_2}; \eta_{\text {A}}  \big\rangle  \, \big\langle d_{\text {BB}} ; \eta_{\text {B}} \, \big|  \, n_{\text{B}} \, m_{n_\text{B}} \, m_{\text{B}_1} \, m_{\text{B}_2} ; \eta_{\text {B}}   \big\rangle \\ 
     \bigg\{  
   \big\langle m_{\text{A}_1} \, m_{\text{B}_1}  \, \big| \, d_{\text {AB}_1} \big\rangle \, \big\langle m_{\text{A}_2} \, m_{\text{B}_2} \, \big| \, d_{\text {AB}_2} \big\rangle   
   + \eta \, \big\langle m_{\text{A}_2} \, m_{\text{B}_2}  \, \big| \, d_{\text {AB}_1} \big\rangle \, \big\langle m_{\text{A}_1} \, m_{\text{B}_1} \, \big| \, d_{\text {AB}_2} \big\rangle 
     \bigg\}  \\
 \    \delta_{m_{\text{A}_1}+m_{\text{B}_1}, m_1} 
\ \delta_{m_{\text{A}_2}+m_{\text{B}_2}, m_1} .
 \label{eqA12}
\end{multline}

\subsection*{Compact form for molecules with same projection quantum numbers $m_1 = m_2$}

Both Eq.~\eqref{eqA7} and Eq.~\eqref{eqA12} can be written in an even more compact form,
whether the molecules, with same projection quantum numbers $m_1 = m_2$, 
are indistinguishable or not
\begin{multline}
\langle  \, d_{\text {AA}}, d_{\text {BB}} ; \eta_{\text {A}}, \eta_{\text {B}}  \big|  \, d_{\text {AB}_1}, d_{\text {AB}_2} ; \eta \big\rangle^\text{ns} = \delta_{\eta_{\text {A}} \eta_{\text {B}}, \eta} \times 
  \sum_{ m_{n_\text{A}}} \sum_{m_{n_\text{B}}}  \delta_{m_{n_\text{A}} + m_{n_\text{B}}, m_{l_r} - m_{l_p}} \\
\frac{1}{\sqrt{\Delta_d}} \,
 \sum_{ m_{\text{A}_1} = -i_{\text{A}}}^{+i_{\text{A}}} \sum_{m_{\text{A}_2} \ge m_{\text{A}_1} }   \, \sum_{ m_{\text{B}_1} = -i_{\text{B}}}^{+i_{\text{B}}}
 \sum_{m_{\text{B}_2} \le m_{\text{B}_1} }
  \, \big\langle d_{\text {AA}} ; \eta_{\text {A}} \, \big| \, n_{\text{A}} \, m_{n_\text{A}} \, m_{\text{A}_1} \, m_{\text{A}_2}; \eta_{\text {A}}  \big\rangle  \, \big\langle d_{\text {BB}} ; \eta_{\text {B}} \, \big| \, n_{\text{B}} \, m_{n_\text{B}} \, m_{\text{B}_1} \, m_{\text{B}_2} ; \eta_{\text {B}}   \big\rangle \\ 
     \bigg\{  
   \big\langle m_{\text{A}_1} \, m_{\text{B}_1}  \, \big| \, d_{\text {AB}_1} \big\rangle \, \big\langle m_{\text{A}_2} \, m_{\text{B}_2} \, \big| \, d_{\text {AB}_2} \big\rangle   
   + \eta \, \big\langle m_{\text{A}_2} \, m_{\text{B}_2}  \, \big| \, d_{\text {AB}_1} \big\rangle \, \big\langle m_{\text{A}_1} \, m_{\text{B}_1} \, \big| \, d_{\text {AB}_2} \big\rangle 
     \bigg\}  \\
 \    \delta_{m_{\text{A}_1}+m_{\text{B}_1}, m_1} 
\ \delta_{m_{\text{A}_2}+m_{\text{B}_2}, m_1}  .
 \label{eqA13}
\end{multline}

\section*{Appendix C: Sum of the probabilities over the combined dressed states of the products}

We consider here 
the case of reactants in $n_1 = n_2 = m_{n_1} = m_{n_2} = 0$
but we explicitly include the quantum numbers. 
We consider first Eq.~\eqref{eqA7} for 
the case of indistinguishable molecules 
as the expression is easier to work for the derivation,
but we use the intermediate form in Eq.~\eqref{eqA2}, 
explicitly including the term
$\big\langle {n_\text{A}} \, m_{n_\text{A}} \, {n_\text{B}} \, m_{n_\text{B}} \big| \, n_{1}  \, m_{n_1}  \,  n_{2}  \, m_{n_2} \big\rangle $ inside the sum. We use the fact that it
is the same regardless of the order of the rotational quantum numbers
but we don't necessarily use the fact that it is independent of the projection quantum numbers.
Therefore we keep explicit the projection quantum numbers for the purpose of the derivation.
To recall, we have $d_{\text {AB}_1} = d_{\text {AB}_2} = d_{\text {AB}}$
and $m_1 = m_2$.
The total probability summed over all the final combined dressed states of the products
is given by
\begin{multline}
\sum_j P_{i \to j}  = \sum_{\big|  d_{\text {AA}}, d_{\text {BB}} ; \eta_{\text {A}}, \eta_{\text {B}} \big\rangle} \bigg| \langle  \, d_{\text {AA}}, d_{\text {BB}} ; \eta_{\text {A}}, \eta_{\text {B}}  \big|  \, d_{\text {AB}}, d_{\text {AB}} ; +1 \big\rangle \bigg|^2 \\
= \sum_{\big|  d_{\text {AA}} ; \eta_{\text {A}} \big\rangle} \sum_{\big|   d_{\text {BB}} ; \eta_{\text {B}} \big\rangle} \,  \bigg| \langle  \, d_{\text {AA}}, d_{\text {BB}} ; \eta_{\text {A}}, \eta_{\text {B}}  \big|  \, d_{\text {AB}}, d_{\text {AB}} ; +1 \big\rangle \bigg|^2  \\
 = \delta_{\eta_{\text {A}} \eta_{\text {B}}, +1} \times 
\sum_{\big|  d_{\text {AA}} ; \eta_{\text {A}} \big\rangle} \sum_{\big|   d_{\text {BB}} ; \eta_{\text {B}} \big\rangle} \,
 \bigg|   \sum_{ m_{n_\text{A}}} \sum_{m_{n_\text{B}}} \, \sum_{m_{\text{A}_1}} \sum_{m_{\text{A}_2} \ge m_{\text{A}_1} }   \, \sum_{m_{\text{B}_1}} \sum_{m_{\text{B}_2} \le m_{\text{B}_1} }  \\
 \big\langle d_{\text {AA}} ; \eta_{\text {A}} \, \big| \, n_{\text{A}} \, m_{n_\text{A}} \, m_{\text{A}_1} \, m_{\text{A}_2}; \eta_{\text {A}}  \big\rangle  \, \big\langle d_{\text {BB}} ; \eta_{\text {B}} \, \big| \, n_{\text{B}} \, m_{n_\text{B}} \, m_{\text{B}_1} \, m_{\text{B}_2} ; \eta_{\text {B}}   \big\rangle \\
 \big\langle {n_\text{A}} \, m_{n_\text{A}} \, {n_\text{B}} \, m_{n_\text{B}} \big| \, n_{1}  \, m_{n_1}  \,  n_{2}  \, m_{n_2} \big\rangle \, 
 \big\langle m_{\text{A}_1} \, m_{\text{B}_1}  \, \big| \, d_{\text {AB}} \big\rangle \, \big\langle m_{\text{A}_2} \, m_{\text{B}_2} \, \big| \, d_{\text {AB}} \big\rangle  \\
\delta_{m_{n_\text{A}} + m_{n_\text{B}}, m_{l_r} - m_{l_p}} 
\   \delta_{m_{\text{A}_1}+m_{\text{B}_1}, m_1} 
\ \delta_{m_{\text{A}_2}+m_{\text{B}_2}, m_1} \bigg|^2   \\ \nonumber
\end{multline}
\begin{multline}
=  \delta_{\eta_{\text {A}} \eta_{\text {B}}, +1} \times \sum_{\big|  d_{\text {AA}} ; \eta_{\text {A}} \big\rangle} \sum_{\big|   d_{\text {BB}} ; \eta_{\text {B}} \big\rangle} \,
 \sum_{ m_{n_\text{A}}'} \sum_{m_{n_\text{B}}'} \, 
\sum_{m_{\text{A}_1}'} \sum_{m_{\text{A}_2}' \ge m_{\text{A}_1}' }   \, \sum_{m_{\text{B}_1}'} \sum_{m_{\text{B}_2}' \le m_{\text{B}_1}' } \, 
\sum_{ m_{n_\text{A}}} \sum_{m_{n_\text{B}}} \,
\sum_{m_{\text{A}_1}} \sum_{m_{\text{A}_2} \ge m_{\text{A}_1} }   \, \sum_{m_{\text{B}_1}} \sum_{m_{\text{B}_2} \le m_{\text{B}_1} } \\
\big\langle  n_{\text{A}} \, m_{n_\text{A}}' \, m_{\text{A}_1}' \, m_{\text{A}_2}'; \eta_{\text {A}}   \, \big| \,  d_{\text {AA}} ; \eta_{\text {A}} \big\rangle  \,  \big\langle d_{\text {AA}} ; \eta_{\text {A}} \, \big| \,  n_{\text{A}} \, m_{n_\text{A}} \, m_{\text{A}_1} \, m_{\text{A}_2}; \eta_{\text {A}}  \big\rangle 
\\ 
\big\langle  n_{\text{B}} \, m_{n_\text{B}}' \, m_{\text{B}_1}' \, m_{\text{B}_2}' ; \eta_{\text {B}}  \, \big| \, d_{\text {BB}} ; \eta_{\text {B}}    \big\rangle \,
\, \big\langle d_{\text {BB}} ; \eta_{\text {B}} \, \big| \, n_{\text{B}} \, m_{n_\text{B}} \, m_{\text{B}_1} \, m_{\text{B}_2} ; \eta_{\text {B}}   \big\rangle  \\
\big\langle n_{1}  \, m_{n_1}  \,  n_{2}  \, m_{n_2}  \big| \, {n_\text{A}} \, m_{n_\text{A}}' \, {n_\text{B}} \, m_{n_\text{B}}' \big\rangle \,
 \big\langle {n_\text{A}} \, m_{n_\text{A}} \, {n_\text{B}} \, m_{n_\text{B}} \big| \, n_{1}  \, m_{n_1}  \,  n_{2}  \, m_{n_2} \big\rangle \\
\big\langle d_{\text {AB}} \, \big| \,  m_{\text{A}_1}' \, m_{\text{B}_1}'  \big\rangle \, \big\langle  d_{\text {AB}} \, \big| \, m_{\text{A}_2}' \, m_{\text{B}_2}'  \big\rangle   \, 
\big\langle m_{\text{A}_1} \, m_{\text{B}_1}  \, \big| \, d_{\text {AB}} \big\rangle \, \big\langle m_{\text{A}_2} \, m_{\text{B}_2} \, \big| \, d_{\text {AB}} \big\rangle  \\
\shoveright{ 
 \ \delta_{m_{n_\text{A}}' + m_{n_\text{B}}', m_{l_r} - m_{l_p}}  
  \ \delta_{m_{n_\text{A}} + m_{n_\text{B}}, m_{l_r} - m_{l_p}}  
 \   \delta_{m_{\text{A}_1}' +m_{\text{B}_1}', m_1} 
\ \delta_{m_{\text{A}_2}'+m_{\text{B}_2}', m_1}
\   \delta_{m_{\text{A}_1}+m_{\text{B}_1}, m_1} 
\ \delta_{m_{\text{A}_2}+m_{\text{B}_2}, m_1} } \\ \nonumber
\end{multline}
\begin{multline}
 =  \delta_{\eta_{\text {A}} \eta_{\text {B}}, +1} \times  \,
 \sum_{ m_{n_\text{A}}'} \sum_{m_{n_\text{B}}'} \, 
\sum_{m_{\text{A}_1}'} \sum_{m_{\text{A}_2}' \ge m_{\text{A}_1}' }   \, \sum_{m_{\text{B}_1}'} \sum_{m_{\text{B}_2}' \le m_{\text{B}_1}' } \, 
\sum_{ m_{n_\text{A}}} \sum_{m_{n_\text{B}}} \,
\sum_{m_{\text{A}_1}} \sum_{m_{\text{A}_2} \ge m_{\text{A}_1} }   \, \sum_{m_{\text{B}_1}} \sum_{m_{\text{B}_2} \le m_{\text{B}_1} } \\
\big\langle n_{\text{A}} \, m_{n_\text{A}}' \, m_{\text{A}_1}' \, m_{\text{A}_2}'; \eta_{\text {A}}   \,  \big| \bigg( \sum_{\big|  d_{\text {AA}} ; \eta_{\text {A}} \big\rangle} \big| \,  d_{\text {AA}} ; \eta_{\text {A}} \big\rangle  \,
\big\langle d_{\text {AA}} ; \eta_{\text {A}} \, \big| \bigg) \big| \, n_{\text{A}} \, m_{n_\text{A}} \, m_{\text{A}_1} \, m_{\text{A}_2}; \eta_{\text {A}}  \big\rangle \\
\big\langle  n_{\text{B}} \, m_{n_\text{B}}' \, m_{\text{B}_1}' \, m_{\text{B}_2}' ; \eta_{\text {B}}  \, \big| \bigg( \sum_{\big|   d_{\text {BB}} ; \eta_{\text {B}} \big\rangle} \big| \, d_{\text {BB}} ; \eta_{\text {B}}    \big\rangle \,
\big\langle d_{\text {BB}} ; \eta_{\text {B}} \, \big| \bigg) \big| \,  n_{\text{B}} \, m_{n_\text{B}} \, m_{\text{B}_1} \, m_{\text{B}_2} ; \eta_{\text {B}}   \big\rangle \\
\big\langle n_{1}  \, m_{n_1}  \,  n_{2}  \, m_{n_2}  \big| \, {n_\text{A}} \, m_{n_\text{A}}' \, {n_\text{B}} \, m_{n_\text{B}}' \big\rangle \,
 \big\langle {n_\text{A}} \, m_{n_\text{A}} \, {n_\text{B}} \, m_{n_\text{B}} \big| \, n_{1}  \, m_{n_1}  \,  n_{2}  \, m_{n_2} \big\rangle \\
  \big\langle d_{\text {AB}} \, \big| \,  m_{\text{A}_1}' \, m_{\text{B}_1}'  \big\rangle \, \big\langle  d_{\text {AB}} \, \big| \, m_{\text{A}_2}' \, m_{\text{B}_2}'  \big\rangle  \,
\big\langle m_{\text{A}_1} \, m_{\text{B}_1}  \, \big| \, d_{\text {AB}} \big\rangle \, \big\langle m_{\text{A}_2} \, m_{\text{B}_2} \, \big| \, d_{\text {AB}} \big\rangle  \\
 \delta_{m_{n_\text{A}}' + m_{n_\text{B}}', m_{l_r} - m_{l_p}}  
  \ \delta_{m_{n_\text{A}} + m_{n_\text{B}}, m_{l_r} - m_{l_p}}  
 \   \delta_{m_{\text{A}_1}' +m_{\text{B}_1}', m_1} 
\ \delta_{m_{\text{A}_2}'+m_{\text{B}_2}', m_1}
\   \delta_{m_{\text{A}_1}+m_{\text{B}_1}, m_1} 
\ \delta_{m_{\text{A}_2}+m_{\text{B}_2}, m_1}  .  \nonumber
\end{multline}
We can use the closure relations $\sum_{\big|  d_{\text {AA}} ; \eta_{\text {A}} \big\rangle} \big| \,  d_{\text {AA}} ; \eta_{\text {A}} \big\rangle  \,
\big\langle d_{\text {AA}} ; \eta_{\text {A}} \, \big| = 1$  and
$\sum_{\big|  d_{\text {BB}} ; \eta_{\text {B}} \big\rangle} \big| \,  d_{\text {BB}} ; \eta_{\text {B}} \big\rangle  \,
\big\langle d_{\text {BB}} ; \eta_{\text {B}} \, \big| = 1$. We get
\begin{multline}
\sum_j P_{i \to j}  =  \delta_{\eta_{\text {A}} \eta_{\text {B}}, +1} \times 
 \sum_{ m_{n_\text{A}}'} \sum_{m_{n_\text{B}}'} \, 
\sum_{m_{\text{A}_1}'} \sum_{m_{\text{A}_2}' \ge m_{\text{A}_1}' }   \, \sum_{m_{\text{B}_1}'} \sum_{m_{\text{B}_2}' \le m_{\text{B}_1}' } \, 
\sum_{ m_{n_\text{A}}} \sum_{m_{n_\text{B}}} \,
\sum_{m_{\text{A}_1}} \sum_{m_{\text{A}_2} \ge m_{\text{A}_1} }   \, \sum_{m_{\text{B}_1}} \sum_{m_{\text{B}_2} \le m_{\text{B}_1} } \\
\big\langle  n_{\text{A}} \, m_{n_\text{A}}' \, m_{\text{A}_1}' \, m_{\text{A}_2}'; \eta_{\text {A}}   \,  \big| \,  n_{\text{A}} \, m_{n_\text{A}} \, m_{\text{A}_1} \, m_{\text{A}_2}; \eta_{\text {A}}  \big\rangle \,
\big\langle  n_{\text{B}} \, m_{n_\text{B}}' \, m_{\text{B}_1}' \, m_{\text{B}_2}' ; \eta_{\text {B}}  \, \big| \,  n_{\text{B}} \, m_{n_\text{B}} \, m_{\text{B}_1} \, m_{\text{B}_2} ; \eta_{\text {B}}   \big\rangle \\
\big\langle n_{1}  \, m_{n_1}  \,  n_{2}  \, m_{n_2}  \big| \, {n_\text{A}} \, m_{n_\text{A}}' \, {n_\text{B}} \, m_{n_\text{B}}' \big\rangle \,
 \big\langle {n_\text{A}} \, m_{n_\text{A}} \, {n_\text{B}} \, m_{n_\text{B}} \big| \, n_{1}  \, m_{n_1}  \,  n_{2}  \, m_{n_2} \big\rangle \\
  \big\langle d_{\text {AB}} \, \big| \,  m_{\text{A}_1}' \, m_{\text{B}_1}'  \big\rangle \, \big\langle  d_{\text {AB}} \, \big| \, m_{\text{A}_2}' \, m_{\text{B}_2}'  \big\rangle  \,
\big\langle m_{\text{A}_1} \, m_{\text{B}_1}  \, \big| \, d_{\text {AB}} \big\rangle \, \big\langle m_{\text{A}_2} \, m_{\text{B}_2} \, \big| \, d_{\text {AB}} \big\rangle  \\
 \delta_{m_{n_\text{A}}' + m_{n_\text{B}}', m_{l_r} - m_{l_p}}  
  \ \delta_{m_{n_\text{A}} + m_{n_\text{B}}, m_{l_r} - m_{l_p}}  
 \   \delta_{m_{\text{A}_1}' +m_{\text{B}_1}', m_1} 
\ \delta_{m_{\text{A}_2}'+m_{\text{B}_2}', m_1}
\   \delta_{m_{\text{A}_1}+m_{\text{B}_1}, m_1} 
\ \delta_{m_{\text{A}_2}+m_{\text{B}_2}, m_1}  . \nonumber
\end{multline}
We now use the fact that
$\big\langle n_{\text{A}} \, m_{n_\text{A}}' \, m_{\text{A}_1}' \, m_{\text{A}_2}'; \eta_{\text {A}}   \,  \big| \, n_{\text{A}} \, m_{n_\text{A}} \, m_{\text{A}_1} \, m_{\text{A}_2}; \eta_{\text {A}}  \big\rangle = \delta_{m_{n_\text{A}}, m_{n_\text{A}}'} \, \delta_{m_{\text{A}_1}, m_{\text{A}_1}'} \, \delta_{m_{\text{A}_2}, m_{\text{A}_2}'}$,
$\big\langle n_{\text{B}} \, m_{n_\text{B}}' \, m_{\text{B}_1}' \, m_{\text{B}_2}' ; \eta_{\text {B}}  \, \big| \, n_{\text{B}} \, m_{n_\text{B}} \, m_{\text{B}_1} \, m_{\text{B}_2} ; \eta_{\text {B}}   \big\rangle = \delta_{m_{n_\text{B}}, m_{n_\text{B}}'} \, \delta_{m_{\text{B}_1}, m_{\text{B}_1}'} \, \delta_{m_{\text{B}_2}, m_{\text{B}_2}'}$.
Then
\begin{multline}
\sum_j P_{i \to j} =  \delta_{\eta_{\text {A}} \eta_{\text {B}}, +1} \times  \\
\sum_{m_{\text{A}_1}} \sum_{m_{\text{A}_2} \ge m_{\text{A}_1} }   \, \sum_{m_{\text{B}_1}} \sum_{m_{\text{B}_2} \le m_{\text{B}_1} }  \,
\big\langle n_{1}  \, m_{n_1}  \,  n_{2}  \, m_{n_2}  \big| \, \bigg( \sum_{ m_{n_\text{A}}} \sum_{m_{n_\text{B}}} \,  \big| {n_\text{A}} \, m_{n_\text{A}} \, {n_\text{B}} \, m_{n_\text{B}} \big\rangle   \,
 \big\langle {n_\text{A}} \, m_{n_\text{A}} \, {n_\text{B}} \, m_{n_\text{B}} \big|    \bigg)
\big| \, n_{1}  \, m_{n_1}  \,  n_{2}  \, m_{n_2} \big\rangle \\
  \big\langle d_{\text {AB}} \, \big| \,  m_{\text{A}_1}' \, m_{\text{B}_1}'  \big\rangle \, \big\langle  d_{\text {AB}} \, \big| \, m_{\text{A}_2}' \, m_{\text{B}_2}'  \big\rangle  \,
\big\langle m_{\text{A}_1} \, m_{\text{B}_1}  \, \big| \, d_{\text {AB}} \big\rangle \, \big\langle m_{\text{A}_2} \, m_{\text{B}_2} \, \big| \, d_{\text {AB}} \big\rangle \\ 
 \delta_{m_{n_\text{A}} + m_{n_\text{B}}, m_{l_r} - m_{l_p}}  
 \  \delta_{m_{\text{A}_1}+m_{\text{B}_1}, m_1} 
\ \delta_{m_{\text{A}_2}+m_{\text{B}_2}, m_1}  . 
\end{multline}
We use the closure relation
$\sum_{ m_{n_\text{A}}} \sum_{m_{n_\text{B}}} \,  \big|  m_{n_\text{A}} \,  m_{n_\text{B}} \big\rangle   \,
 \big\langle  m_{n_\text{A}} \, m_{n_\text{B}} \big| = 1 $ 
(the term $\delta_{m_{n_\text{A}} + m_{n_\text{B}}, m_{l_r} - m_{l_p}} $ disappears), so that
\begin{multline}
\sum_j P_{i \to j} =  \delta_{\eta_{\text {A}} \eta_{\text {B}}, +1} \times  
\sum_{m_{\text{A}_1}} \sum_{m_{\text{A}_2} \ge m_{\text{A}_1} }   \, \sum_{m_{\text{B}_1}} \sum_{m_{\text{B}_2} \le m_{\text{B}_1} }  \,
\big\langle n_{1}  \, m_{n_1}  \,  n_{2}  \, m_{n_2}  \big| \, {n_\text{A}} \, {n_\text{B}} \, \big\rangle   \,
 \big\langle {n_\text{A}}  \, {n_\text{B}} \big| \, n_{1}  \, m_{n_1}  \,  n_{2}  \, m_{n_2} \big\rangle \\
  \big\langle d_{\text {AB}} \, \big| \,  m_{\text{A}_1}' \, m_{\text{B}_1}'  \big\rangle \, \big\langle  d_{\text {AB}} \, \big| \, m_{\text{A}_2}' \, m_{\text{B}_2}'  \big\rangle  \,
\big\langle m_{\text{A}_1} \, m_{\text{B}_1}  \, \big| \, d_{\text {AB}} \big\rangle \, \big\langle m_{\text{A}_2} \, m_{\text{B}_2} \, \big| \, d_{\text {AB}} \big\rangle \\ 
 \delta_{m_{n_\text{A}} + m_{n_\text{B}}, m_{l_r} - m_{l_p}}  
 \  \delta_{m_{\text{A}_1}+m_{\text{B}_1}, m_1} 
\ \delta_{m_{\text{A}_2}+m_{\text{B}_2}, m_1}  . 
\end{multline}
Using the fact that $ \big\langle n_{1}  \, m_{n_1}  \,  n_{2}  \, m_{n_2}  \big| \, {n_\text{A}} \, {n_\text{B}} \, \big\rangle   \,
 \big\langle {n_\text{A}}  \, {n_\text{B}} \big| \, n_{1}  \, m_{n_1}  \,  n_{2}  \, m_{n_2} \big\rangle \equiv \big| \big\langle {n_\text{A}} \, {n_\text{B}} \big| \, n_{1}  \,  n_{2}  \big\rangle \big|^2 = P^\text{rot} $ (see arguments in Appendix A), we get at the end
\begin{eqnarray}
\sum_j P_{i \to j} =  P^\text{rot} \times \sum_j P_{i \to j}^\text{ns}  
\end{eqnarray}
with
\begin{multline}
\sum_j P_{i \to j}^\text{ns} = \delta_{\eta_{\text {A}} \eta_{\text {B}}, +1} \times \\
\sum_{m_{\text{A}_1}} \sum_{m_{\text{A}_2} \ge m_{\text{A}_1} }   \, \sum_{m_{\text{B}_1}} \sum_{m_{\text{B}_2} \le m_{\text{B}_1} }  \,
 \bigg|  \big\langle m_{\text{A}_1} \, m_{\text{B}_1}  \, \big| \, d_{\text {AB}} \big\rangle \, \big\langle m_{\text{A}_2} \, m_{\text{B}_2} \, \big| \, d_{\text {AB}} \big\rangle 
\bigg|^2  \,  \delta_{m_{\text{A}_1}+m_{\text{B}_1}, m_1} 
\, \delta_{m_{\text{A}_2}+m_{\text{B}_2}, m_1}  . 
\label{sumproba-indistcase}
\end{multline}
The previous derivation remains similar for the other cases,
as the element $\big\langle d_{\text {AA}} ; \eta_{\text {A}} \, \big| \, n_{\text{A}} \, m_{n_\text{A}} \, m_{\text{A}_1} \, m_{\text{A}_2}; \eta_{\text {A}}  \big\rangle  \, \big\langle d_{\text {BB}} ; \eta_{\text {B}} \, \big| \, n_{\text{B}} \, m_{n_\text{B}}  \, m_{\text{B}_1} \, m_{\text{B}_2} ; \eta_{\text {B}}   \big\rangle $ 
factorizes as well in front of the full expression. The same overall procedure is used
and this element disappears due to the closure relations. The case $m_1 \ne m_2$ in Eq.~\eqref{eqA3bis}
would give
\begin{multline}
\sum_j P_{i \to j}^\text{ns}   =  
\delta_{\eta_{\text {A}} \eta_{\text {B}}, \eta} \times 
\sum_{m_{\text{A}_1}} \sum_{m_{\text{A}_2} \ge m_{\text{A}_1} }   \, \sum_{m_{\text{B}_1}} \sum_{m_{\text{B}_2} \le m_{\text{B}_1} }  \,
  \frac{1}{{\Delta_\text{A}}} \, \frac{1}{{\Delta_\text{B}}} \, \frac{1}{{\Delta_d}} \\
\shoveright{     \bigg|   
   \big\langle m_{\text{A}_1} \, m_{\text{B}_1}  \, \big| \, d_{\text {AB}_1} \big\rangle \, \big\langle m_{\text{A}_2} \, m_{\text{B}_2} \, \big| \, d_{\text {AB}_2} \big\rangle   \bigg[ 1 + \eta_\text{A} \, \eta_\text{B} \, \eta  \bigg] 
\ \delta_{m_{\text{A}_1}+m_{\text{B}_1}, m_1} 
\ \delta_{m_{\text{A}_2}+m_{\text{B}_2}, m_2} } \\
\shoveright{     +    \big\langle m_{\text{A}_2} \, m_{\text{B}_2}  \, \big| \, d_{\text {AB}_1} \big\rangle \, \big\langle m_{\text{A}_1} \, m_{\text{B}_1} \, \big| \, d_{\text {AB}_2} \big\rangle   \bigg[ \eta_\text{A} \, \eta_\text{B} +\eta \bigg] 
\ \delta_{m_{\text{A}_2}+m_{\text{B}_2}, m_1} 
\ \delta_{m_{\text{A}_1}+m_{\text{B}_1}, m_2} } \\
\shoveright{  +     \big\langle m_{\text{A}_1} \, m_{\text{B}_2}  \, \big| \, d_{\text {AB}_1} \big\rangle \, \big\langle m_{\text{A}_2} \, m_{\text{B}_1} \, \big| \, d_{\text {AB}_2} \big\rangle   \bigg[  \eta_\text{B} + \eta_\text{A} \, \eta \bigg] 
\ \delta_{m_{\text{A}_1}+m_{\text{B}_2}, m_1} 
\ \delta_{m_{\text{A}_2}+m_{\text{B}_1}, m_2} } \\
 \shoveright{   +    \big\langle m_{\text{A}_2} \, m_{\text{B}_1}  \, \big| \, d_{\text {AB}_1} \big\rangle \, \big\langle m_{\text{A}_1} \, m_{\text{B}_2} \, \big| \, d_{\text {AB}_2} \big\rangle  \bigg[ \eta_\text{A} + \eta_\text{B} \, \eta \bigg]  
\ \delta_{m_{\text{A}_2}+m_{\text{B}_1}, m_1} 
\ \delta_{m_{\text{A}_1}+m_{\text{B}_2}, m_2}    \bigg|  ^2 ,  }  \\
\label{sumproba-generalcase}
\end{multline}
and the case of molecules with same values $m_1 = m_2$, indistinguishable or not, in Eq.~\eqref{eqA13}
would give
\begin{multline}
\sum_j P_{i \to j}^\text{ns}   =  
\delta_{\eta_{\text {A}} \eta_{\text {B}}, \eta} \times 
\sum_{m_{\text{A}_1}} \sum_{m_{\text{A}_2} \ge m_{\text{A}_1} }   \, \sum_{m_{\text{B}_1}} \sum_{m_{\text{B}_2} \le m_{\text{B}_1} }  \,
     \frac{1}{{\Delta_d}} \\
   \bigg|  
   \big\langle m_{\text{A}_1} \, m_{\text{B}_1}  \, \big| \, d_{\text {AB}_1} \big\rangle \, \big\langle m_{\text{A}_2} \, m_{\text{B}_2} \, \big| \, d_{\text {AB}_2} \big\rangle   
   + \eta \, \big\langle m_{\text{A}_2} \, m_{\text{B}_2}  \, \big| \, d_{\text {AB}_1} \big\rangle \, \big\langle m_{\text{A}_1} \, m_{\text{B}_1} \, \big| \, d_{\text {AB}_2} \big\rangle 
      \bigg|  ^2  \\
  \delta_{m_{\text{A}_1}+m_{\text{B}_1}, m_1} 
\ \delta_{m_{\text{A}_2}+m_{\text{B}_2}, m_1} . 
\label{sumproba-indistdistcase}
\end{multline}

\twocolumngrid

\end{document}